\documentclass[masters]{osudissert96}
\usepackage{epsf,psfrag,amsfonts,amsmath,amsthm,amssymb,latexsym,cite}
\newcommand{\vectornorm}[1]{\left|\left|#1\right|\right|} 
\newcommand{\field}[1]{\mathbb{#1}} 
\newcommand{\of}[1]{^{\scriptscriptstyle (#1)}}
\newcommand{\mc}[1]{\ensuremath{\mathcal{#1}}}

\newcommand{\TP}{\frac{T}{P}}
\newcommand{\tht}{\frac{\hat{\theta\of{l}}}{\theta\of{l}}}
\newcommand{\ahat}{\hat{a}\of{l}}
\newcommand{\tauhat}{\hat{\tau}\of{l}}
\newcommand{\al}{a\of{l}}
\newcommand{\taul}{\tau\of{l}}

\newcommand{\wl}{w\of{l}}

\newcommand{\deltal}{\delta\of{l}}
\newcommand{\zetal}{\zeta\of{l}}
\newcommand{\chil}{\chi\of{l}}
\newcommand{\chils}{\chi\of{l}_S}
\newcommand{\chilr}{\chi\of{l}_R}
\newcommand{\varthetal}{\vartheta\of{l}}
\newcommand{\varthetals}{\vartheta\of{l}_S}
\newcommand{\varthetalr}{\vartheta\of{l}_R}
\newcommand{\sigmal}{\sigma\of{l}}

\newcommand{\sigmals}{\sigma\of{l}_S}
\newcommand{\sigmalr}{\sigma\of{l}_R}
\newcommand{\phil}{\phi\of{l}}
\newcommand{\vlv}{v\of{l}_v}
\newcommand{\vlh}{v\of{l}_h}

\newcommand{\gl}{g\of{l}}

\newcommand{\fl}{f\of{l}}
\newcommand{\glk}{g\of{l}_k}

\newcommand{\gammal}{\gamma\of{l}}
\newcommand{\gammahat}{\hat{\gamma}\of{l}}

\newcommand{\thetal}{\theta\of{l}}

\newcommand{\betal}{\hat{\beta}\of{l}}

\newcommand{\lambdal}{\hat{\lambda}\of{l}}
\newcommand{\omegal}{\hat{\omega}\of{l}}

\newcommand{\alphal}{\alpha\of{l}}

\newcommand{\dl}{d\of{l}}
\newcommand{\dhat}{\hat{d}\of{l}}

\newcommand{\ssl}{s\of{l}}
\newcommand{\shat}{\hat{s}\of{l}}

\newcommand{\thetahat}{\hat{\theta}\of{l}}

\newcommand{\Cdopp}{\tilde{C}\of{l}}
\newcommand{\Ddopp}{\tilde{D}\of{l}}
\newcommand{\Edopp}{\tilde{E}\of{l}}
\newcommand{\deltaldopp}{\tilde{\delta}\of{l}}

\newcommand{\Bdopp}{B}

\newcommand{\gammadopp}{\tilde{\gamma}\of{l}}
\newcommand{\adopp}{\tilde{a}\of{l}}
\newcommand{\fdopp}{\tilde{f}\of{l}}
\newcommand{\sdopp}{\tilde{s}\of{l}}
\newcommand{\ddopp}{\tilde{d}\of{l}}
\newcommand{\thetadopp}{\tilde{\theta}\of{l}}
\newcommand{\ydopp}{\tilde{y}}
\newcommand{\eydopp}{\tilde{e}_y}
\newcommand{\metricdopp}{\frac{1}{M}\Expec{\vectornorm{\eydopp(t)}^2}}
\newcommand{\thtdopp}{\frac{\tilde{\theta}\of{l}}{\theta\of{l}}}
\newcommand{\ddoppc}{\tilde{d}^{(l)*}}
\newcommand{\sdoppc}{\tilde{s}^{(l)*}}
\newcommand{\betaldopp}{\tilde{\beta}\of{l}}
\newcommand{\lambdaldopp}{\tilde{\lambda}\of{l}}
\newcommand{\omegaldopp}{\tilde{\omega}\of{l}}

\newcommand{\shatc}{\hat{s}^{(l)*}}
\newcommand{\dlc}{d^{(l)*}}
\newcommand{\dhatc}{\hat{d}^{(l)*}}
\newcommand{\thetalc}{{\theta}^{(l)*}}
\newcommand{\thetahatc}{\hat{\theta}^{(l)*}}

\newcommand{\Expec}[1]{\text{E} \left\{ #1 \right\} }
\newcommand{\Expecl}[1]{\text{E} \left\{ #1 \right. }
\newcommand{\Expecr}[1]{\left. #1 \right\} }
\newcommand{\real}[1]{\text{Re} \left\{ #1 \right\} }
\newcommand{\reall}[1]{\text{Re} \left\{ #1 \right. }
\newcommand{\realr}[1]{\left. #1 \right\} }
 
\newcommand{\Real}{{\mathbb{R}}}

\newcommand{\abss}[1]{\left| #1 \right|^2}
\newcommand{\abs}[1]{\left| #1 \right|}

\newcommand{\metric}{\frac{1}{M}\Expec{\vectornorm{e_y(t)}^2}}
\newcommand{\psyms}{\{ b_n \}^{N/2-1}_{n=-N/2}}
\newcommand{\dsyms}{\{ c_m \}^{M/2-1}_{m=-M/2}}
\renewcommand{\vec}[1]{\ensuremath{\boldsymbol{#1}}}
 
 \DeclareMathOperator{\sinc}{sinc}

 \newcommand{\defn}{\triangleq}
 \renewcommand{\eqref}[1]{(\ref{eq:#1})}
 
 \newcommand{\Figref}[1]{Figure~\ref{fig:#1}}
 \newcommand{\figref}[1]{Fig.~\ref{fig:#1}}
 
 \newcommand{\secref}[1]{Section~\ref{sec:#1}}
 \newcommand{\appref}[1]{Appendix~\ref{app:#1}}

\newcommand{\chref}[1]{Chapter~\ref{ch:#1}}  

\begin{document}

\author{Brian Carroll}
\title{Analysis of Sparse Channel Estimation}
\unit{Graduate Program in Electrical Engineering}
\advisorname{Philip Schniter}
\member{Lee Potter}
\maketitle

\disscopyright

\begin{abstract}
Channel Estimation is an essential component in applications such as radar and data communication.  In multi path time varying environments, it is necessary to estimate time-shifts, scale-shifts (the wideband equivalent of Doppler-shifts), and the gains/phases of each of the multiple paths.  With recent advances in sparse estimation (or ``compressive sensing"), new estimation techniques have emerged which yield more accurate estimates of these channel parameters than traditional strategies.  These estimation strategies, however, restrict potential estimates of time-shifts and scale-shifts to a finite set of values separated by a choice of grid spacing.  A small grid spacing increases the number of potential estimates, thus lowering the quantization error, but also increases complexity and estimation time.  Conversely, a large grid spacing lowers the number of potential estimates, thus lowering the complexity and estimation time, but increases the quantization error.  In this thesis, we derive an expression which relates the choice of grid spacing to the mean-squared quantization error.  Furthermore, we consider the case when scale-shifts are approximated by Doppler-shifts, and derive a similar expression relating the choice of the grid spacing and the quantization error.  Using insights gained from these expressions, we further explore the effects of the choice and grid spacing, and examine when a wideband model can be well approximated by a narrowband model.
\end{abstract}





\tableofcontents
\listoffigures

\chapter{Introduction}
\label{ch:bg}
Accurate channel estimation is an important component of digital communication and radar.  For our research, we concentrate on the estimation of multi path channels with sparse impulse responses.  Such channels are encountered in many applications such as high definition television (HDTV), communication near a hilly terrain, and underwater acoustic communication near the surf zone \cite{C2002} \cite{L2007}.   Due to the sparse impulse responses of these channels, traditional estimation techniques such as least-squares result in over-parameterization and thus poor performance of the estimator \cite{C2007}.  Fortunately, the structure of these channels can be exploited using sparse reconstruction algorithms such as Matching Pursuit (MP) \cite{M1993} \cite{C2002}, Orthogonal Matching Pursuit (OMP) \cite{P1993} \cite{K2004},  Basis Pursuit (BP) \cite{C1998}, or Fast Bayesian Matching Pursuit (FBMP) \cite{S2009}.  Since these algorithms are better suited for the channels we consider, they give more accurate channel estimates than the traditional least squares method \cite{C2007}.
   
In this thesis, we narrow our focus to the case where the signals used for channel probing are wideband and the channel is rapidly time varying.  In this case, the narrowband Doppler approximation used by other authors may not be valid \cite{W1994} \cite{L2007}.   Our goal is to find the cost of using the narrowband Doppler approximation, as well gain insight on the error resulting from this estimation strategy.  In \chref{chan} we present a model for the channels we consider as well as a few channel examples. In \chref{estimation_strategy} we present detailed explanation of the estimation strategy we use.  In particular, we explain how in order to use a sparse reconstruction algorithm, a linearized model is constructed composed of either the effects of time-shifts and Doppler-shifts or the effects of time-shifts and scale-shifts (By scale-shifts, we mean the wideband equivalent of Doppler-shifts \cite{W1994}).  A design parameter of this linearized model is something we define as the grid spacing, which we describe in detail in \secref{grid}. In general, the larger the grid spacing, the fewer the number of potential estimates the sparse algorithm has to choose from.  In \chref{error} we derive an analytical expression which relates the estimation error to the choice of grid spacing when time-shifts and time-scales are used.  Then in \chref{doppler} we derive an analytical expression which relates the estimation error to the choice of grid spacing when time-shifts and Doppler-shifts are used.  With these two expressions, we gain insight on the effects of the Doppler approximation as well as the effects of the choice of grid spacing.  Finally, in \chref{matlab} we present computer simulations of the estimation strategy.  We focus on the application of digital communication, but we keep the problem general enough so that the results provide insight to radar systems as well.

\chapter{Channel Model} 
\label{ch:chan}
\begin{figure}[htbp]
	\begin{center}
		\psfrag{A}{$\tau\of{1}(t)$}
		\psfrag{B}{$\tau\of{2}(t)$}
		\psfrag{C}{$\tau\of{3}(t)$}
        \epsfxsize=5.5in
        \epsfbox{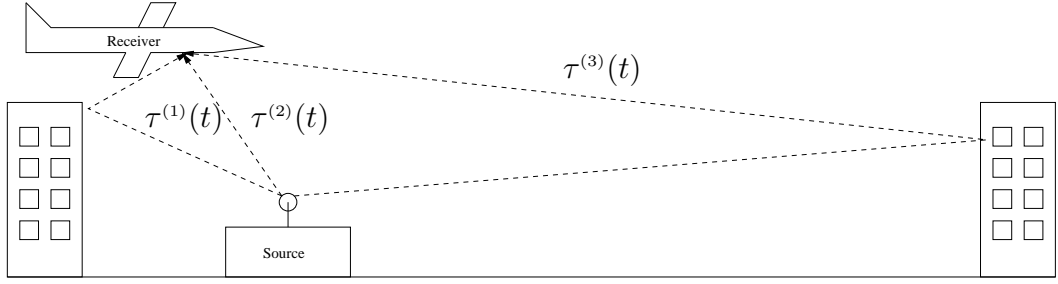}
    \end{center}
	\caption{Multi path channel example}
    \label{fig:channel}
\end{figure}

In general, the channels we consider are multi path channels with time varying delays and gains (see \Figref{channel} for a simple example).  We define the channel impulse response as follows
\begin{equation}
h(t,\tau) \defn \displaystyle\sum_{l=1}^{L} \fl(t) \delta(\tau-\tau\of{l}(t)), \label{eq:chanold}
\end{equation}
where $L$ is number of channel paths, $\taul(t)$ is the delay in seconds of the $l^{th}$ path at time $t$, and $\fl(t)$ is the complex gain of the $l^{th}$ path at time $t$.  Over short periods of time, $\taul(t)$ can be approximated by a first degree polynomial as follows
\begin{equation}
\taul(t) = \taul(0) +\al t. \label{eq:tau}
\end{equation} 
In \eqref{tau}, $\al$ is the rate of change of the channel delay $\frac{\partial \taul(t)}{\partial t}$, and $\taul(0)$ is the delay when $t=0$.
By denoting $c$ as the speed of propagation of the signal and $\chil(t)$ as the length of the $l^{th}$ path in meters, \eqref{tau} becomes
\begin{equation}
\taul(t) = \frac{\chil(t)}{c} = \frac{\chil(0)}{c} + \frac{\zetal}{c} t,
\end{equation}
where $\zetal$ is the rate at which the length of the $l^{th}$ path changes.  

One must be careful with joint estimation of the parameters $\taul(0)$ and $\al$.  As shown in \cite{G1998}, the parameters are coupled, and as a result, an estimation error in one parameter may result in an estimation error in the other.  This motivates us to use a different model for $\taul(t)$ composed of uncoupled parameters.  To do this, we will first note that at time $t$, the signal travels a distance of $c t$, while the path length $\chil(t)$ is equal to $\chil(0) + \zetal t$.  By setting these two distances equal and solving for $t$, we can denote $\gammal$ as the time it takes the signal to travel from the transmitter to the receiver along the $l^{th}$ path
\begin{eqnarray}
& &c t = \chil(0) + \zetal t |_{t=\gammal} \\
& &\Rightarrow \gammal = \frac{\chil(0)}{c-\zetal}. \label{eq:tau_2}
\end{eqnarray}
Using the fact that $\zetal = c \al$ and $\chil(t)= c \taul(t)$, we can rewrite \eqref{tau_2} as follows
\begin{equation}
\gammal = \frac{c \taul(0)}{c - c \al} =\frac{\taul(0)}{1-\al}. \label{eq:tau_3}
\end{equation}
Now by combining the results of \eqref{tau} and \eqref{tau_3}, we can model the time varying delay in the following way
\begin{equation}
\taul(t) = (1-\al) \gammal +\al t. \label{eq:taunew}
\end{equation}
Since it has been shown that there is no coupling between the parameters $\gammal$ and $\al$ in \cite{G1998}, the goal of the estimator will therefore be to find $\al$ (the rate of change of the delay) and $\gammal$ (the time it takes the signal to travel from the transmitter to the receiver).  We can now rewrite the channel impulse response \eqref{chanold} as
\begin{equation}
h(t,\tau) = \displaystyle\sum_{l=1}^{L} g\of{l}(t) \sqrt{1-\al} \delta(\tau-(1-\al)\gammal-\al t ). \label{eq:chan}
\end{equation}
By replacing $\fl$ with $\gl(t) \sqrt{1-\al}$ in \eqref{chan}, we in effect have normalized the model so that for any $\al$, the $l^{th}$ path is energy preserving when $\gl(t)=1$ \cite{W1994}.  Now we will present two examples of time varying channels to further illustrate this model.  
\section{Moving Receiver and Fixed Transmitter} \label{sec:chan_examp_1}
\begin{figure}[htbp]
	\begin{center}
		\psfrag{A}{$\vlv$}
		\psfrag{B}{$\vlh$}
		\psfrag{C}{$\varthetal(0)$}
		\psfrag{D}{$\chil(0)$}
		\psfrag{E}{$\phil(0)$}
		\psfrag{F}{$\sigmal(0)$}
        \epsfxsize=5.5in
        \epsfbox{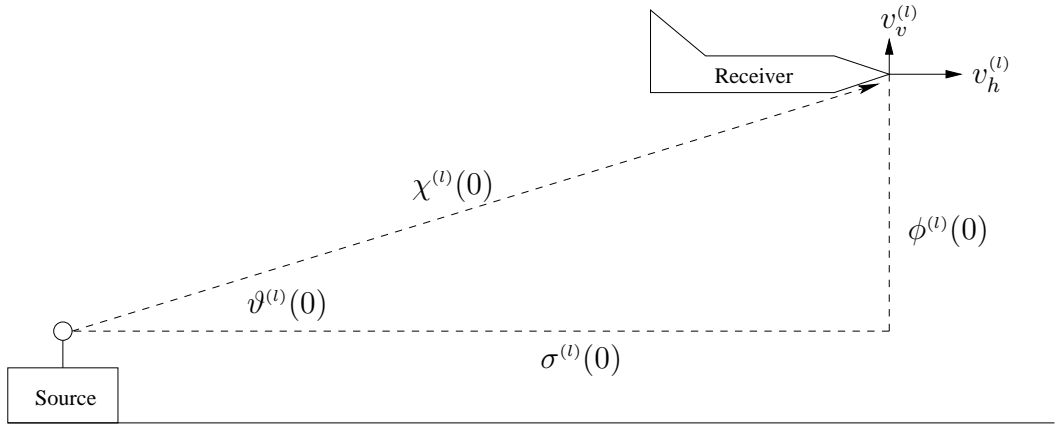}
    \end{center}
	\caption{Channel example (moving receiver and fixed transmitter)}
    \label{fig:chan_examp_1}
\end{figure}

The first example is a single path channel with a fixed transmitter and a moving receiver (see \Figref{chan_examp_1}).  In this example, the receiver is traveling with a vertical velocity of $\vlv$ and a horizontal velocity of $\vlh$.  The horizontal distance between the transmitter and the receiver is $\sigmal(t)$, and the vertical distance is $\phil(t)$.  The path delay, $\taul(t)$, can be expressed as follows
\begin{equation}
\taul(t) = \frac{\chil(t)}{c} = \frac{\sqrt{\sigma^{(l)2}(t)+\phi^{(l)2}(t)}}{c} = \frac{\sqrt{(\sigmal(0)+\vlh t)^2 + (\phil(0)+\vlv t)^2}}{c} \label{eq:tau_examp_1}
\end{equation}

We would like to approximate \eqref{tau_examp_1} to fit the model given in \eqref{taunew}.  To do this we will first take the derivative of \eqref{tau_examp_1} with respect to $t$.
\begin{equation}
\frac{\partial \taul(t)}{\partial t} = \frac{(\sigmal(0)+\vlh t)\vlh+(\phil(0)+\vlv t)\vlv}{ c \sqrt{(\sigmal(0)+\vlh t)^2+(\phil(0)+\vlv t)^2}}
\end{equation}
Now we will use the Taylor Series Approximation around $t=0$.  
\begin{eqnarray}
\taul(t) &\approx &\taul(0)+ \frac{\partial \taul(0)}{\partial t} \cdot t \\
& = & \frac{\chil(0)}{c} + \frac{\sigmal(0) \vlh+ \phil(0) \vlv}{c \sqrt{\sigma^{(l)2}(0)+\phi^{(l)2}}} \cdot t \\
& = & \frac{\chil(0)}{c} + \frac{\sigmal(0) \vlh+ \phil(0) \vlv}{c \chil(0)} \cdot t \\
& = & \frac{\chil(0)}{c} +\frac{\vlh \cos{\varthetal(0)}+ \vlv \sin{\varthetal(0)}}{c} \cdot t \label{eq:tau_examp_1_2}
\end{eqnarray}
Equation \eqref{tau_examp_1_2} follows from the facts that $\cos{\varthetal(t)} = \frac{\sigmal(t)}{\chil(t)}$ and $\sin{\varthetal(t)} = \frac{\phil(t)}{\chil(t)}$.  We can recognize that if we express $\al$ and $\gammal$ as follows
\begin{eqnarray}
\al &=& \frac{\vlh \cos{\varthetal(0)}+ \vlv \sin{\varthetal(0)}}{c} \\
\gammal &=& \frac{\chil(0)}{c(1-\al)},
\end{eqnarray}
then \eqref{tau_examp_1_2} fits the model given in \eqref{taunew}, with $\al$ being approximately the rate the channel delay of the $l^{th}$ path changes and $\gammal$ being approximately how long it takes the message to travel from the transmitter to the receiver along the $l^{th}$ path.  

\section{Underwater Acoustic Channel} \label{sec:chan_examp_2}
\begin{figure}[htbp]
	\begin{center}
		\psfrag{A}{$\vlv$}
		\psfrag{B}{$\vlh$}
		\psfrag{C}{$\varthetal_S(0)$}
		\psfrag{D}{$\phil(0)$}
		\psfrag{E}{$\chils(0)$}
		\psfrag{F}{$\chilr(0)$}
		\psfrag{G}{$\sigmals(0)$}
		\psfrag{H}{$\sigmalr(0)$}
		\psfrag{I}{$\varthetal_R(0)$}
        \epsfxsize=5.5in
        \epsfbox{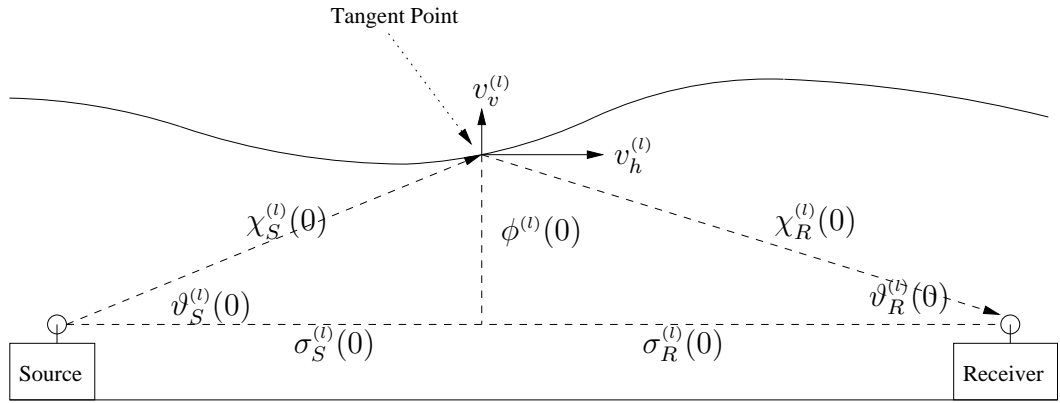}
    \end{center}
	\caption{Acoustic channel model}
    \label{fig:chan_examp_2}
\end{figure}

\begin{figure}[htbp]
	\begin{center}
		\psfrag{A}{$\vartheta\of{l}$}
		\psfrag{B}{$\vartheta\of{l}$}
        \epsfxsize=5.5in
        \epsfbox{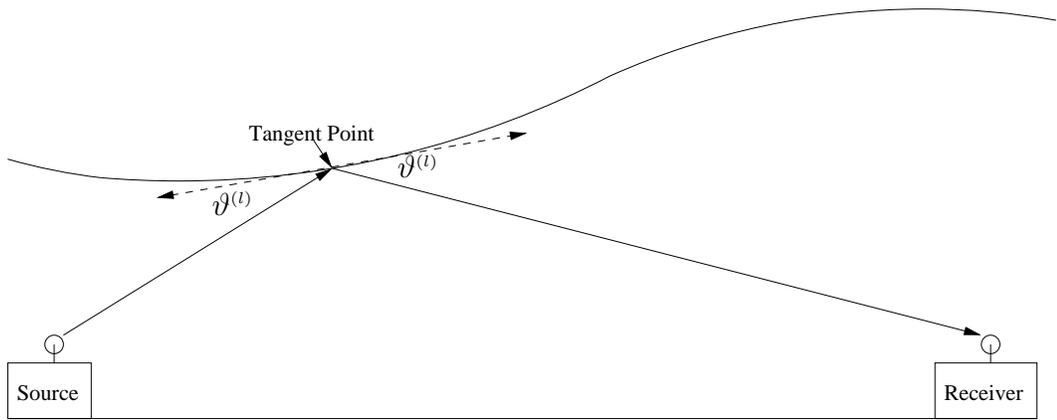}
    \end{center}
	\caption{Signal bouncing off the tangent point of the wave surface.}
    \label{fig:tan_point}
\end{figure}

Underwater acoustic channels near the surf zone provide a challenging application for sparse channel estimation techniques.  Such channels are especially challenging due to rapid fluctuations in per-path gain and delay \cite{P2004}.  An example of a single path in such a channel is given in \Figref{chan_examp_2}.  We define the tangent point (shown in \Figref{tan_point}) as the point  on the water surface where the signal bounces on its path to the receiver.  With the exception of the line of sight path, each of the other $L-1$ paths has its own unique tangent point.  As the water surface moves, the tangent points move as well.  We therefore expect each path delay to be time varying even if the transmitter and receiver are fixed. 

We now would like to find an expression for a single path delay $\taul(t)$ that fits the model given in \eqref{taunew}.  To aid us in this task, we will use the model drawn in \Figref{chan_examp_2}.  We assume that the tangent point moves with a horizontal velocity of $\vlh$ and a vertical velocity of $\vlv$.  We can therefore express the horizontal distance between the transmitter and the tangent point as $\sigmals(t)= \sigmals(0)+ \vlh t$, and the vertical distance as $\phil(t)=\phil(0)+\vlv t$.   Similarly, the horizontal distance between the tangent point and the receiver can be expressed as $\sigmalr(t)=\sigmalr(0)-\vlh t$.  We can thus express the channel delay as follows
\begin{eqnarray}
\taul(t) &=& \frac{\chil(t)}{c}  \\
&=& \frac{\chils(t)+\chilr(t)}{c}  \\
&=& \frac{\sqrt{{\sigmals}^2(t)+{\phil}^2(t)}+\sqrt{{\phil}^2(t)+{\sigmalr}^2(t)}}{c} \\
&=& \frac{\sqrt{(\sigmals(0)+\vlh t)^2+(\phil(0)+\vlv t)^2}}{c} \nonumber \\
& & + \mbox{ } \frac{\sqrt{(\phil(0)+\vlv t)^2+(\sigmalr(0)-\vlh t)^2}}{c}. \label{eq:tau2}
\end{eqnarray}
In order to approximate \eqref{tau2} to fit the model expressed in \eqref{taunew}, we will use the Taylor Series Approximation as was done in \secref{chan_examp_1}.
\begin{eqnarray}
\taul(t) &\approx &\taul(0)+ \frac{\partial \taul(0)}{\partial t} t \\
&= & \frac{\chils(0)}{c}+\frac{\chilr(0)}{c} \nonumber \\
& & + \mbox{ } \frac{\sigmals(0) \vlh + \phil(0) \vlv}{c \chils(0)}t+\frac{\phil(0) \vlv + \sigmalr(0) \vlh}{c \chilr(0)}t \\
&=& \frac{\chils(0)+\chilr(0)}{c} \nonumber \\
& & + \mbox{ } \frac{\cos{\varthetals(0)} \vlh+\sin{\varthetals(0)} \vlv+\sin{\varthetalr(0)}\vlv+\cos{\varthetalr(0)}v_h}{c} t \label{eq:tau_examp_2_2}
\end{eqnarray}
Equation \eqref{tau_examp_2_2} follows from the facts that $\cos{\varthetals(t)}=\frac{\sigmals(t)}{\chils(t)}$, $\sin{\varthetals(t)}=\frac{\phil(t)}{\chils(t)}$,\\ $\sin{\varthetalr(t)}=\frac{\phil(t)}{\chilr(t)}$, and $\cos{\varthetalr(t)}=\frac{\sigmalr(t)}{\chilr(t)}$.  We now recognize that if  $\al$ and $\gammal$ are expressed as follows
\begin{eqnarray}
\al &=& \frac{\cos{\varthetalr(0)} \vlh+\sin{\varthetalr(0)} \vlv+\sin{\varthetals(0)}\vlv+\cos{\varthetalr(0)}\vlh}{c} \\
\gammal &=& \frac{\chils(0)+\chilr(0)}{c(1-\al)},
\end{eqnarray}
then \eqref{tau_examp_2_2} fits the model given in \eqref{taunew}.  From the figures of experimental data collected in \cite{P2004}, the linear approximation does appear to be valid over short periods of time.

\chapter{Estimation Strategy} 
\label{ch:estimation_strategy}
The objective of the channel estimator is to estimate the unknown channel parameters $\{ \gammal$, $\al$, and $\gl(t) \}^L_{l=1}$ accurately.  We now present a general strategy to achieve this goal.  The first step, which is outlined in \secref{training}, is to send a training signal $x(t)$ through the channel.  In \secref{grid} we describe in detail how the system model is linearized in the unknown parameters $\{ \gammal$ and $ \al \}^L_{l=1}$.  Then in \secref{cs} we demonstrate how a sparse reconstruction algorithm can use the received training signal and the linearized model to estimate the unknown channel parameters.  For this chapter and \chref{error}, we assume that the linearized model is composed of the effects of time-shifts and scale-shifts.  Later in \chref{doppler}, we present modifications to this strategy to create a linearized model composed of the effects of time-shifts and Doppler-shifts instead.

\section{Training Sequence} \label{sec:training}
In order to estimate the channel, a sequence of $N$ pseudo-random QPSK pilot symbols is modulated, sent through the channel, and observed at the receiver.  By pseudo-random, we mean that the values the pilot symbols are chosen to be random \text{i.i.d.}, but they are known in advance at the receiver.  The $N$ pilot symbols, which we denote as $\{b_n \in ( \pm 1 \pm j ) / \sqrt{2} \}^{N/2-1}_{n=-N/2}$, are transmitted using single carrier modulation.  By using $T$ to represent the symbol period, $f_c$ to represent the carrier frequency, and $p(t)$ to represent a square-root raised-cosine pulse, the passband transmitted signal becomes
\begin{equation}
x(t)= \real{  e^{j 2 \pi f_c t} \displaystyle\sum_{n=-N/2}^{N/2-1} b_n p(t -n T) }.
\end{equation}
The transmitted signal is then sent through the channel and observed at the receiver.  The received passband signal $r(t)$ can be expressed as follows
\begin{eqnarray}
r(t) &=& \int x(\tau) h(t,t- \tau) \,d\tau + n(t) \\
	&=& \int x(\tau) \displaystyle\sum_{l=1}^{L} \fl(t) \delta(t-\tau-\tau\of{l}(t)) \,d\tau + n(t) \\
	&=& \displaystyle\sum_{l=1}^{L} \fl(t) x(t- \tau\of{l}(t)) +n(t) \\
	&=& \displaystyle\sum_{l=1}^{L} \fl(t) \real{ e^{j 2 \pi f_c (t-\taul(t) )} \displaystyle\sum_{n=-N/2}^{N/2-1} b_n p(t -\taul(t) -n T) } + n(t), \label{eq:rt}
\end{eqnarray}
where $n(t)$ denotes white Gaussian noise.  At the receiver, $r(t)$ is low pass filtered and converted to baseband.  We express the complex baseband received signal $z(t)$ as
\begin{eqnarray}
z(t) &=& \text{LPF} \left\{ 2 r(t) e^{-j 2 \pi f_c t} \right\} \\ 
&=& \displaystyle\sum_{l=1}^{L} \fl(t) \displaystyle\sum_{n=-N/2}^{N/2-1} b_n p(t-nT - \tau\of{l}(t) ) e^{-j 2 \pi f_c \tau\of{l}(t)} +w(t), \label{eq:z}
\end{eqnarray}
where $w(t)$ denotes the noise after down conversion and filtering.  Using equation \eqref{taunew}, the complex baseband received signal $z(t)$ becomes
\begin{eqnarray}
z(t) &=& \displaystyle\sum_{l=1}^{L} \gl(t) \sqrt{1-\al} \displaystyle\sum_{n=-N/2}^{N/2-1} b_n p((1-\al) t-nT - (1-\al) \gammal) \nonumber \\
& & \times e^{-j 2 \pi f_c ((1-\al)\gammal + \al t)} +w(t). \label{eq:zt}
\end{eqnarray}

In practice, the complex baseband received signal is sampled at $P$ times the symbol rate.  Over the estimation period, $G$ samples are yielded $\{ z_k \}^{G+H-1}_{k=H}$
\begin{eqnarray}
z_k &=& z(k \TP) \\
	&=& \displaystyle\sum_{l=1}^{L} \glk \sqrt{1-\al} \displaystyle\sum_{n=-N/2}^{N/2-1} b_n p((1-\al) k \TP-nT - (1-\al)\gammal)  \nonumber \\
 & & \times e^{-j 2 \pi f_c((1-\al)\gammal + \al k \TP)} +w_k, \label{eq:z2}
\end{eqnarray}
where $w_k = w(k \TP)$, $\glk=\gl(k \TP)$, and $H$ is the index of the first sample.  

We model $\gl(t)$ using an $N_b^{th}$ order basis expansion model so that
\begin{equation}
\gl(t) = \displaystyle\sum_{i=1}^{N_b} \theta\of{l}_i \rho_i(t),
\end{equation}
or in the discrete case
\begin{equation}
\glk = \displaystyle\sum_{i=1}^{N_b} \theta\of{l}_i \rho_{i,k}. \label{eq:bem}
\end{equation}
Equation \eqref{bem} can also be written in vector form as follows \label{eq:bem2}
\begin{equation}
\vec{\gl} = \displaystyle\sum_{i=1}^{N_b} \theta\of{l}_i \vec{\rho_i},
\end{equation}
where $\vec{\gl} $ and $\vec{\rho_i}$ are $G \times 1 $ vectors.  Using this basis expansion model, \eqref{z2} becomes
\begin{eqnarray}
z_k & = & \displaystyle\sum_{i=1}^{N_b} \displaystyle\sum_{l=1}^{L} \theta\of{l}_i \rho_{i,k} \sqrt{1-\al} \displaystyle\sum_{n=-N/2}^{N/2-1} b_n p((1-\al) k \TP-nT - (1-\al)\gammal) \nonumber \\
& & \times \mbox{} e^{-j 2 \pi f_c ((1-\al)\gammal + \al k \TP)} +w_k. \label{eq:z3}
\end{eqnarray}
\section{Linearization of the Model} \label{sec:grid}
Even though they are uncoupled, estimation of the unknown channel parameters $ \{ \gammal \}_{l=1}^L$ and $ \{ \al \}_{l=1}^L$ is still challenging, because they are nonlinearly related to the complex baseband received signal samples $\{ z_k \}^{G+H-1}_{k=H}$.  However, a linear approximation can be made by modeling $\{ z_k \}^{G+H-1}_{k=H}$ as the summation of contributions from many possible combinations of $(\gamma,a)$, but with only $L$ of these contributions being ``active" or non-zero.  In particular, we consider values of $\gamma$ and $a$ restricted to a uniform grid:
\begin{eqnarray}
\gamma_p \defn p \Delta_{\gamma} \label{eq:gammagrid} \\
a_q \defn q \Delta_a \label{eq:agrid}
\end{eqnarray}
where $p \in \{ N_0,...,N_0 + N_{\gamma}-1 \}  $ and $q \in \{ -N_a/2,...,N_a/2-1 \}$.  The linear model is constructed of these $N_{\gamma} N_a$ possible combinations of $\gamma_p$ and $a_q$.
Before describing the model further, it should be noted that $N_0$, $N_{\gamma}$, $N_a$, $\Delta_{\gamma}$, and $\Delta_a$ are chosen so that
\begin{eqnarray*}
N_0 \Delta_{\gamma} &\leq& \gamma_{min} \\
(N_0+N_{\gamma}-1) \Delta_{\gamma} &\geq& \gamma_{max} \\
-\frac{N_a}{2} \Delta_a &\leq & -a_{max} \\
(\frac{N_a}{2}-1)\Delta_a &\geq & a_{max}
\end{eqnarray*}
where $\gamma_{min}$ and $\gamma_{max}$ are the minimum and maximum values of $\gamma$ expected in the channel, respectively, and $a_{max}$ is the maximum value of $a$ expected in the channel.  Using the newly defined ``grid" of restricted values ($\left\{ \gamma_p \right\}^{N_0+N_{\gamma}-1}_{p=N_0}$ and $\left\{ a_q \right\}^{N_a/2-1}_{q=-N_a/2}$), we can now write the linearized \eqref{z3} as follows
\begin{eqnarray}
z_k &=& \displaystyle\sum_{p=N_0}^{N_0+N_{\tau}-1} \displaystyle\sum_{q=-N_a/2}^{N_a/2-1} \displaystyle\sum_{i=1}^{N_b} \theta_{p,q,i} \rho_{i,k} \sqrt{1-a_q}e^{-j 2 \pi f_c( (1-a_q)\gamma_p + a_q k \TP )} \nonumber \\
& &\times \mbox{ } \displaystyle\sum_{n=-N/2}^{N/2-1} b_n p((1-a_q) k \TP-nT - (1-a_q)\gamma_p) + a_q k \TP) +w_k-e_k, \label{eq:z4}
\end{eqnarray}
where $e_k$ is the error due to the linear approximation and $\theta_{p,q,i}$ is the channel gain corresponding to $\gamma_p$, $a_q$, and the basis vector $\vec{\rho_i}$.  We now define $\vec{\theta}$ as a $N_{\gamma} N_a N_b \times 1$ vector composed of all the values of $\theta_{p,q,i}$ concatenated together.  Equation \eqref{z4} can now be written as follows
\begin{equation}
\vec{z} = \vec{A} \vec{\theta} +\vec{w} -\vec{e}. \label{eq:zvect2}
\end{equation}
In \eqref{zvect2}, $\vec{A}$ is an appropriately defined  $G \times N_{\gamma} N_a N_b $ matrix, where each column of $\vec{A}$ corresponds to a particular $\gamma_p$, $a_q$, and $\vec{\rho_i}$.  The quantities $\vec{z}$, $\vec{w}$, $\vec{e}$ are $G\times 1$ vectors composed of the output samples, noise samples, and the linear approximation error samples, respectively.
It should be noted that since the channel has $L$ paths, we expect only $L N_b$ of the $N_{\gamma} N_a N_b$ elements in $\vec{\theta}$ to be non-zero.  Since $\vec{\theta}$ is mostly composed of zeros, we consider $\vec{\theta}$ ``sparse".  In \secref{cs}, we show how sparsity is used in the estimation of $\vec{\theta}$. 

\section{Channel Parameter Estimation via Sparse Reconstruction} \label{sec:cs}
Our interest now turns to the sparse algorithm which is used to estimate the unknown channel parameters $\vec{\theta}$ in \eqref{zvect2} using the complex baseband received signal $\vec{z}$.  In order to accomplish this estimation, the algorithm has complete knowledge, either implicitly or explicitly, of the $G \times N_{\gamma} N_a N_b$ matrix $\vec{A}$, because $\vec{A}$ is composed of pilot symbols known at the receiver.  A wide variety of sparse estimation algorithms can be used including Matching Pursuit \cite{M1993}, Orthogonal Matching Pursuit \cite{P1993}, Basis Pursuit \cite{C1998}, Greedy Basis Pursuit \cite{H2007} and Fast Bayesian Matching Pursuit \cite{S2009}.  As explained in the introduction, these algorithms have an advantage of being more accurate than traditional estimation methods.  This is because these algorithms exploit the fact that $\vec{\theta}$ is sparse (composed mostly of zeros).  Additionally, these algorithms have the advantage being able to work when the problem is underdetermined (i.e. $G < N_{\gamma} N_a N_b$).  

To allow further analysis of the estimation strategy, we now outline analytically how we expect the sparse estimator to behave.  We assume that the channel gains are time invariant (i.e. $N_b =1$ yielding the simpler notation $\gl(t) =\thetal$).  In this case, the estimator makes estimates of the channel parameters $\{ \gammal$, $\al, \thetal \}^L_{l=1}$.  We assume that $\sum_{l=1}^L |\thetal|^2 =1$ so that the channel is energy preserving.  We also assume that the number of significant paths the estimator finds $\hat{L}$ is equivalent to the true number of paths $L$.  For each of the $L$ estimated paths, parameter estimates are made which we denote as $\{\gammahat$, $\ahat$, and $\thetahat) \}^{L}_{l=1}$.  Using these parameter estimates, a reconstructed complex baseband received training signal $\hat{z}(t)$ is made, which is expressed as
\begin{eqnarray}
\hat{z}(t) &=& \displaystyle\sum_{l=1}^{L} \thetahat \sqrt{1-\ahat} \displaystyle\sum_{n=-N/2}^{N/2-1} b_n p((1-\ahat) t-nT - (1-\ahat) \gammahat) \nonumber \\
& & \times e^{-j 2 \pi f_c ((1-\ahat)\gammahat + \ahat t)}. \label{eq:recon_z}
\end{eqnarray}
We define the reconstruction error $e_z(t)$ as the difference between the actual complex baseband received training signal $z(t)$ and the reconstructed complex baseband received training signal $\hat{z}(t)$:
\begin{equation}
e_z(t) \defn \hat{z}(t)-z(t) = \displaystyle\sum_{l=1}^{L} \thetahat  \shat(t) -  \thetal \ssl(t) -w(t),
\end{equation}
where $\ssl(t)$ and $\shat(t)$ are defined as 
\begin{eqnarray}
\ssl(t) &\defn&  e^{-j 2 \pi f_c ((1-\al)\gammal+\al t)}  \sqrt{1-\al} \nonumber \\
& & \times \mbox{ } \displaystyle\sum_{n=-N/2}^{N/2-1} b_n p((1-\al)t -n T -(1-\al)\gammal)  \label{eq:sdef} \\ 
\shat(t) &\defn& e^{-j 2 \pi f_c ((1-\ahat)\gammahat+\ahat t)} \sqrt{1-\ahat} \nonumber \\
& & \times \mbox{ } \displaystyle\sum_{n=-N/2}^{N/2-1} b_n p((1-\ahat)t -n T -(1-\ahat) \gammahat). \label{eq:shatdef}
\end{eqnarray}
The quantity $\vectornorm{e_z(t)}^2$ can be upper bounded as follows 
\begin{eqnarray}
\vectornorm{e_z(t)}^2 &=& \vectornorm{\displaystyle\sum_{l=1}^{L} \thetahat  \shat(t) -  \thetal \ssl(t) -\wl(t)}^2 \\
&\leq& \displaystyle\sum_{l=1}^{L} \vectornorm{\thetahat  \shat(t) -  \thetal \ssl(t)}^2+ \vectornorm{w(t)}^2,\label{eq:thetahatub}
\end{eqnarray}
where the operator $\vectornorm{\mbox{ }}^2$ is defined as follows
\begin{equation}
\vectornorm{a(t)}^2 \defn \int \abss{a(t)} \,dt.
\end{equation}

We assume that the sparse estimation algorithm selects channel estimates\\ $\{ \gammahat , \ahat , \thetahat \}^L_{l=1}$ that approximately minimize $\vectornorm{e_z(t)}^2$.  In particular, we assume that for each channel path, $\gammahat$ and $\ahat$ are selected as the values closest to $\gammal$ and $\al$, respectively, on the uniform grid \eqref{gammagrid}-\eqref{agrid}.  Therefore, we can say that $\ahat$ and $\gammahat$ are at most half a grid space away from $\al$ and $\gammal$, respectively, or in other words, 
\begin{eqnarray}
|\ahat-\al| &\leq& \frac{\Delta_a}{2} \\
|\gammahat-\gammal| &\leq& \frac{\Delta_\gamma}{2}.
\end{eqnarray}

For each channel path, the component $\shat(t)$ is yielded from $\gammahat$ and $\ahat$.  Given this $\shat(t)$, we assume that the sparse estimation algorithm minimizes the upper bound of \eqref{thetahatub} by choosing each value of $\thetahat$ as follows
\begin{equation}
\thetahat =  \frac{\langle \ssl(t),\shat(t)\rangle \thetal} {\langle \shat(t),\shat(t) \rangle}, \label{eq:thetahat1_old} 
\end{equation}
where the operator $\langle \mbox{ } , \mbox{ } \rangle $ denotes the complex inner product
\begin{equation}
\langle a(t),b(t)\rangle \defn \int a(t) b^*(t) \,dt. \label{eq:innerprod}
\end{equation}
Equation \eqref{thetahat1_old} can be simplified further by noticing that $\langle \shat(t),\shat(t) \rangle$ reduces as follows
\begin{eqnarray}
\langle \shat(t),\shat(t) \rangle&=& \int \sqrt{1-\ahat} \displaystyle\sum_{r=-N/2}^{N/2-1} b_r p((1-\ahat)t -r T -(1-\ahat)\gammahat ) \nonumber \\
& &  \times \mbox{} \sqrt{1-\ahat} \displaystyle\sum_{q=-N/2}^{N/2-1} b_q^* p((1-\ahat)t -q T -(1-\ahat)\gammahat) \,dt \nonumber \\
&=& \frac{1-\ahat}{1-\ahat} \displaystyle\sum_{r=-N/2}^{N/2-1} \displaystyle\sum_{q=-N/2}^{N/2-1}  b_r b_q^* \int  p(v)   p(v+rT-qT) \,dv  \\
&=&  \displaystyle\sum_{r=-N/2}^{N/2-1} \displaystyle\sum_{q=-N/2}^{N/2-1}  b_r b_q^* R((q-r)T) \label{eq:rc_step}\\
&=& \displaystyle\sum_{r=-N/2}^{N/2-1} \displaystyle\sum_{q=-N/2}^{N/2-1} b_r b_q^* \delta_{r-q} \label{eq:dd_step} \\
&=& N,\label{eq:last_step}
\end{eqnarray}
where $R(\tau)$ is the raised-cosine impulse response and $\delta_{n}$ is the Kronecker delta function.  Equation \eqref{rc_step} follows from the fact that two square-root raised-cosine pulses integrate to raised-cosine pulse as follows $\int p(t) p(t-\tau) \,dt = R(\tau) $.  Equation \eqref{dd_step} follows from the fact that $R(\tau)$ is equal to $0$ when $\tau$ is a nonzero multiple of $T$, and equal to $1$ when $\tau=0$.  Finally, \eqref{last_step} follows from the fact $b_r b_q^* =1$ if $r=q$.  Plugging in the results of \eqref{last_step} into \eqref{thetahat1_old}, we find that
\begin{equation}
\thetahat =  \frac{\langle \ssl(t),\shat(t)\rangle \thetal} {N}. \label{eq:thetahat1} 
\end{equation}

\chapter{Estimation Error Using a Time-Shift/Scale-Shift Model} 
\label{ch:error}
\section{Error Metric} \label{sec:error_metric}
To analyze the performance of the estimator, we need to measure the estimation error.  There are many ways to do this, one of which is to directly compare the true channel parameters $\{ \taul , \al ,  \thetal \}^L_{l=1}$ with the parameter estimates $\{ \tauhat , \ahat , \thetahat \}^{\hat{L}}_{l=1}$.  However, since it is unclear how to assign the relative importance of these parameter estimation errors, this may not be the best way to proceed.  Instead, we measure the estimation error by evaluating how well on average the channel parameter estimates allow us to reconstruct a random data signal.  To do this, we consider a series of $M$ QPSK random i.i.d.\ data symbols $\{ c_n \}^{M/2-1}_{m=-M/2}$ modulated the the same way as the series of $N$ training symbols $\{ b_n \}^{N/2-1}_{n=-N/2}$.  A passband signal modulated from these data symbols travels through the same channel as the training symbols so that the actual baseband received data signal $y(t)$ can be written as
\begin{eqnarray}
y(t) &=&  \displaystyle\sum_{l=1}^{L} \thetal \sqrt{1-\al} \displaystyle\sum_{m=-M/2}^{M/2-1} c_m p((1-a\of{l}) t-nT - (1-\al)\gammal)  \nonumber \\
& & \times e^{-j 2 \pi f_c((1-\al)\gammal + \al t)}.
\end{eqnarray}
Using the parameter estimates, the reconstructed baseband received signal $\hat{y}(t)$ can be expressed as
\begin{eqnarray}
\textstyle \hat{y}(t) &=& \sum_{l=1}^{\hat{L}} \thetahat \sqrt{1-\ahat} \sum_{m=-M/2}^{M/2-1} c_m p((1-\ahat) t -nT - (1-\ahat )\gammahat ) \nonumber \\
& & \times \mbox{ } e^{-j 2 \pi f_c((1-\ahat)\gammahat + \ahat t)}.
\end{eqnarray}
We define $e_y(t)$ as the difference between the baseband received data signal and the reconstructed baseband received data signal
\begin{equation}
e_y(t) \defn \hat{y}(t)-y(t) ~=~\sum_{l=1}^{L} \thetahat  \dhat(t) - \thetal \dl(t), \label{eq:edata1} 
\end{equation}
where
\begin{eqnarray}
\dl(t) &\defn &  e^{-j 2 \pi f_c ((1-\al)\gammal+\al t)} \sqrt{1-\al} \nonumber \\ 
& & \times \mbox{ } \displaystyle\sum_{m=-M/2}^{M/2-1} c_m p((1-\al)t -n T -(1-\al) \gammal) \label{eq:ddef} \\
\dhat(t) &\defn & e^{-j 2 \pi f_c ((1-\ahat)\gammahat+\ahat t)} \sqrt{1-\ahat} \nonumber \\
& & \times \mbox{ } \displaystyle\sum_{m=-M/2}^{M/2-1} c_m p((1-\ahat)t -n T -(1-\ahat)\gammahat). \label{eq:dhatdef}
\end{eqnarray}

Since we are interested in how well the parameter estimates allow us to reconstruct an unknown data signal, we use $\metric$ as the error metric, where $\Expec{\mbox{ }}$ denotes the expected value.  The reason we divide by $M$ is because we expect $\vectornorm{e_y(t)}^2$ to increase as $M$ increases.  Thus, $\metric$ can be thought of as a metric of the estimation error per data symbol.  Generally the larger the value of $\metric$, the worse the estimator performed.  In the following sections, we find a relationship between the error metric $\metric$ and the grid spacing ($\Delta_a$ and $\Delta_{\gamma}$) defined in \secref{grid}.  We first examine the case when the channel is time invariant in \secref{error_lti}, while in \secref{error_tv} we consider the more complicated time varying case.  Then, in \secref{error_doppler}, we consider the resulting error when the narrowband approximation is used.

\section{Average Estimation Error of LTI Channels} \label{sec:error_lti}
The objective of this sections is to find a relationship between the grid spacing ($\Delta_a$ and $\Delta_{\gamma}$) described in \secref{grid} and the estimation error metric $\metric$ described in \secref{error_metric}.  Considering a LTI channel (as opposed to a time varying channel) greatly simplifies this problem.  In particular, the channel impulse response becomes
\begin{equation}
h(\tau)=\displaystyle\sum_{l=1}^{L} \thetal \delta(\tau-\gammal),
\end{equation}
and the complex baseband received training signal becomes
\begin{equation}
z(t)= \displaystyle\sum_{l=1}^{L} \thetal e^{-j 2 \pi f_c \gammal}  \displaystyle\sum_{n=-N/2}^{N/2-1} b_n p(t -n T -\gammal) +w(t). \label{eq:z_lti}
\end{equation} 
Additionally, in the time invariant case, \eqref{ddef} and \eqref{dhatdef} can be simplified as follows
\begin{eqnarray}
\dhat(t)&=& e^{-j 2 \pi f_c \gammahat}  \displaystyle\sum_{m=-M/2}^{M/2-1} c_m p(t -n T -\gammahat) \label{eq:dhatlti} \\
\dl(t) &=&  e^{-j 2 \pi f_c \gammal}  \displaystyle\sum_{m=-M/2}^{M/2-1} c_m p(t -n T  -\gammal). \label{eq:dlti}
\end{eqnarray}
From \eqref{edata1}, if follows that the error metric can be upper bounded as follows
\begin{eqnarray}
\metric &=&\frac{1}{M}\Expec{\vectornorm{\displaystyle\sum_{l=1}^{L} \hat{\theta}\of{l} \hat{d}\of{l}(t) - \theta\of{l} d\of{l}(t) }^2} \\
&\leq & \frac{1}{M} \displaystyle\sum_{l=1}^{L} \Expec{\vectornorm{\hat{\theta}\of{l}  \hat{d}\of{l}(t) - \theta\of{l} d\of{l}(t)}^2}. \label{eq:enorm}
\end{eqnarray}

Instead of solving for $\metric$ in terms of $\Delta_{\gamma}$ directly, we instead solve for the upper bound in \eqref{enorm}, thus finding an upper bound on the error metric.  Focusing on a particular value of $l \in \{ 1,2,...,L \}$, we find (as shown in \appref{3to1}) 
\begin{eqnarray}
\frac{1}{M}\Expec{\vectornorm{\thetahat  \dhat(t) - \thetal \dl(t)}^2} 
&=& \frac{\abss{\thetal}}{M} \Expec{\int \abss{\tht} \abss{\dhat(t)} \,dt} \nonumber \\
& & -\mbox{} \frac{2\abss{\thetal}}{M} \Expec{\int \real{ \tht \dhat(t) \dlc(t)  } \,dt} \nonumber \\
& & +\mbox{} \frac{\abss{\thetal}}{M} \Expec{\int \abss{\dl(t)} \,dt}. \label{eq:1error}
\end{eqnarray}
We now evaluate each of the three terms in \eqref{1error} separately, and then combine the results. 

The first term of \eqref{1error} becomes
\begin{eqnarray}
\lefteqn{\frac{\abss{\thetal}}{M}E \left\{ \int \abss{\tht} \abss{\dhat(t)} \,dt \right\} } \nonumber \\
&=& \frac{\abss{\thetal}}{M}\Expec{\abss{\tht} } \Expec{\int \abss{\dhat(t)} \,dt } \label{eq:expec_split_x} \\
&=& \frac{\abss{\thetal}}{M} \Expec{ \frac{\abss{ \langle \ssl(t) , \shat(t) \rangle }}{\abss{N}}   } \nonumber \\
& &\times \Expec{ \int \displaystyle\sum_{r=-M/2}^{M/2-1} c_r p(t -r T -\gammahat) \displaystyle\sum_{q=-M/2}^{M/2-1} c_q^* p(t -q T -\gammahat) \,dt }. \label{eq:first_term_2}
\end{eqnarray}
In \eqref{expec_split_x}, we were able to split the expectation into the product of two expectations, because the values of the pilot symbols $\psyms$ are independent of the values of the data symbols $\dsyms$.  To simplify the second expectation in \eqref{first_term_2}, we note that, as was explained in \secref{cs}, the integral of two square-root raised-cosine pulse can be reduced as follows
\begin{equation}
\int  p(t -r T -\gammahat) p(t -q T -\gammahat) \,dt =  R((q-r)T)=  \delta_{r-q}. \label{eq:trick2}
\end{equation}
Using \eqref{trick2}, we can simplify the following expression 
\begin{eqnarray}
\lefteqn{ \int \displaystyle\sum_{r=-M/2}^{M/2-1} c_r p(t -r T -\gammahat) \displaystyle\sum_{r=-M/2}^{M/2-1} c_q^* p(t -q T -\gammahat) \,dt  } \nonumber \\
&=& \displaystyle\sum_{r=-M/2}^{M/2-1} \displaystyle\sum_{r=-M/2}^{M/2-1} c_r c_q^* \int p(t -r T -\gammahat) p(t -q T -\gammahat) \,dt  \\
&=& \displaystyle\sum_{r=-M/2}^{M/2-1} \displaystyle\sum_{r=-M/2}^{M/2-1} c_r c_q^* \delta_{r-q} \\
&=&  M, \label{eq:right_half}
\end{eqnarray}
where \eqref{right_half} follows from the fact that $c_r c_q^* =1$ if $r=q$.  From \eqref{right_half}, it follows that the second expectation of \eqref{first_term_2} is equal to $M$.  Before reducing the first expectation of \eqref{first_term_2}, we note that $\ssl(t)$ and $\shat(t)$ (defined in \eqref{sdef} and \eqref{shatdef}, respectively) can be reduced as follows for the case of a LTI channel 
\begin{eqnarray}
\ssl(t)=  e^{-j 2 \pi f_c \gammal} \displaystyle\sum_{n=-N/2}^{N/2-1} b_n p(t -n T -\gammal)  \\ 
\shat(t)= e^{-j 2 \pi f_c \gammahat} \displaystyle\sum_{n=-N/2}^{N/2-1} b_n p(t -n T - \gammahat).
\end{eqnarray}
Keeping this in mind, we express the first expectation of \eqref{first_term_2} as
\begin{eqnarray}
\lefteqn{\Expec{ \frac{\abss{ \langle \ssl(t), \shat(t) \rangle }}{\abss{N}}} } \nonumber \\
&=& \frac{1}{N^2}\displaystyle\sum_{q,r,m,n=-N/2}^{N/2-1} \Expecl{ b_r b_q^* b_n^* b_m  \int p(t -rT -\gammal) p(t-qT -\gammahat) \,dt } \nonumber   \\
& &  \Expecr{ \times \mbox{} \left\{ \int p(t -nT -\gammal) p(t-mT -\gammahat) \,dt \right\}^* } \\
&=& \frac{1}{N^2}\displaystyle\sum_{q,r,m,n=-N/2}^{N/2-1} \nonumber \\
& &  \Expec{ b_r b_q^* b_n^* b_m R((q-r)T +\gammal - \gammahat ) R((m-n)T +\gammal - \gammahat) }. \label{eq:left_half}
\end{eqnarray}
To simplify \eqref{left_half} further, the expected value of the product of four random variables $ b_r b_q^* b_n^* b_m$ has to be evaluated. To do this, we note that because the pilot symbols $\{b_n \}^{N/2-1}_{n=-N/2}$ are pseudo-random QPSK i.i.d.\ symbols, $\Expec{b_n}= 0$, $\Expec{b_n b_m} = 0$, and $\Expec{b_n b_m^*} = \delta_{n-m}$.  It therefore follows that $\Expec{b_r b_q^* b_n^* b_m}$ is equal to $1$ if $q=r=m=n$, if $q =r \neq m = n$, or if $q=m \neq r =n $.  In all other cases $\Expec{b_q^* b_r b_m b_n^*}$ is equal to zero.  Also notice that in \eqref{left_half}, the case $q=r=m=n$ occurs $N$ times, the case $q =r \neq m = n$ occurs $N(N-1)$ times, and the case $q=m \neq r =n $ occurs $N(N-1)$ times.  Therefore, \eqref{left_half} can be reduced as follows 
\begin{eqnarray}
\lefteqn{\frac{1}{N^2}\displaystyle\sum_{q,r,m,n=-N/2}^{N/2-1} \Expec{  b_r b_q^* b_n^* b_m R((q-r)T +\gammal - \gammahat ) R((m-n)T +\gammal - \gammahat) }  } \nonumber \\
&=& \frac{1}{N^2} \{ N R^2(\gammal-\gammahat) +N(N-1) R^2 ( \gammal - \gammahat)  \} +\alphal \\ 
&=& \frac{1}{N^2} \{ N^2 R^2(\gammal-\gammahat)  \} +\alphal  \\
&=& R^2(\gammal-\gammahat) +\alphal, \label{eq:left_half_3}
\end{eqnarray}
where
\begin{equation}
\alphal \defn \frac{1}{N^2}\sum_{q=-N/2}^{N/2-1} \sum_{q \neq r=-N/2}^{N/2-1} R^2((q-r)T+\gammal-\gammahat). \label{eq:alpha}
\end{equation}
By combining the results of \eqref{right_half} and \eqref{left_half_3}, we find that first term of \eqref{1error} can be expressed as
\begin{equation}
\frac{\abss{\thetal}}{M} \Expec{\int \abss{\tht} \abss{\dhat(t)} \,dt} = \abss{\thetal} R^2(\gammal-\gammahat) + \abss{\thetal} \alphal. \label{eq:first_term_final}
\end{equation}

We now simplify the second term of \eqref{1error} using similar arguments.
\begin{eqnarray}
\lefteqn{ \frac{2\abss{\thetal}}{M} \Expec{ \int \real{ \tht \dhat(t) \dlc(t) } \,dt} } \\
&=& \frac{2\abss{\thetal}}{M} \real{ \Expec{ \tht } \Expec{ \int \dhat(t) \dlc(t)  \,dt  } } \label{eq:expec_split}\\
&=& \frac{2\abss{\thetal}}{M N} \real{ \Expec{  \langle \ssl(t),\shat(t) \rangle   } \Expec{\int  \dhat(t) \dlc(t)  \,dt }  } \\
&=& \frac{2\abss{\thetal}}{M N} \reall{ \Expec{  \displaystyle\sum_{m,n=-N/2}^{N/2-1}  b_m b_n^* R((m-n)T + \gammal -\gammahat)  } } \nonumber \\
& & \realr{ \times \mbox{} \Expec{  \displaystyle\sum_{q,r=-M/2}^{M/2-1} c_q c_r^* R((q-r)T + \gammal -\gammahat)  } } \\
&=& \frac{2\abss{\thetal} M N}{M N}  R(\gammal -\gammahat)  R(\gammal -\gammahat) \label{eq:second_termxx} \\
&=& 2\abss{\thetal} R^2(\gammal -\gammahat) \label{eq:second_term_final} 
\end{eqnarray}
Equation \eqref{second_termxx} follows from the facts that $\Expec{c_q c_r^*} = \delta_{q-r}$, $\Expec{b_m b_n^*} = \delta_{m-n}$, and $R(\tau)$ is real for all $\tau$.  

Using the arguments we used to reduce the first two terms \eqref{1error}, the third term of \eqref{1error} reduces as follows
\begin{equation}
\frac{\abss{\thetal}}{M} \Expec{\int \abss{\dl(t)} \,dt} = \frac{\abss{\thetal}}{M} \Expec{\displaystyle\sum_{q,r} c_q c_r^* R((q-r)T)} = \frac{M\abss{\thetal}}{M} = \abss{\thetal}. \label{eq:third_term_final}
\end{equation}

\begin{figure}[htbp]
	\begin{center}
		\psfrag{x01}{$0$}
		\psfrag{x02}[c][c][1][0]{$0.05$}
		\psfrag{x03}[c][c][1][0]{$0.10$}
		\psfrag{x04}[c][c][1][0]{$0.15$}
		\psfrag{x05}[c][c][1][0]{$0.20$}
		\psfrag{x06}[c][c][1][0]{$0.25$}
		\psfrag{x07}[c][c][1][0]{$0.30$}
		\psfrag{x08}[c][c][1][0]{$0.35$}
		\psfrag{x09}[c][c][1][0]{$0.40$}
		\psfrag{x10}[c][c][1][0]{$0.45$}
		\psfrag{x11}[c][c][1][0]{$0.50$}
		\psfrag{v01}[r][c][1][0]{ }
		\psfrag{v02}[r][c][1][0]{$0.05$}
		\psfrag{v03}[r][c][1][0]{$0.10$}
		\psfrag{v04}[r][c][1][0]{$0.15$}
		\psfrag{v05}[r][c][1][0]{$0.20$}
		\psfrag{v06}[r][c][1][0]{$0.25$}
		\psfrag{s03}{}
		\psfrag{s04}[c][b][1][0]{$\frac{\Delta_{\gamma}}{T}$ }

        \epsfxsize=4.0in
        \epsfbox{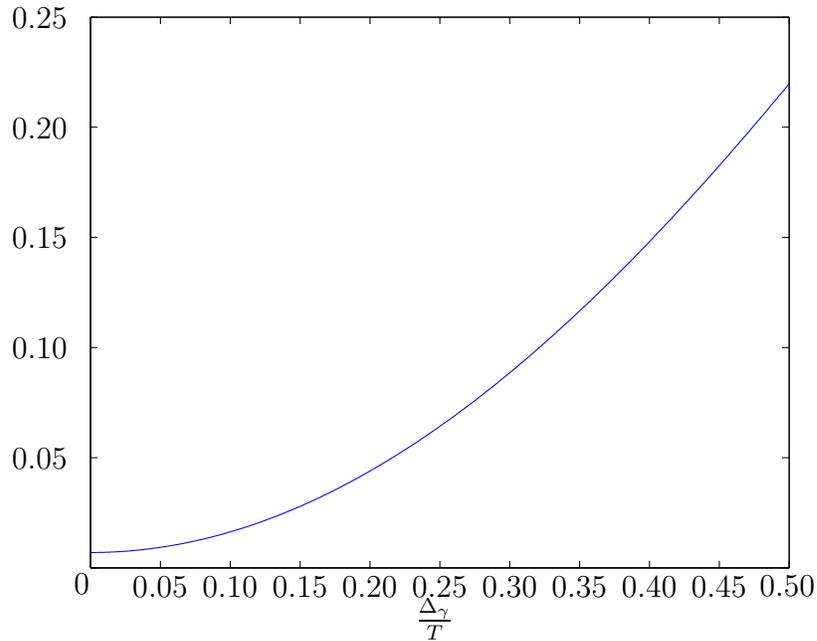}
    \end{center}
	\caption{Plot of $1-R^2( \frac{\Delta_{\gamma}}{2})+\frac{0.694}{N}$ vs $\frac{\Delta_{\gamma}}{T}$ }
    \label{fig:error_vs_delta_tau}
\end{figure}

Finally, combining the results of \eqref{first_term_final}, \eqref{second_term_final}, and \eqref{third_term_final} into \eqref{1error}, we find that
\begin{equation} 
\frac{1}{M} \Expec{ \vectornorm{\thetahat  \dhat(t) - \thetal \dl(t) }^2 } = \abss{\thetal} (1-R^2(\gammal-\gammahat) +\alphal ). \label{eq:one_branch_lti}
\end{equation}
In \appref{alphabound}, it is shown that by assuming $\abs{ \gammahat-\gammal} \leq \frac{T}{4}$, $\alphal$ can be upper bounded as follows, $\alphal \leq \frac{0.694}{N}$.  Using this fact and keeping in mind the channel is energy preserving $\sum_{l=1}^L | \thetal|^2=1 $, we can combine \eqref{one_branch_lti} with \eqref{enorm} to find
\begin{eqnarray}
\metric &\leq & \displaystyle\sum_{l=1}^{L} \abss{\thetal} \left( 1-R^2(\gammal-\gammahat ) +\frac{0.694}{N} \right)  \\
&=& 1-R^2(\gammal-\gammahat ) +\frac{0.694}{N}.  \label{eq:almost_bound_lti}
\end{eqnarray}
Recalling our earlier assumptions that  $ |\gammal-\gammahat | \leq \frac{\Delta_{\gamma}}{2}$ (assumed in \secref{cs}) and $ |\gammal-\gammahat | \leq \frac{T}{4}$ (assumed in \appref{alphabound}), it follows that $\Delta_{\gamma}$ must be selected so that $\Delta_{\gamma} \leq \frac{T}{2}$.  Keeping in mind that $R(x)=R(|x|)$ and $R(|x|)$ decreases as $|x|$ increases for $|x| \leq \frac{T}{2}$, \eqref{almost_bound_lti} can be upper bounded as follows
\begin{eqnarray}
\metric &\leq &  1-R^2(\gammal-\gammahat ) +\frac{0.694}{N}  \\
&\leq & 1-R^2( \frac{\Delta_{\gamma}}{2})+\frac{0.694}{N}.  \label{eq:error_result}
\end{eqnarray}
This is the expression we were looking for to relate the choice of grid spacing $\Delta_{\gamma}$ with the estimation error metric $\metric$.  \figref{error_vs_delta_tau} shows a plot of $1-R^2( \frac{\Delta_{\gamma}}{2} )+\frac{0.694}{N}$ vs. $\Delta_{\gamma}$ on a $T$ normalized scale, when $N=100$. As can be seen, the finer the grid spacing, the lower the average estimation error. 

\subsection{Simulation of LTI Channel Estimation Error}

\begin{figure}[htbp]
	\begin{center}
		\psfrag{x01}{$0$}
		\psfrag{x02}[c][c][1][0]{$0.05$}
		\psfrag{x03}[c][c][1][0]{$0.10$}
		\psfrag{x04}[c][c][1][0]{$0.15$}
		\psfrag{x05}[c][c][1][0]{$0.20$}
		\psfrag{x06}[c][c][1][0]{$0.25$}
		\psfrag{x07}[c][c][1][0]{$0.30$}
		\psfrag{x08}[c][c][1][0]{$0.35$}
		\psfrag{x09}[c][c][1][0]{$0.40$}
		\psfrag{x10}[c][c][1][0]{$0.45$}
		\psfrag{x11}[c][c][1][0]{$0.50$}
		\psfrag{v01}{$ $}
		\psfrag{v02}[r][c][1][0]{$0.05$}
		\psfrag{v03}[r][c][1][0]{$0.10$}
		\psfrag{v04}[r][c][1][0]{$0.15$}
		\psfrag{v05}[r][c][1][0]{$0.20$}
		\psfrag{v06}[r][c][1][0]{$0.25$}
		\psfrag{s10}{}
		\psfrag{s11}{}
		\psfrag{s13}{Upper Bound}
		\psfrag{s14}{$\gammahat=\gammal+\Delta_{\gamma}/2$}
		\psfrag{s15}{$\gammahat=\gammal+\Delta_{\gamma}/3$}
		\psfrag{s16}{$\gammahat=\gammal+\Delta_{\gamma}/4$}
		\psfrag{s17}{$\gammahat=\gammal+\Delta_{\gamma}/5$}
		\psfrag{s05}[c][b][1][0]{$\frac{\Delta_{\gamma}}{T}$}
		\psfrag{s06}[c][c][1][0]{Estimation Error $\frac{1}{M}\vectornorm{e_y(t)}^2$ vs. $\frac{\Delta_{\gamma}}{T}$}
        \epsfxsize=4.0in
        \epsfbox{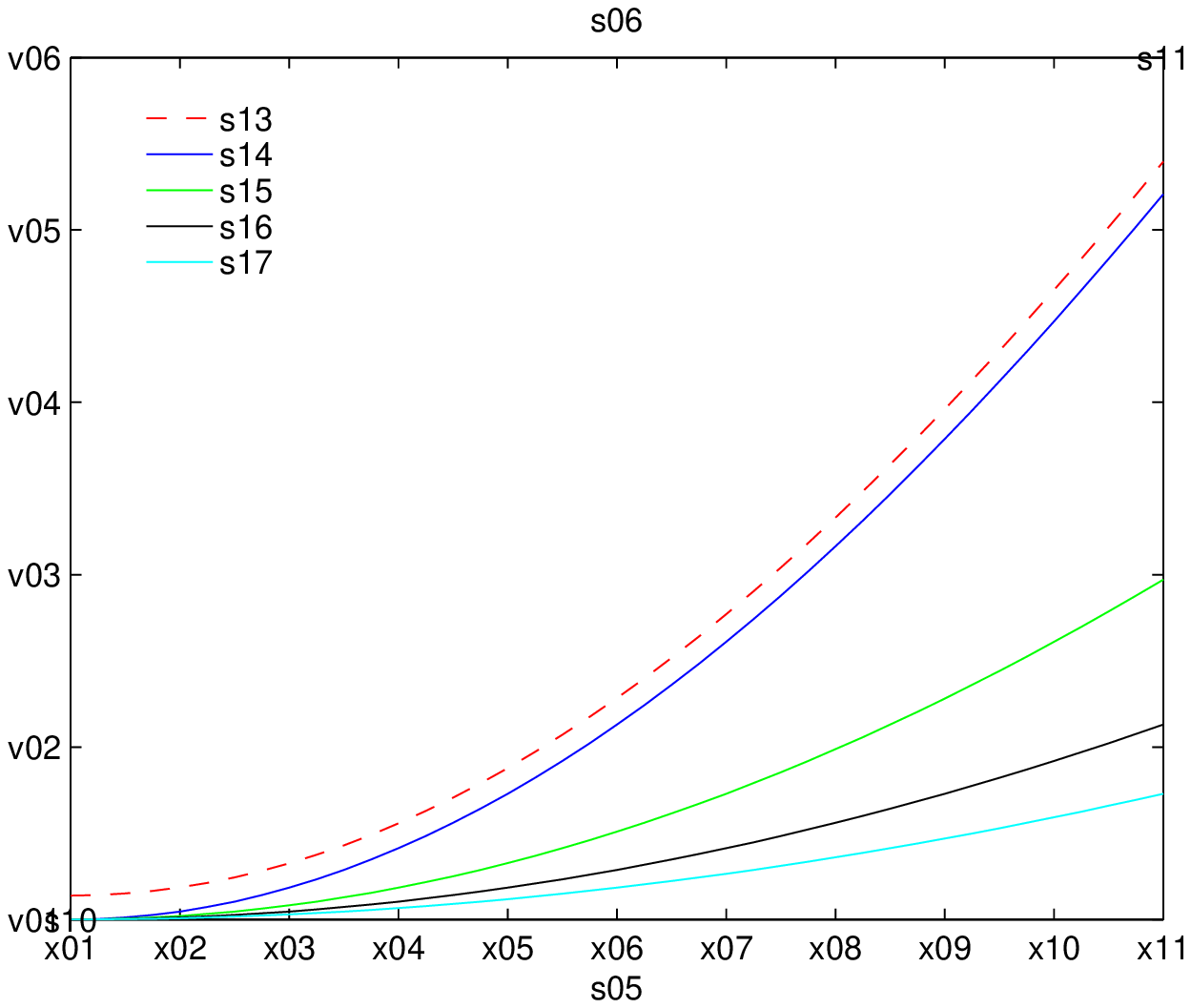}
    \end{center}
	\caption{Derived upper bound and average simulated error for the LTI case}
    \label{fig:error_lti}
\end{figure}

To verify the results in \eqref{error_result}, we ran a simulation in Matlab of a one path LTI channel using a carrier frequency of 18 kHz, a symbol rate of 24 kHz, and a sampling rate of 72 kHz (3 times the symbol rate).  The channel had unity gain and a time invariant delay of 7 ms.  The simulation followed what was outlined in \chref{estimation_strategy}.  Specifically, a signal composed of a series of 100 training symbols ($N=100$) was sent through the channel and an estimate of $\theta\of{1}$ was made based on \eqref{thetahat1}.  After that a signal composed of $100$ data symbols ($M=100$) was sent through the channel.  The error per data symbol $\frac{1}{M}\vectornorm{e_y(t)}^2$ was calculated for 4 different estimates.  The simulation was repeated 30 times and the average results are shown in \figref{error_lti}.  As expected, the estimation error increases when the the grid spacing $\Delta_{\gamma}$ increases.  It can also be seen that the upper bound of \eqref{error_result} was satisfied for all values of $\Delta_{\gamma}$

\section{Average Estimation Error of Time Varying Channels} \label{sec:error_tv}
In \secref{error_lti}, we found a relationship between the estimation error metric \\$\metric$ and the grid spacing $\Delta_{\gamma}$ for the case of linear time invariant channels.  We now consider channels with path delays that vary with time (i.e. $\al$ may not be equal to zero for $l \in \{1,2,...,L \}$ ).  We still assume that the channel path gains are time invariant (i.e. $\gl(t) = \thetal$ for $l=1,2,...L$), the channel is energy preserving $\sum_{l=1}^L |\thetal|^2 =1$,  and that the number of estimated paths $\hat{L}$ is equal to the true number of paths $L$.  Additionally, we now assume that the number pilot symbols $N$ is equal to the number of data symbols $M$.   

First we recall, as noted in \secref{error_lti}, that the error metric can be upper bounded as follows
\begin{equation}
\metric \leq \frac{1}{M} \displaystyle\sum_{l=1}^{L} \Expec{ \vectornorm{\thetahat  \dhat(t) - \thetal \dl(t) }^2 } \label{eq:error_sum_tv}.
\end{equation}
As we did in \secref{error_lti}, we evaluate this upper bound of instead of evaluating $\metric$ directly.  We again focus on a particular value of $l \in \{ 1,2,...,L \}$ and use the fact that (as shown in \appref{3to1}) 
\begin{eqnarray}
\frac{1}{M}\Expec{\vectornorm{\thetahat  \dhat(t) - \thetal \dl(t)}^2} 
&=& \frac{\abss{\thetal}}{M} \Expec{\int \abss{\tht} \abss{\dhat(t)} \,dt} \nonumber \\
& & -\mbox{} \frac{2\abss{\thetal}}{M} \Expec{\int \real{ \tht \dhat(t) \dlc(t) } \,dt } \nonumber \\
& & +\mbox{} \frac{\abss{\thetal}}{M} \Expec{\int \abss{\dl(t)} \,dt}. \label{eq:2error}
\end{eqnarray}
Now we evaluate the 3 terms in \eqref{2error} separately and later combine the results.  

Using \eqref{thetahat1} and the assumption $N=M$, the first term of \eqref{2error} can be expressed as follows
\begin{eqnarray}
\lefteqn{ \frac{\abss{\thetal}}{M} \Expec{\int \abss{\tht} \abss{\dhat(t)} \,dt} } \nonumber  \\
&=& \frac{\abss{\thetal}}{M} \Expec{\abss{\tht} } \Expec{\int \dhat(t) \dhatc(t)  \,dt }  \label{eq:2first_term}\\
&=& \frac{\abss{\thetal}}{M} \Expec{ \left\{ \frac{ \langle \ssl(t) , \shat(t)) \rangle }{M} \right\} \left\{ \frac{ \langle \ssl(t) , \shat(t) \rangle }{M} \right\}^* }  \nonumber \\
& &\times \Expecl{ \int \sqrt{1-\ahat} \displaystyle\sum_{r=-M/2}^{M/2-1} c_r p((1-\ahat)t -r T -(1-\ahat)\gammahat)  } \nonumber \\
& & \Expecr{ \times \mbox{} \sqrt{1-\ahat} \displaystyle\sum_{q=-M/2}^{M/2-1} c_q^* p((1-\ahat)t -q T -(1-\ahat)\gammahat) \,dt }. \label{eq:2first_term_2}
\end{eqnarray}
To simplify the second expectation of \eqref{2first_term_2}, we note that by making the substitution $v=(1-\ahat)t -rT-(1-\ahat)\gammahat$, we can reduce the following expression
\begin{eqnarray}
\lefteqn{  \int \sqrt{1-\ahat} \displaystyle\sum_{r=-M/2}^{M/2-1} c_r p((1-\ahat)t -r T -(1-\ahat)\gammahat)    }  \nonumber \\
\lefteqn{  \times \mbox{} \sqrt{1-\ahat} \displaystyle\sum_{q=-M/2}^{M/2-1} c_q^* p((1-\ahat)t -q T -(1-\ahat)\gammahat) \,dt   }  \nonumber \\
&=& \frac{1-\ahat}{1-\ahat}   \displaystyle\sum_{r=-M/2}^{M/2-1} \displaystyle\sum_{q=-M/2}^{M/2-1} c_r c_q^*   \int   p(v)   p(v+rT-qT) \,dv \\
&=& \displaystyle\sum_{r=-M/2}^{M/2-1} \displaystyle\sum_{q=-M/2}^{M/2-1} c_r c_q^* \delta_{r-q} \label{eq:astep1}\\
&=& M. \label{eq:2rightsidefinal} 
\end{eqnarray}
It therefore follows that the second expectation of \eqref{2first_term_2} is equal to $M$.  Equation \eqref{astep1} follows from \eqref{trick2}, and \eqref{2rightsidefinal} follows from the fact that $c_r c_q^* = 1$ if $q=r$. Before simplifying the first expectation of \eqref{2first_term_2}, we use the fact that $\langle \ssl(t), \shat(t) \rangle $, as shown in \appref{sshat}, can be approximated as follows
\begin{eqnarray}
\langle \ssl(t), \shat(t) \rangle &=& \displaystyle\sum_{r=-M/2}^{M/2-1} \displaystyle\sum_{q=-M/2}^{M/2-1} b_r b_q^* \hat{F}_{q,r}\of{l} e^{j 2 \pi f_c ((1-\ahat )\gammahat -(1-\al )\gammal )} \nonumber \\
 & & \times W(qT - \frac{rT}{\betal}+(1-\ahat )(\gammahat-\gammal ),\betal ), \label{eq:thingxxx}
\end{eqnarray}
where $W(\mbox{ })$ is the wideband ambiguity function defined in \eqref{wbdef}, $\hat{F}_{q,r}\of{l}$ is defined in \eqref{Fhatdef}, and $\betal \defn \frac{1-\al}{1-\ahat}$.  Using \eqref{thingxxx}, we can express the first expectation of \eqref{2first_term_2} as follows
\begin{eqnarray}
\lefteqn{ \Expec{ \left\{ \frac{ \langle \ssl(t) , \shat(t)) \rangle }{M} \right\} \left\{ \frac{ \langle \ssl(t) , \shat(t) \rangle }{M} \right\}^* }  } \nonumber \\
&=& \frac{1}{M^2} \Expec{ \displaystyle\sum_{r,q,m,n=-M/2}^{M/2-1} b_r b_q^* b_n^* b_m \hat{F}_{q,r}\of{l} \hat{F}_{m,n}^{(l)*} W(\lambdal_{r,q},\betal) W(\lambdal_{n,m},\betal) }, \label{eq:evilthing}
\end{eqnarray}
where
\begin{equation}
\lambdal_{r,q} \defn qT-\frac{rT}{\betal}+(1-\ahat )(\gammahat-\gammal ). \label{eq:lambda}
\end{equation}
As we did in \secref{error_lti}, we use the fact that $\Expec{b_r b_q^* b_n^* b_m}$ is equal to $1$ if $q=r=n=m$, if $q =r \neq n = m$, or if $q=m \neq r =n $.  In all other cases $\Expec{b_r b_q^* b_n^* b_m}$ is equal to zero.  We also use the fact that $\hat{F}_{q,r}\of{l} \hat{F}_{m,n}^{(l)*} =1$ if $q=m$ and $r=n$ to simplify \eqref{evilthing} as follows   
\begin{eqnarray}
\lefteqn{ \frac{1}{M^2} \Expec{ \sum_{r,q,m,n=-M/2}^{M/2-1} b_r b_q^* b_n^* b_m \hat{F}_{q,r}\of{l} \hat{F}_{m,n}^{(l)*}  W(\lambdal_{r,q},\betal) W(\lambdal_{n,m},\betal) }   } \nonumber \\
&=& \omegal +\frac{1}{M^2} \left\{ \sum_{r=-M/2}^{M/2-1} W^2(\lambdal_{r,r},\betal ) \right. \nonumber \\
& & \left. + \mbox{ } \sum_{m =-M/2}^{M/2-1} \sum_{n\neq m =-M/2}^{M/2-1} \hat{F}_{m,m}\of{l} \hat{F}_{n,n}^{(l)*} W(\lambdal_{m,m},\betal ) W(\lambdal_{n,n},\betal )   \right\} \label{eq:evilthing2} \\
&=& \frac{1}{M^2}  \left\{ \sum_{r=-M/2}^{M/2-1} W^2(\lambdal_{r,r},\betal ) - \sum_{q=-M/2}^{M/2-1} W^2(\lambdal_{q,q},\betal )  \right.  \\
& & \left. +\mbox{ } \sum_{m,n=-M/2}^{M/2-1} \hat{F}_{m,m}\of{l} \hat{F}_{n,n}^{(l)*} W(\lambdal_{m,m},\betal ) W(\lambdal_{n,n},\betal ) \right\} +\omegal \nonumber \\
&=&\frac{1}{M^2} \left\{ \sum_{m,n=-M/2}^{M/2-1} \hat{F}_{m,m}\of{l} \hat{F}_{n,n}^{(l)*} W(\lambdal_{m,m},\betal ) W(\lambdal_{n,n},\betal )  \right\} +\omegal \\
&=&\frac{1}{M^2} \abss{ \sum_{n=-M/2}^{M/2-1} \hat{F}_{n,n}\of{l} W(\lambdal_{n,n},\betal ) } +\omegal , \label{eq:evilthing3}
\end{eqnarray}
where
\begin{equation}
\omegal \defn \frac{1}{M^2} \displaystyle\sum_{e \neq r=-M/2}^{M/2-1} W^2(\lambdal_{e,r},\betal ). \label{eq:omegadef}
\end{equation}
Now by combining the results of \eqref{2rightsidefinal} and \eqref{evilthing3}, we can express the first term of \eqref{2error} as follows  
\begin{eqnarray}
\lefteqn{ \frac{\abss{\thetal}}{M} \Expec{\int \abss{\tht} \abss{\dhat(t)} \,dt} } \nonumber \\
&=&  \frac{M\abss{\thetal} }{M M^2} \abss{ \displaystyle\sum_{n=-N/2}^{N/2-1} \hat{F}_{n,n}\of{l} W(\lambdal_{n,n},\betal )  } +  \frac{\abss{\thetal} M \omegal}{M}  \\ 
&=& \frac{\abss{\thetal} }{M^2} \abss{ \displaystyle\sum_{n=-M/2}^{M/2-1} \hat{F}_{n,n}\of{l} W(\lambdal_{n,n},\betal )  } +  \abss{\thetal} \omegal .   \label{eq:2firsttermfinal} 
\end{eqnarray}

Using some of the arguments we used to reduce the first term of \eqref{2error}, the second term of \eqref{2error} can be simplified as follows
\begin{eqnarray}
\lefteqn{ \frac{2\abss{\thetal}}{M} \Expec{\int \real{ \tht \dhat(t) \dlc(t) } \,dt}  } \nonumber \\
&=& \frac{2\abss{\thetal}}{M} \real{ \Expec{\int \tht \dhat(t) \dlc(t)   \,dt  }    } \\
&=& \frac{2\abss{\thetal}}{M} \real{ \Expec{  \tht } \Expec{ \int \dhat(t) \dlc(t) \,dt  } }  \\
&=& \frac{2\abss{\thetal}}{M} \real{ \Expec{  \frac{ \langle \ssl(t) , \shat(t) \rangle }{M} }  \Expec{ \int \dhat(t) \dlc(t) \,dt } } \\
&=& \frac{2\abss{\thetal}}{M^2} \real{ \displaystyle\sum_{n=-M/2}^{M/2-1} \hat{F}_{n,n}\of{l} W(\lambdal_{n,n},\betal ) \displaystyle\sum_{m=-M/2}^{M/2-1} \hat{F}_{m,m}^{(l)*} W(\lambdal_{m,m},\betal ) } \label{eq:2secondtermx} \\
&=& \frac{2\abss{\thetal}}{M^2} \real{ \abss{ \displaystyle\sum_{n=-M/2}^{M/2-1} \hat{F}_{n,n}\of{l} W(\lambdal_{n,n},\betal )  } }   \\
&=&   \frac{2\abss{\thetal}}{M^2}  \abss{ \displaystyle\sum_{n=-M/2}^{M/2-1} \hat{F}_{n,n}\of{l} W(\lambdal_{n,n},\betal )  }.       \label{eq:2secondtermfinal}
\end{eqnarray}
Equation \eqref{2secondtermx} follows from \eqref{thingxxx}, \eqref{lambda}, and the facts that $\Expec{b_n b_m^*}= \delta_{n-m}$,\\ and $\Expec{c_n c_m^*}=\delta_{n-m}$.   

Finally the third expectation of \eqref{2error} can be reduced as follows
\begin{equation}
\frac{\abss{\thetal}}{M} \Expec{\int \abss{\dl(t)} \,dt} = \frac{\abss{\thetal} M}{M} = \abss{\thetal}. \label{eq:2thirdtermfinal}
\end{equation}

Thus, combining the results of \eqref{2firsttermfinal}, \eqref{2secondtermfinal}, and \eqref{2thirdtermfinal}, we can express \eqref{2error} as follows
\begin{eqnarray}
\lefteqn{ \frac{1}{M} \Expec{\vectornorm{\thetahat  \dhat(t) - \thetal \dl(t) }^2} } \nonumber  \\
&=& \abss{\thetal} - \frac{\abss{\thetal}}{M^2} \abss{ \displaystyle\sum_{n=-M/2}^{M/2-1} \hat{F}_{n,n}\of{l} W(\lambdal_{n,n},\betal ) } +\abss{\thetal} \omegal .\label{eq:one_branch_tv}
\end{eqnarray}
Furthermore, if we merge the results of \eqref{error_sum_tv} and \eqref{one_branch_tv},  $\metric$ can be upper bounded as 
\begin{eqnarray}
\lefteqn{\metric} \nonumber \\
&\leq&  \displaystyle\sum_{l=1}^{L} \abss{\thetal} - \frac{\abss{\thetal}}{M^2} \abss{ \displaystyle\sum_{n=-M/2}^{M/2-1} \hat{F}_{n,n}\of{l} W(\lambdal_{n,n},\betal ) } + \abss{\thetal} \omegal. \label{eq:final_ub_tv}
\end{eqnarray}
Finally, by applying the results of \appref{omegabound} and \appref{fwbound}, \eqref{final_ub_tv} becomes
\begin{eqnarray}
\lefteqn{\metric} \\
&\leq& \displaystyle\sum_{l=1}^{L} \abss{\thetal} -\abss{\thetal}~\frac{M-1}{M+1} \cdot \frac{1}{4} \textstyle \cos^2 ( \frac{B\Delta_\gamma}{2T}(1+\frac{1}{M}) ) \nonumber \\
&     & \times \mbox{ } \left( \sinc\left( \left( \frac{\pi f_c T}{1-\frac{1}{2M}}+\frac{\pi f_c T + B}{1-\al} \right) \frac{M \Delta_a}{4\pi} \right)  \right. \nonumber \\
&     & \left.  + \mbox{ } \sinc\left( \left( \frac{\pi f_c T}{1-\frac{1}{2M}}+\frac{\pi f_c T - B}{1-\al} \right) \frac{M \Delta_a}{4\pi} \right)  \right)^2 + \frac{1.2011\abss{\thetal}}{M}, \label{eq:ts_final}
\end{eqnarray}
where $B \leq \pi$ is chosen to make $\sinc(x) \gtrsim \cos(Bx)$ for $|x| \leq \frac{1}{2}$.  Although this is a rather complicated expression, we can use it to gain insight on the performance of the estimator.  For instance, if we assume that $M\geq100$ and that $\al = .005$ for all $l \in \{1,...,L\}$, and plug in a value of $B=1.8138$, we find
\begin{eqnarray}
\lefteqn{\metric} \\
&\leq&  1 -0.2450 \textstyle \cos^2 ( \frac{0.9160 \Delta_\gamma}{T}) \nonumber \\
&     & \times \mbox{ } \Big( \sinc\left( .2488 M \Delta_a \left( 2 \pi f_c T+ .5774 \right)  \right)   \nonumber \\
&     &   + \mbox{ } \sinc\left( .2488 M \Delta_a \left( 2 \pi f_c T- .5774 \right) \right)  \Big)^2 + \frac{1.2011}{M}, \label{eq:timescalebound}
\end{eqnarray}
which follows from the assumption $\sum_{l=1}^L | \thetal|^2=1$.  It can now be seen that \\$\metric$ decreases for decreasing  $\Delta_\gamma$ and $\Delta_a$.  Additionally,  as the fractional bandwidth $\frac{1}{f_cT}$ decreases, $\metric$ increases.  Next we examine how tight the upper bound of \eqref{ts_final} is through simulation.

\subsection{Simulation of Time Varying Channel Estimation Error} \label{sec:error_sim_tv}
To examine the tightness of the bound in \eqref{timescalebound}, we ran a simulation of a one path channel with parameters  $a\of{1}=-0.001$, and $\gamma\of{1}=0$.  We sent a training signal modulated of 300 random symbols through the channel.  This training signal had a fractional bandwidth $\frac{1}{f_cT}=1$.  Parameter estimates were given by $\hat{\gamma}\of{1}=\gamma\of{1}+\frac{\Delta_\gamma}{2}$ and $\hat{a}\of{1}=a\of{1}+\frac{\Delta_a}{2}$.  Thus, we simulated the worst case scenario where the parameter estimates were half a grid spacing away from the true parameter values.  The estimate of $\theta\of{1}$ was made based on \eqref{thetahat1}.
Next a data signal modulated from a series of 300 random data symbols was sent through the channel.  The error metric $\metric$ was calculated for different many choices of grid spacing.  This simulation was repeated 10 times and the results were averaged.  We set $B=1.8138$ and plugged in the simulations parameters into \eqref{timescalebound} to calculate the error upper bound as
\begin{eqnarray}
\lefteqn{\metric} \nonumber \\
&\leq&  1.004 -0.2483 \textstyle \cos^2 ( \frac{0.9099 \Delta_\gamma}{T}) \left( \sinc\left(  193.32 \Delta_a   \right)  +  \sinc\left( 106.8 \Delta_a  \right)  \right)^2. 
\end{eqnarray}

\begin{figure}[htbp]
	\begin{center}
		\psfrag{s11}{}
		\psfrag{s10}{}
		\psfrag{s07}{}
		\psfrag{v01}[l][c][1][0]{$0.02$}
		\psfrag{v02}[l][c][1][0]{$0.04$}
		\psfrag{v03}[l][c][1][0]{$0.05$}
		\psfrag{v04}[l][c][1][0]{$0.08$}
		\psfrag{v05}[l][c][1][0]{$0.10$}
		\psfrag{v06}[l][c][1][0]{$0.12$}
		\psfrag{v07}[l][c][1][0]{$0.14$}
		\psfrag{v08}[l][c][1][0]{$0.16$}
		\psfrag{v09}[l][c][1][0]{$0.18$}
		\psfrag{v10}[l][c][1][0]{$0.20$}
		\psfrag{v11}[l][c][1][0]{$0.22$}
		\psfrag{x04}[c][c][1][0]{$0$}
		\psfrag{x05}[c][c][1][0]{$0.2$}
		\psfrag{x06}[c][c][1][0]{$0.4$}
		\psfrag{x07}[c][c][1][0]{$0.6$}
		\psfrag{x08}[c][c][1][0]{$0.8$}
		\psfrag{x09}[c][c][1][0]{$1.0$}
		\psfrag{x10}[c][c][1][0]{$1.2$}
		\psfrag{x11}[c][c][1][0]{$1.4$}
		\psfrag{x12}[c][c][1][0]{$1.6$}
		\psfrag{v12}[r][c][1][0]{$0$}
		\psfrag{v13}[r][c][1][0]{$0.05$}
		\psfrag{v14}[r][c][1][0]{$0.10$}
		\psfrag{v15}[r][c][1][0]{$0.15$}
		\psfrag{v16}[r][c][1][0]{$0.20$}
		\psfrag{v17}[r][c][1][0]{$0.25$}
		\psfrag{s06}[r][c][1.2][270]{$\frac{\Delta_\gamma}{T}$}
		\psfrag{s05}[c][b][1.2][0]{$\Delta_a \times 10^{3}$}
        \epsfxsize=4.0in
        \epsfbox{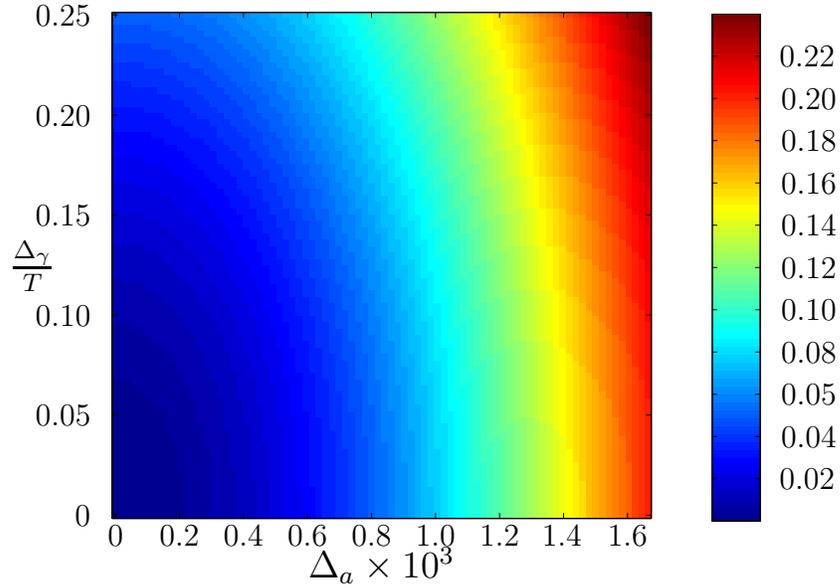}
    \end{center}
	\caption{Average simulated error using a time-shift/scale-shift model}
    \label{fig:ts_simulated}
\end{figure}

\begin{figure}[htbp]
	\begin{center}
		
		\psfrag{s07}{}
		\psfrag{s11}{}
		\psfrag{s10}{}
		\psfrag{x04}[c][c][1][0]{$0$}
		\psfrag{x05}[c][c][1][0]{$0.2$}
		\psfrag{x06}[c][c][1][0]{$0.4$}
		\psfrag{x07}[c][c][1][0]{$0.6$}
		\psfrag{x08}[c][c][1][0]{$0.8$}
		\psfrag{x09}[c][c][1][0]{$1.0$}
		\psfrag{x10}[c][c][1][0]{$1.2$}
		\psfrag{x11}[c][c][1][0]{$1.4$}
		\psfrag{x12}[c][c][1][0]{$1.6$}
		\psfrag{v01}[l][c][1][0]{$0.05$}
		\psfrag{v02}[l][c][1][0]{$0.10$}
		\psfrag{v03}[l][c][1][0]{$0.15$}
		\psfrag{v04}[l][c][1][0]{$0.20$}
		\psfrag{v05}[l][c][1][0]{$0.25$}
		\psfrag{v06}[r][c][1][0]{$0$}
		\psfrag{v07}[r][c][1][0]{$0.05$}
		\psfrag{v08}[r][c][1][0]{$0.10$}
		\psfrag{v09}[r][c][1][0]{$0.15$}
		\psfrag{v10}[r][c][1][0]{$0.20$}
		\psfrag{v11}[r][c][1][0]{$0.25$}
		\psfrag{s06}[r][c][1.2][270]{$\frac{\Delta_\gamma}{T}$}
		\psfrag{s05}[c][b][1.2][0]{$\Delta_a \times 10^{3}$}
        \epsfxsize=4.0in
        \epsfbox{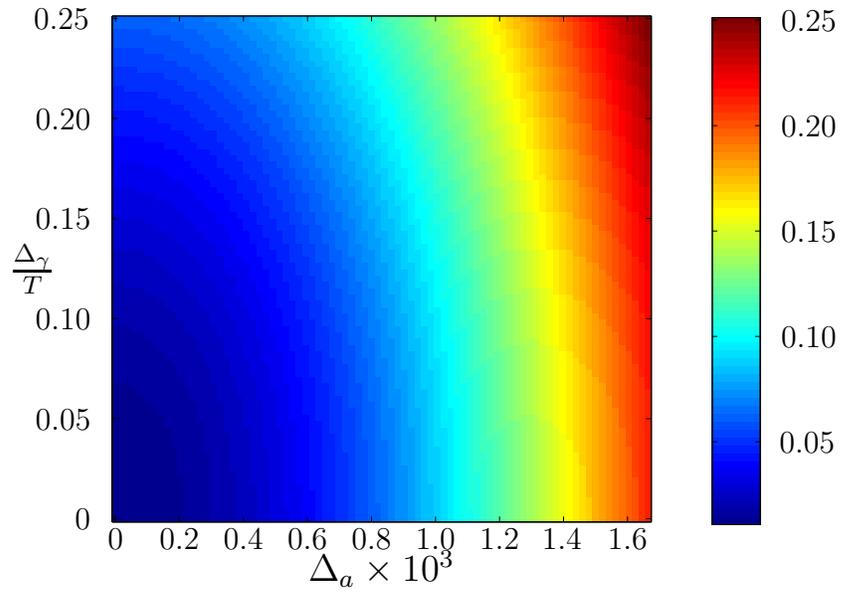}
    \end{center}
	\caption{Derived upper bound for time-shift/scale-shift model}
    \label{fig:ts_ub}
\end{figure}

\begin{figure}[htbp]
	\begin{center}
		\psfrag{x04}[c][c][1][0]{$0$}
		\psfrag{x05}[c][c][1][0]{$0.2$}
		\psfrag{x06}[c][c][1][0]{$0.4$}
		\psfrag{x07}[c][c][1][0]{$0.6$}
		\psfrag{x08}[c][c][1][0]{$0.8$}
		\psfrag{x09}[c][c][1][0]{$1.0$}
		\psfrag{x10}[c][c][1][0]{$1.2$}
		\psfrag{x11}[c][c][1][0]{$1.4$}
		\psfrag{x12}[c][c][1][0]{$1.6$}
		\psfrag{v01}[l][c][1][0]{$0.05$}
		\psfrag{v02}[l][c][1][0]{$0.10$}
		\psfrag{v03}[l][c][1][0]{$0.15$}
		\psfrag{v04}[l][c][1][0]{$0.20$}
		\psfrag{v05}[l][c][1][0]{$0.25$}
		\psfrag{v06}[r][c][1][0]{$0$}
		\psfrag{v07}[r][c][1][0]{$0.05$}
		\psfrag{v08}[r][c][1][0]{$0.10$}
		\psfrag{v09}[r][c][1][0]{$0.15$}
		\psfrag{v10}[r][c][1][0]{$0.20$}
		\psfrag{s18}[l][c][1][0]{Simulated}
		\psfrag{s19}[l][c][1][0]{Derived}
		\psfrag{s10}[r][c][1.2][270]{$\frac{\Delta_\gamma}{T}$}
		\psfrag{s09}[c][b][1.2][0]{$\Delta_a \times 10^{3}$}
		\psfrag{s11}{}
		\psfrag{s14}{}
		\psfrag{s15}{}
        \epsfxsize=4.0in
        \epsfbox{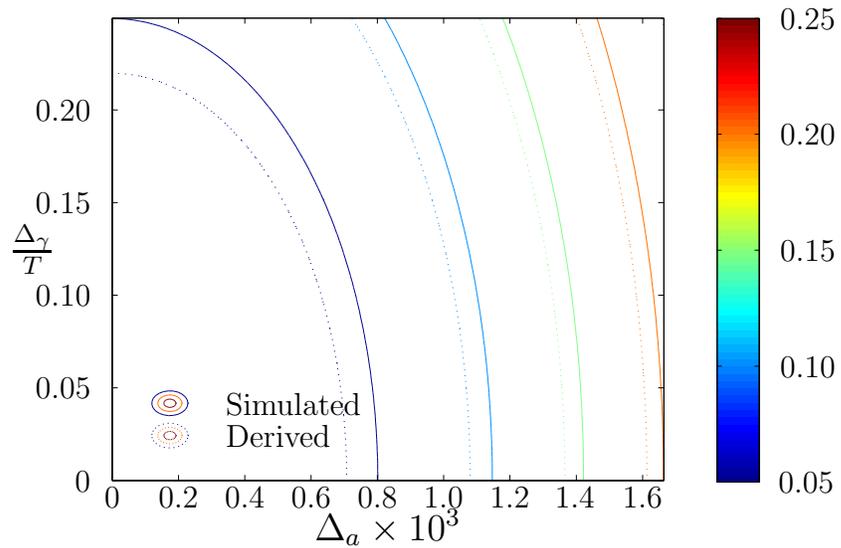}
    \end{center}
	\caption{Derived upper bound vs. average simulated error for time-shift/scale-shift model}
    \label{fig:ts_contour}
\end{figure}

\begin{figure}[htbp]
	\begin{center}
		\psfrag{x01}[c][c][1][0]{$0$}
		\psfrag{x02}[c][c][1][0]{$0.05$}
		\psfrag{x03}[c][c][1][0]{$0.10$}
		\psfrag{x04}[c][c][1][0]{$0.15$}
		\psfrag{x05}[c][c][1][0]{$0.20$}
		\psfrag{x06}[c][c][1][0]{$0.25$}
		\psfrag{v01}[r][c][1][0]{$0$ }
		\psfrag{v02}[r][c][1][0]{$0.05$}
		\psfrag{v03}[r][c][1][0]{$0.10$}
		\psfrag{v04}[r][c][1][0]{$0.15$}
		\psfrag{v05}[r][c][1][0]{$0.20$}
		\psfrag{v06}[r][c][1][0]{$0.25$}
		\psfrag{v07}[r][c][1][0]{$0.30$}
		\psfrag{v08}[r][c][1][0]{$0.35$}
		\psfrag{s10}{ }
		\psfrag{s11}{ }
		
		\psfrag{s13}[l][l][1][0]{$\Delta_a= 0$ }
		\psfrag{s14}[l][l][1][0]{$\Delta_a= 0.8011 \times10^{-3}$ }
		\psfrag{s15}[l][l][1][0]{$\Delta_a= 1.6639 \times10^{-3}$ }
		\psfrag{s05}[c][b][1.2][0]{$\frac{\Delta_\gamma}{T}$}

        \epsfxsize=4.0in
        \epsfbox{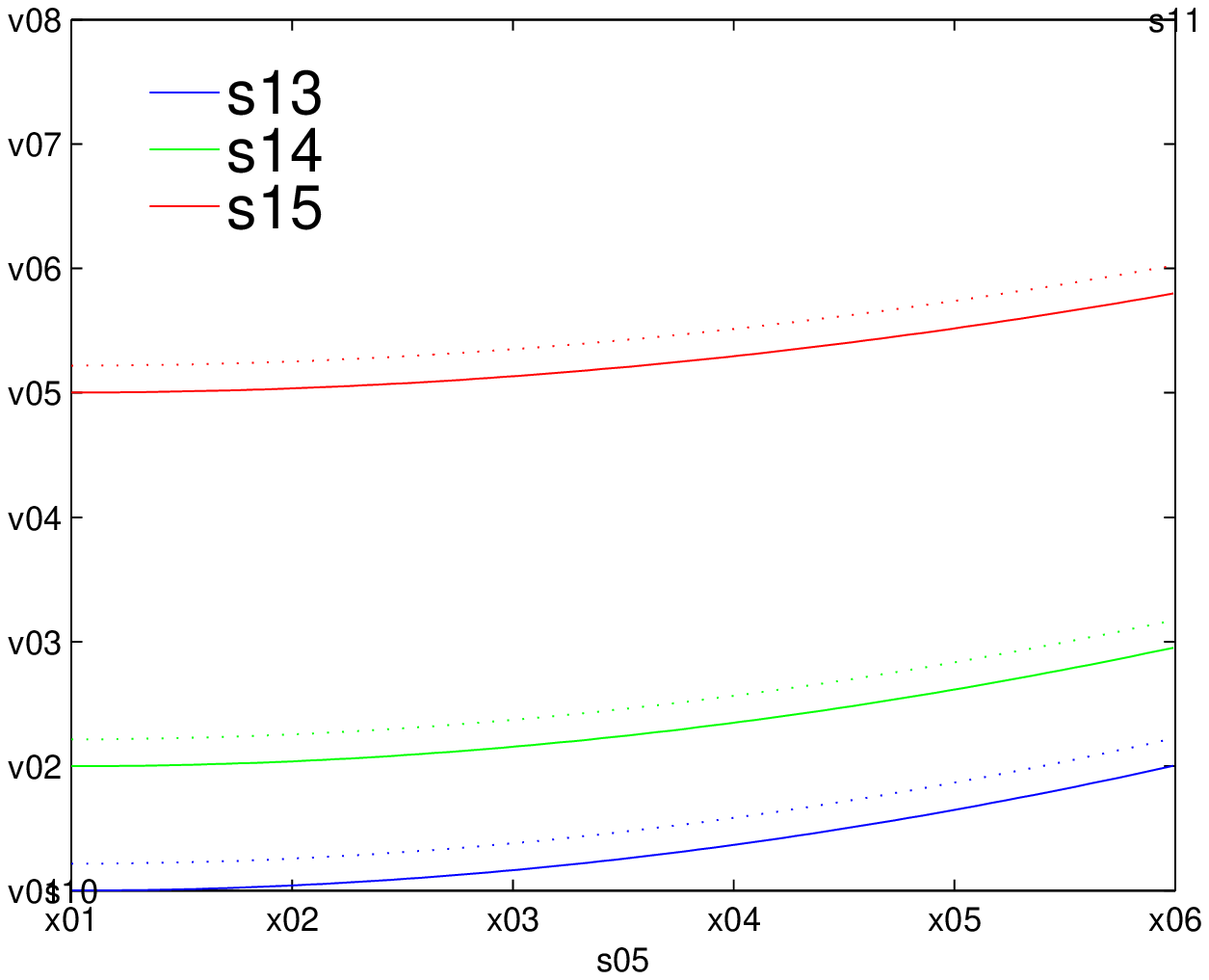}
    \end{center}
	\caption{Simulated error and derived upper bound vs. $\frac{\Delta_\gamma}{T}$ for fixed values of $\Delta_a$ using time-shift/scale-shift model.  Simulated error is plotted with solid lines while the derived bound is plotted with dotted lines.}
    \label{fig:scale_vs_gamma}
\end{figure}

\begin{figure}[htbp]
	\begin{center}
		\psfrag{x01}[c][c][1][0]{$0$}
		\psfrag{x02}[c][c][1][0]{$0.2$}
		\psfrag{x03}[c][c][1][0]{$0.4$}
		\psfrag{x04}[c][c][1][0]{$0.6$}
		\psfrag{x05}[c][c][1][0]{$0.8$}
		\psfrag{x06}[c][c][1][0]{$1.0$}
		\psfrag{x07}[c][c][1][0]{$1.2$}
		\psfrag{x08}[c][c][1][0]{$1.4$}
		\psfrag{x09}[c][c][1][0]{$1.6$}

		\psfrag{v01}[r][c][1][0]{$0$}
		\psfrag{v02}[r][c][1][0]{$0.05$}
		\psfrag{v03}[r][c][1][0]{$0.10$}
		\psfrag{v04}[r][c][1][0]{$0.15$}
		\psfrag{v05}[r][c][1][0]{$0.20$}
		\psfrag{v06}[r][c][1][0]{$0.25$}

		\psfrag{s10}{ }
		\psfrag{s11}{ }
	
		\psfrag{s13}[l][l][1][0]{$\Delta_\gamma/T= 0$ }
		\psfrag{s14}[l][l][1][0]{$\Delta_\gamma/T=0.1674$ }
		\psfrag{s15}[l][l][1][0]{$\Delta_\gamma/T=0.2496$ }
		\psfrag{s05}[c][b][1][0]{$\Delta_a \times 10^3$}

        \epsfxsize=4.0in
        \epsfbox{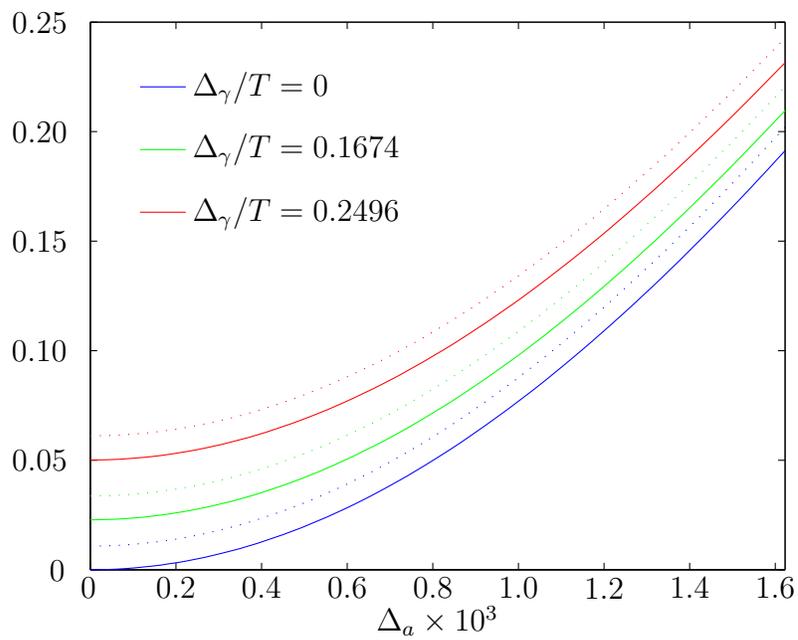}
    \end{center}
	\caption{Simulated error and derived upper bound vs. $\Delta_a$ for fixed values of $\frac{\Delta_\gamma}{T}$ using time-shift/scale-shift model.  Simulated error is plotted with solid lines while the derived bound is plotted with dotted lines.}
    \label{fig:scale_vs_a}
\end{figure}

\Figref{ts_ub} plots this upper bound while \Figref{ts_simulated} plots the average simulated error.  \Figref{ts_contour} shows the two together, and it can be seen that the simulated error is less than the upper bound we derived.  These results are also shown in \Figref{scale_vs_gamma} and \Figref{scale_vs_a}.  \Figref{scale_vs_gamma} plots the simulated error and the derived bound for fixed values of $\Delta_a$, while \Figref{scale_vs_a} plots the simulated error and the derived bound  for fixed values of $\Delta_\gamma$.  It all cases it can be seen that the bound derived in \eqref{ts_final} does indeed hold and appears to be relatively tight.

\chapter{Estimation Error Using a Time-Shift/Doppler-Shift Model} 
\label{ch:doppler}
\section{Background} \label{sec:doppbackground}
We now consider the case when a simplifying narrowband Doppler approximation is used to construct the linearized model described in \secref{grid}.  Using this approximation, the effects of time scaling by the quantity $1-a$ are approximated by a Doppler-shift by the quantity $-2 \pi f_c a$ as described in \cite{W1994}.  For example, consider a transmitted passband signal $x(t)$ with a carrier frequency $f_c$.  If $x(t)$ travels along an energy preserving single path channel with a time varying delay $\tau(t) = (1-a)\gamma + a t$, the resulting received signal $r(t)= \sqrt{1-a} x( (1-a)(t-\gamma) )$ is approximated as $e^{- 2 \pi f_c a (t-\gamma)} x (t-\gamma)$. 

As we did in \chref{error}, we assume that the path gains are time invariant\\ ($\{f\of{l}(t)=\sqrt{1-\al}\thetal \}_{l=1}^L$), the number of estimated paths is equal to the true number of paths ($\hat{L}=L$), and the number of pilot symbols is equal to the number of data symbols ($N=M$).  In the previous sections, we said that the sparse estimator makes estimates $\{ \ahat \}^L_{l=1}$ of the true channel values $\{ \al \}^L_{l=1}$, where $\al$ is the rate of change of the $l^{th}$ channel path.  In this chapter, we say the sparse estimator instead makes Doppler-shift estimates denoted as $\{ \fdopp \}^L_{l=1}$.   We continue to use the tilde symbol to denote the Doppler approximated estimates.  Thus, the sparse estimator makes parameter estimates $\{ \gammadopp , \fdopp , \thetadopp  \}^{L}_{l=1} $, where $\gammadopp$ and $\thetadopp $ are the estimates of $\gammal $ and $\thetal$, respectively.  Equivalently, we can say the sparse estimator makes parameter estimates $\{ \gammadopp , \adopp , \thetadopp  \}^{L}_{l=1} $ where $\adopp \defn \frac{\fdopp}{f_c}$.  As outlined in \secref{grid}, a linearized model is constructed and parameter estimates $\{ \gammadopp , \adopp  \}^{L}_{l=1} $ are grid restricted with grid spacing $\Delta_\gamma$ and $\Delta_a$.  Using these parameter estimates, the pass band signal is approximated as follows.

\begin{eqnarray}
r(t) &=& \displaystyle\sum_{l=1}^{L} \thetal \sqrt{1-\al} x( (1-\al)(t-\gammal ) ) +n(t) \\
\tilde{r}(t) &=& \displaystyle\sum_{l=1}^{L} \thetadopp e^{-2 \pi f_c \adopp (t - \gammadopp) } x(t-\gammadopp)  +n(t) \label{eq:rdopp1} \\
&=& \displaystyle\sum_{l=1}^{L} \thetadopp e^{-2 \pi f_c \adopp (t - \gammadopp) } \real{ e^{j 2 \pi f_c (t-\gammahat ) }  \displaystyle\sum_{n=-N/2}^{N/2-1} b_n p(t - nT- \gammahat) } \nonumber \\
& &  + \mbox{ }n(t)
\end{eqnarray}
We express the Doppler approximated estimate of the complex baseband received signal $z(t)$ as
\begin{eqnarray}
\tilde{z}(t) &=& \text{LPF} \left\{ 2 \tilde{r}(t) e^{-j 2 \pi f_c t} \right\} \\ 
&=& \displaystyle\sum_{l=1}^{L} \thetadopp e^{ -j 2 \pi f_c ( (1-\adopp ) \gammadopp + \adopp t )} \displaystyle\sum_{n=-N/2}^{N/2-1} b_n p(t - nT- \gammadopp)+w(t) \\
&=& \displaystyle\sum_{l=1}^{L} \thetadopp \sdopp(t) + w(t),
\end{eqnarray}
where
\begin{equation}
\sdopp(t) \defn e^{-j 2 \pi f_c ((1-\adopp)\gammadopp+\adopp t)} \displaystyle\sum_{n=-N/2}^{N/2-1} b_n p(t -n T -\gammadopp). \label{eq:sdopp}
\end{equation}

For each estimate pair $\gammadopp$ and $\adopp$, $\sdopp(t)$ is yielded.  Using similar arguments which lead to \eqref{thetahat1} of \secref{cs}, we assume that the estimate of $\thetal$ can be expressed as
\begin{equation}
\thetadopp =\frac{\langle \ssl(t),\sdopp(t)\rangle \thetal} {\langle \sdopp(t),\sdopp(t) \rangle} =\frac{\langle \ssl(t),\sdopp(t)\rangle \thetal} {N}. \label{eq:thetadopp} 
\end{equation}

We now need a way to measure the estimation error resulting from the use of the Doppler approximation with channel parameter estimates $\{ \gammadopp , \adopp , \thetadopp  \}^{L}_{l=1}$.  In \secref{error_metric}, we considered a complex baseband received signal $y(t)$ and another complex baseband received signal $\hat{y}(t)$ reconstructed from the channel parameter estimates.  Both $y(t)$ and $\hat{y}(t)$ were modulated from the same series of $M$ PSK symbols.  We defined $e_y(t) \defn \hat{y}(t)-y(t)$ and used $\metric$ to measure the estimation error.  For this chapter, we define $\ydopp(t)$ as the Doppler approximation estimate of $y(t)$ constructed from the channel parameter estimates $\{ \gammadopp , \adopp , \thetadopp  \}^{L}_{l=1}$ , so that
\begin{equation}
\ydopp(t)= \displaystyle\sum_{l=1}^{L} \thetadopp \ddopp(t),
\end{equation}
where
\begin{equation}
\ddopp(t) \defn e^{-j 2 \pi f_c ((1-\adopp)\gammadopp+\adopp t)} \displaystyle\sum_{n=-N/2}^{N/2-1} b_n p(t -n T -\gammadopp).
\end{equation}
We define $\eydopp(t) \defn \ydopp(t)-y(t)$ and use $\metricdopp$ as the metric of estimation error. In \secref{error_doppler}, we derive an expression which relates this metric to the grid spacing $\Delta_{\gamma}$ and $\Delta_a$.  Then in \secref{dopp_vs_ts} we evaluate how this expression compares to the expression derived in \secref{error_tv}, where the Doppler approximation was not used.  This analysis helps us gain insight on the effects of the grid spacing choice and the effects of the Doppler approximation.

\section{Average Estimation Error} \label{sec:error_doppler}
As we did in \secref{error_lti} and \secref{error_tv}, we evaluate the upper bound of the error metric instead of the metric directly.
\begin{equation}
\metricdopp \leq \frac{1}{M} \displaystyle\sum_{l=1}^{L} \Expec{\vectornorm{\thetadopp  \ddopp(t) - \thetal \dl(t)}^2} \label{eq:error_sum_tv_dopp}
\end{equation}
We again focus on a particular value of $l \in \{ 1,2,...,L \}$ and find, by a similar derivation to the one presented in \appref{3to1}, that
\begin{eqnarray}
\frac{1}{M}\Expec{\vectornorm{\thetadopp  \ddopp(t) - \thetal \dl(t)}^2} 
&=& \frac{\abss{\thetal}}{M} \Expec{\int \abss{\thtdopp} \abss{\ddopp(t)} \,dt} \nonumber \\
& & -\mbox{} \frac{2\abss{\thetal}}{M} \Expec{\int \real{ \thtdopp \ddopp(t) \dlc(t) } \,dt } \nonumber \\
& & +\mbox{} \frac{\abss{\thetal}}{M} \Expec{\int \abss{\dl(t)} \,dt}. \label{eq:2error_dopp}
\end{eqnarray}
The task now is to evaluate the 3 terms in \eqref{2error_dopp} and then combine the results.  

The first term of \eqref{2error_dopp} can be reduced as follows
\begin{eqnarray}
\lefteqn{ \frac{\abss{\thetal}}{M} \Expec{\int \abss{\thtdopp} \abss{\ddopp(t)} \,dt} } \nonumber  \\
&=& \frac{\abss{\thetal}}{M} \Expec{\abss{\thtdopp} } \Expec{\int \ddopp(t) \ddoppc(t)  \,dt }  \label{eq:2first_term_dopp}\\
&=& \frac{\abss{\thetal}}{M} \Expec{ \left\{ \frac{ \langle \ssl(t) , \sdopp(t)) \rangle }{M} \right\} \left\{ \frac{ \langle \ssl(t) , \sdopp(t) \rangle }{M} \right\}^* }  \nonumber \\
& &\times \Expec{ \int \displaystyle\sum_{r=-M/2}^{M/2-1} c_r p(t -r T -\gammahat )  \displaystyle\sum_{q=-M/2}^{M/2-1} c_q^* p( t -q T -\gammahat) \,dt }. \label{eq:2first_term_2_dopp}
\end{eqnarray}
Using \eqref{trick2}, it can be shown that the second expectation in \eqref{2first_term_2_dopp} reduces to $M$.  To simplify the first expectation in \eqref{2first_term_2_dopp}, we use the fact that $\langle \ssl(t) , \sdopp(t) \rangle$ can be approximated as (shown in \appref{ssdopp} )
\begin{eqnarray}
\langle \ssl(t), \sdopp(t) \rangle &=& \displaystyle\sum_{r=-M/2}^{M/2-1} \displaystyle\sum_{q=-M/2}^{M/2-1} b_r b_q^* \tilde{F}_{q,r} e^{j 2 \pi f_c ((1-\adopp )\gammadopp -(1-\al )\gammal )} \nonumber \\
 & & \times W(qT - \frac{rT}{\betaldopp}+\gammadopp-\gammal,\betaldopp ), \label{eq:thingxxx_dopp}
\end{eqnarray}
where $\tilde{F}_{q,r}$ is defined in \eqref{Ftildef}, and $\betaldopp \defn 1-\al$.  Using \eqref{thingxxx_dopp}, we can express the first expectation of \eqref{2first_term_2_dopp} as follows
\begin{eqnarray}
\lefteqn{ \Expec{ \left\{ \frac{ \langle \ssl(t) , \sdopp(t)) \rangle }{M} \right\} \left\{ \frac{ \langle \ssl(t) , \sdopp(t) \rangle }{M} \right\}^* }  } \nonumber \\
&=& \frac{1}{M^2} \Expec{ \displaystyle\sum_{r,q,m,n=-M/2}^{M/2-1} b_r b_q^* b_n^* b_m \tilde{F}_{q,r} \tilde{F}_{m,n}^* W(\lambdaldopp_{r,q},\betaldopp) W(\lambdaldopp_{n,m},\betaldopp) }, \label{eq:evilthing_dopp}
\end{eqnarray}
where
\begin{equation}
\lambdaldopp_{r,q}=qT-\frac{rT}{\betaldopp}+\gammadopp-\gammal. \label{eq:lambda_dopp}
\end{equation}

As we did in \secref{error_lti} and \secref{error_tv}, we use the fact that $\Expec{b_r b_q^* b_n^* b_m}$ is equal to $1$ if $q=r=n=m$, if $q =r \neq n = m$, or if $q=m \neq r =n $.  In all other cases $\Expec{b_r b_q^* b_n^* b_m}$ is equal to zero.  Additionally, we use the fact that $\tilde{F}_{q,r} \tilde{F}_{m,n}^* =1$ if $q=m$ and $r=n$ to simplify \eqref{evilthing_dopp} as follows   
\begin{eqnarray}
\lefteqn{ \frac{1}{M^2} \Expec{ \displaystyle\sum_{r,q,m,n=-M/2}^{M/2-1} b_r b_q^* b_n^* b_m \tilde{F}_{q,r} \tilde{F}_{m,n}^*  W(\lambdaldopp_{r,q},\betaldopp) W(\lambdaldopp_{n,m},\betaldopp) } } \nonumber \\
&=& \frac{1}{M^2} \left\{ \displaystyle\sum_{r=-M/2}^{M/2-1} W^2(\lambdaldopp_{r,r},\betaldopp ) \right. \nonumber \\
& & \left. + \mbox{}\displaystyle\sum_{m \neq n=-M/2}^{M/2-1} \tilde{F}_{m,m} \tilde{F}_{n,n}^* W(\lambdaldopp_{m,m},\betaldopp ) W(\lambdaldopp_{n,n},\betaldopp ) \right\} + \omegaldopp \label{eq:evilthing2_dopp} \\
&=& \frac{1}{M^2} \left\{ \displaystyle\sum_{r=-M/2}^{M/2-1} W^2(\lambdaldopp_{r,r},\betaldopp ) - \displaystyle\sum_{q=-M/2}^{M/2-1} W^2(\lambdaldopp_{q,q},\betaldopp )  \right. \\
& & \left. + \mbox{ } \displaystyle\sum_{m,n=-M/2}^{M/2-1} \tilde{F}_{m,m} \tilde{F}_{n,n}^* W(\lambdaldopp_{m,m},\betaldopp ) W(\lambdaldopp_{n,n},\betaldopp ) \right\} +\omegaldopp \nonumber \\
&=&\frac{1}{M^2} \left\{ \displaystyle\sum_{m,n=-M/2}^{M/2-1} \tilde{F}_{m,m} \tilde{F}_{n,n}^* W(\lambdaldopp_{m,m},\betaldopp ) W(\lambdaldopp_{n,n},\betaldopp )  \right\} +\omegaldopp \\
&=&\frac{1}{M^2} \abss{ \displaystyle\sum_{n=-M/2}^{M/2-1} \tilde{F}_{n,n} W(\lambdaldopp_{n,n},\betaldopp ) } +\omegaldopp, \label{eq:evilthing3_dopp}
\end{eqnarray}
where
\begin{equation}
\omegaldopp \defn \frac{1}{M^2} \displaystyle\sum_{e \neq r=-M/2}^{M/2-1} W^2(\lambdaldopp_{e,r},\betaldopp ). \label{eq:omegadefdopp}
\end{equation}
Using the results of \eqref{evilthing3_dopp}, we can express the first term of \eqref{2error_dopp} as follows  
\begin{eqnarray}
\lefteqn{ \frac{\abss{\thetal}}{M} \Expec{\int \abss{\thtdopp} \abss{\ddopp(t)} \,dt} } \nonumber \\
&=&  \frac{M\abss{\thetal} }{M M^2} \abss{ \displaystyle\sum_{n=-N/2}^{N/2-1} \tilde{F}_{n,n} W(\lambdaldopp_{n,n},\betaldopp )  } +  \frac{\abss{\thetal} M \omegaldopp}{M}  \\ 
&=& \frac{\abss{\thetal} }{M^2} \abss{ \displaystyle\sum_{n=-M/2}^{M/2-1} \tilde{F}_{n,n} W(\lambdaldopp_{n,n},\betaldopp )  } +  \abss{\thetal} \omegaldopp     \label{eq:2firsttermfinal_dopp} 
\end{eqnarray}
Using some of the arguments we used to reduce the first term of \eqref{2error_dopp}, the second term of \eqref{2error_dopp} can be simplified as follows
\begin{eqnarray}
\lefteqn{ \frac{2\abss{\thetal}}{M} \Expec{\int \real{ \thtdopp \ddopp(t) \dlc(t) } \,dt}  } \nonumber \\
&=& \frac{2\abss{\thetal}}{M} \real{ \Expec{  \thtdopp } \Expec{ \int \ddopp(t) \dlc(t) \,dt  } }  \\
&=& \frac{2\abss{\thetal}}{M} \real{ \Expec{  \frac{ \langle \ssl(t) , \sdopp(t) \rangle }{M} }  \Expec{ \int \ddopp(t) \dlc(t) \,dt } } \\
&=& \frac{2\abss{\thetal}}{M^2} \real{ \displaystyle\sum_{n=-N/2}^{N/2-1} \tilde{F}_{n,n} W(\lambdaldopp_{n,n},\betaldopp ) \displaystyle\sum_{m=-M/2}^{M/2-1} \tilde{F}_{m,m}^* W(\lambdaldopp_{m,m},\betaldopp ) } \label{eq:2secondtermx_dopp} \\
&=& \frac{2\abss{\thetal}}{M^2} \real{ \abss{ \displaystyle\sum_{n=-N/2}^{N/2-1} \tilde{F}_{n,n} W(\lambdaldopp_{n,n},\betaldopp )  } }   \\
&=&   \frac{2\abss{\thetal}}{M^2}  \abss{ \displaystyle\sum_{n=-N/2}^{N/2-1} \tilde{F}_{n,n} W(\lambdaldopp_{n,n},\betaldopp )  }       \label{eq:2secondtermfinal_dopp}
\end{eqnarray}
Equation \eqref{2secondtermx_dopp} follows from \eqref{thingxxx_dopp}, \eqref{lambda_dopp}, and the facts that $\Expec{b_n b_m^*}= \delta_{n-m}$\\ and $\Expec{c_n c_m^*}=\delta_{n-m}$.  Finally, the third expectation of \eqref{2error_dopp} can be reduced as follows
\begin{equation}
\frac{\abss{\thetal}}{M} \Expec{\int \abss{\dl(t)} \,dt} = \frac{\abss{\thetal} M}{M} = \abss{\thetal}\label{eq:2thirdtermfinal_dopp}
\end{equation}
Thus, combining the results of \eqref{2firsttermfinal_dopp}, \eqref{2secondtermfinal_dopp}, and \eqref{2thirdtermfinal_dopp}, we can express \eqref{2error_dopp} as
\begin{eqnarray}
\lefteqn{\frac{1}{M} \Expec{\vectornorm{\thetadopp  \ddopp(t) - \thetal \dl(t) }^2}} \nonumber \\
 &=& \abss{\thetal} - \frac{\abss{\thetal}}{M^2} \abss{ \displaystyle\sum_{n=-M/2}^{M/2-1} \tilde{F}_{n,n} W(\lambdaldopp_{n,n},\betaldopp ) } +\abss{\thetal} \omegaldopp. \label{eq:one_branch_tv_dopp}
\end{eqnarray}

Now by combining the results of \eqref{error_sum_tv_dopp} and \eqref{one_branch_tv_dopp},  $\metricdopp$ can be upper bounded as follows
\begin{equation}
\metricdopp \leq  \displaystyle\sum_{l=1}^{L} \abss{\thetal} - \frac{\abss{\thetal}}{M^2} \abss{ \displaystyle\sum_{n=-M/2}^{M/2-1} \tilde{F}_{n,n} W(\lambdaldopp_{n,n},\betaldopp ) } + \abss{\thetal} \omegaldopp. \label{eq:final_ub_tv_dopp}
\end{equation}
By applying the results of \appref{fwdoppbound} and \appref{omegabounddopp}, we find
\begin{eqnarray}
\lefteqn{\metricdopp} \nonumber \\
&\leq& \displaystyle\sum_{l=1}^{L} \abss{\thetal} + \frac{1.2011 \abss{\thetal}}{M} - \frac{\abss{\thetal}(1-\frac{1}{M})}{4} \textstyle \cos^2(\frac{\Bdopp\Delta_\gamma}{2T} ) \nonumber \\
& & \times \mbox{ } \left(  \sinc\left( \left( \frac{\pi f_c \Delta_a T}{2} \left(1+\frac{1}{1-\al} \right) -\frac{\Bdopp \al}{1-\al}  \right) \frac{M}{2\pi} \right)  \right. \nonumber \\
&     & \left.  + \mbox{ }  \sinc\left( \left( \frac{\pi f_c \Delta_a T}{2} \left(1+\frac{1}{1-\al} \right) +\frac{\Bdopp \al}{1-\al}  \right) \frac{M}{2\pi} \right)  \right)^2, \label{eq:doppbound}
\end{eqnarray}
where $B \leq \pi$ is chosen to make $\sinc(x) \gtrsim \cos(Bx)$ for $|x| \leq \frac{1}{2}$.  In \secref{error_sim_dopp}, we examine how tight this bound through simulation.
\subsection{Simulation of Estimation Error Using Doppler Approx.} \label{sec:error_sim_dopp}
To examine the tightness of the bound in \eqref{doppbound}, we ran a simulation of a one path channel similar to the simulation described in \secref{error_sim_tv}.  The single path channel again had parameters given by $a\of{1}=-0.001$, and $\gamma\of{1}=0$.  A training signal modulated of 300 random symbols was sent through the channel.  This training signal had a fractional bandwidth $\frac{1}{f_cT}=1$.  Parameter estimates were given by $\tilde{\gamma}\of{1}=\gamma\of{1}+\frac{\Delta_\gamma}{2}$ and $\tilde{a}\of{1}=a\of{1}+\frac{\Delta_a}{2}$.  Thus, we simulated the worst case scenario where the parameter estimates were half a grid spacing away from the true parameter values.  The estimate of $\theta\of{1}$ was made based on \eqref{thetadopp}.  Next a data signal modulated from a series of 300 random data symbols was sent through the channel.  The error metric $\metricdopp$ was calculated for different many choices of grid spacing.  This simulation was repeated 10 times and the results were averaged.  We set $B=1.8138$ and plugged in the simulations parameters into \eqref{doppbound} to calculate the error upper bound as
\begin{eqnarray}
\lefteqn{\metric} \nonumber \\
&\leq&  1.004 -0.2492 \textstyle \cos^2 ( \frac{0.9069 \Delta_\gamma}{T}) \nonumber \\
& & \times \mbox{ } \left( \sinc\left(  149.9251 \Delta_a +0.0865  \right)  +  \sinc\left( 149.9251 \Delta_a -0.0865 \right)  \right)^2. 
\end{eqnarray}

\begin{figure}[htbp]
	\begin{center}
		\psfrag{x04}[c][c][1][0]{$0$}
		\psfrag{x05}[c][c][1][0]{$0.2$}
		\psfrag{x06}[c][c][1][0]{$0.4$}
		\psfrag{x07}[c][c][1][0]{$0.6$}
		\psfrag{x08}[c][c][1][0]{$0.8$}
		\psfrag{x09}[c][c][1][0]{$1.0$}
		\psfrag{x10}[c][c][1][0]{$1.2$}
		\psfrag{x11}[c][c][1][0]{$1.4$}
		\psfrag{x12}[c][c][1][0]{$1.6$}
		\psfrag{v01}[l][c][1][0]{$0.04$}
		\psfrag{v02}[l][c][1][0]{$0.06$}
		\psfrag{v03}[l][c][1][0]{$0.08$}
		\psfrag{v04}[l][c][1][0]{$0.10$}
		\psfrag{v05}[l][c][1][0]{$0.12$}
		\psfrag{v06}[l][c][1][0]{$0.14$}
		\psfrag{v07}[l][c][1][0]{$0.16$}
		\psfrag{v08}[l][c][1][0]{$0.18$}
		\psfrag{v09}[l][c][1][0]{$0.20$}
		\psfrag{v10}[l][c][1][0]{$0.22$}
		\psfrag{v11}[l][c][1][0]{$0.24$}
		\psfrag{v12}[r][c][1][0]{$0$}
		\psfrag{v13}[r][c][1][0]{$0.05$}
		\psfrag{v14}[r][c][1][0]{$0.10$}
		\psfrag{v15}[r][c][1][0]{$0.15$}
		\psfrag{v16}[r][c][1][0]{$0.20$}
		\psfrag{v17}[r][c][1][0]{$0.25$}
		\psfrag{s06}[r][c][1.2][270]{$\frac{\Delta_\gamma}{T}$}
		\psfrag{s05}[c][b][1.2][0]{$\Delta_a \times 10^{3}$}
		\psfrag{s11}{}
		\psfrag{s07}{}
		\psfrag{s10}{}
        \epsfxsize=4.0in
        \epsfbox{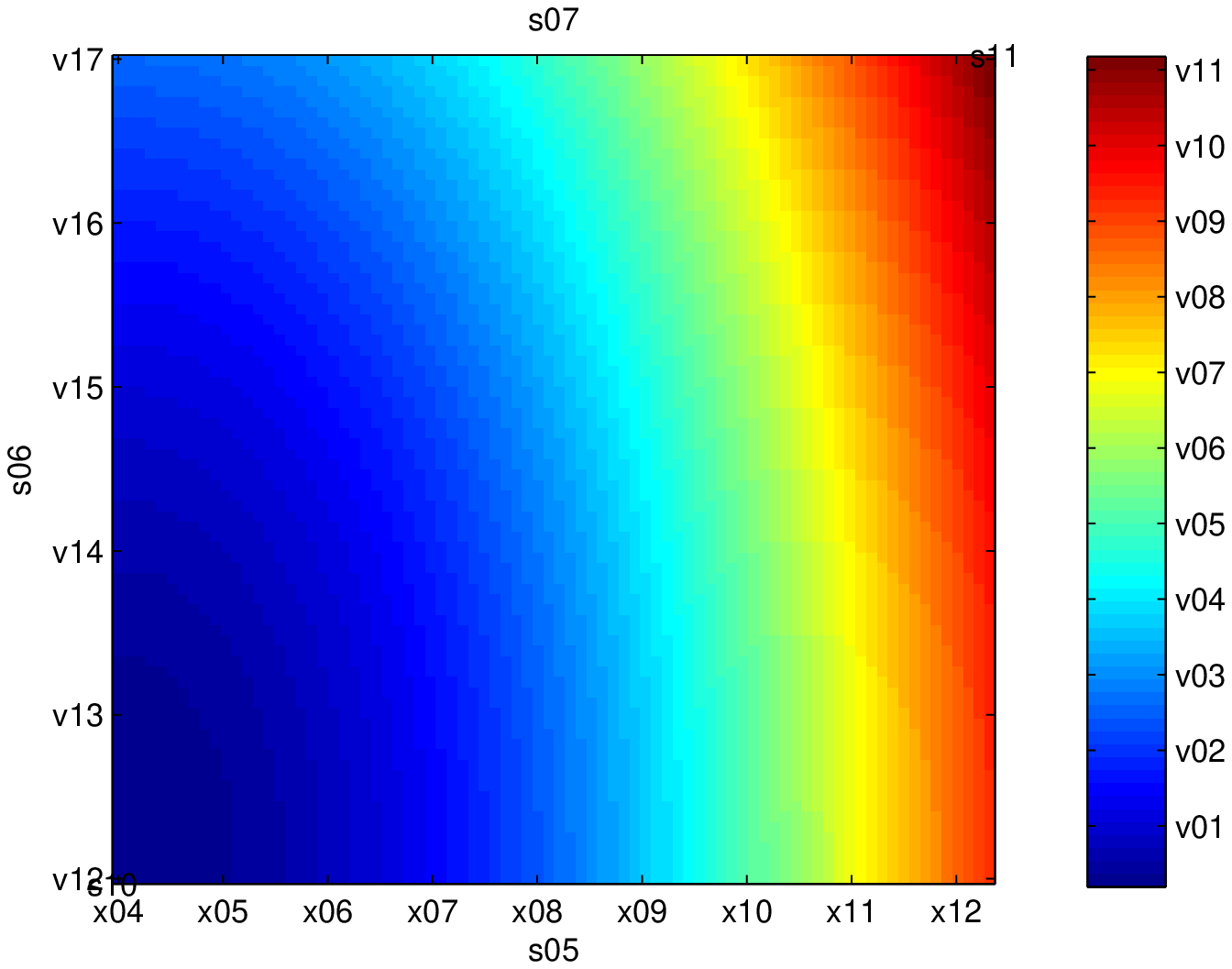}
    \end{center}
	\caption{Average simulated error using a Doppler-shift model}
    \label{fig:dopp_sim}
\end{figure}

\begin{figure}[htbp]
	\begin{center}

		\psfrag{x04}[c][c][1][0]{$0$}
		\psfrag{x05}[c][c][1][0]{$0.2$}
		\psfrag{x06}[c][c][1][0]{$0.4$}
		\psfrag{x07}[c][c][1][0]{$0.6$}
		\psfrag{x08}[c][c][1][0]{$0.8$}
		\psfrag{x09}[c][c][1][0]{$1.0$}
		\psfrag{x10}[c][c][1][0]{$1.2$}
		\psfrag{x11}[c][c][1][0]{$1.4$}
		\psfrag{x12}[c][c][1][0]{$1.6$}
		\psfrag{v01}[l][c][1][0]{$0.04$}
		\psfrag{v02}[l][c][1][0]{$0.06$}
		\psfrag{v03}[l][c][1][0]{$0.08$}
		\psfrag{v04}[l][c][1][0]{$0.10$}
		\psfrag{v05}[l][c][1][0]{$0.12$}
		\psfrag{v06}[l][c][1][0]{$0.14$}
		\psfrag{v07}[l][c][1][0]{$0.16$}
		\psfrag{v08}[l][c][1][0]{$0.18$}
		\psfrag{v09}[l][c][1][0]{$0.20$}
		\psfrag{v10}[l][c][1][0]{$0.22$}
		\psfrag{v11}[l][c][1][0]{$0.24$}
		\psfrag{v12}[r][c][1][0]{$0$}
		\psfrag{v13}[r][c][1][0]{$0.05$}
		\psfrag{v14}[r][c][1][0]{$0.10$}
		\psfrag{v15}[r][c][1][0]{$0.15$}
		\psfrag{v16}[r][c][1][0]{$0.20$}
		\psfrag{v17}[r][c][1][0]{$0.25$}
		\psfrag{s06}[r][c][1.2][270]{$\frac{\Delta_\gamma}{T}$}
		\psfrag{s05}[c][b][1.2][0]{$\Delta_a \times 10^{3}$}
		\psfrag{s11}{}
		\psfrag{s07}{}
		\psfrag{s10}{}
        \epsfxsize=4.0in
        \epsfbox{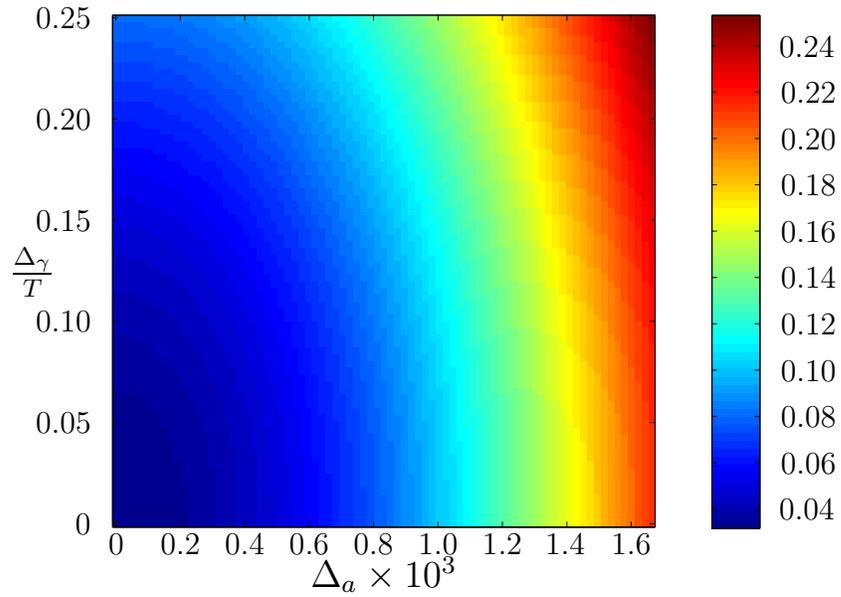}
    \end{center}
	\caption{Derived upper bound for a Doppler-shift model}
    \label{fig:dopp_bound}
\end{figure}

\begin{figure}[htbp]
	\begin{center}
		\psfrag{x04}[c][c][1][0]{$0$}
		\psfrag{x05}[c][c][1][0]{$0.2$}
		\psfrag{x06}[c][c][1][0]{$0.4$}
		\psfrag{x07}[c][c][1][0]{$0.6$}
		\psfrag{x08}[c][c][1][0]{$0.8$}
		\psfrag{x09}[c][c][1][0]{$1.0$}
		\psfrag{x10}[c][c][1][0]{$1.2$}
		\psfrag{x11}[c][c][1][0]{$1.4$}
		\psfrag{x12}[c][c][1][0]{$1.6$}
		\psfrag{v01}[l][c][1][0]{$0.05$}
		\psfrag{v02}[l][c][1][0]{$0.10$}
		\psfrag{v03}[l][c][1][0]{$0.15$}
		\psfrag{v04}[l][c][1][0]{$0.20$}
		\psfrag{v05}[l][c][1][0]{$0.25$}
		\psfrag{v06}[r][c][1][0]{$0$}
		\psfrag{v07}[r][c][1][0]{$0.05$}
		\psfrag{v08}[r][c][1][0]{$0.10$}
		\psfrag{v09}[r][c][1][0]{$0.15$}
		\psfrag{v10}[r][c][1][0]{$0.20$}
		\psfrag{s18}[l][c][1][0]{Simulated}
		\psfrag{s19}[l][c][1][0]{Derived}
		\psfrag{s10}[r][c][1.2][270]{$\frac{\Delta_\gamma}{T}$}
		\psfrag{s09}[c][b][1.2][0]{$\Delta_a \times 10^{3}$}
		\psfrag{s11}{}
		\psfrag{s14}{}
		\psfrag{s15}{}
        \epsfxsize=4.0in
        \epsfbox{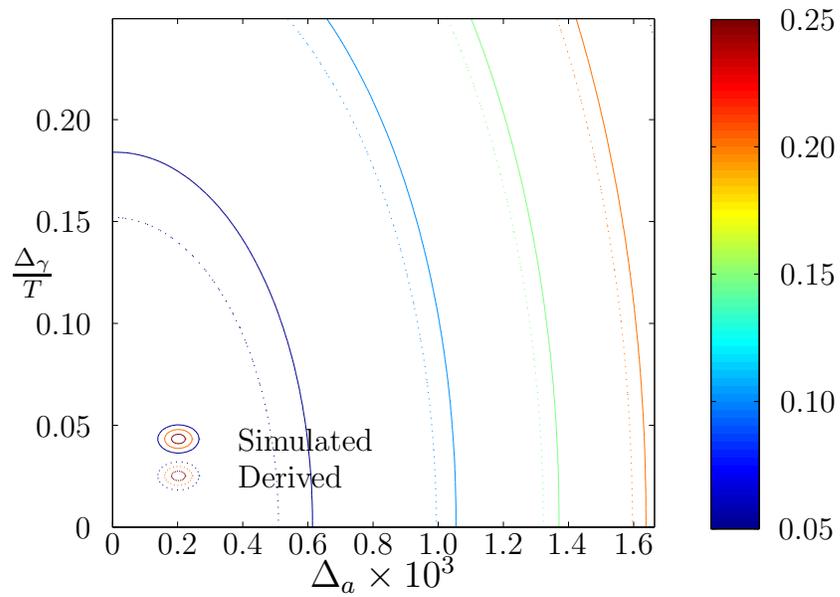}
    \end{center}
	\caption{Derived upper bound vs. simulated error for a Doppler-shift model}
    \label{fig:dopp_contour}
\end{figure}

\begin{figure}[htbp]
	\begin{center}
		\psfrag{x01}[c][c][1][0]{$0$}
		\psfrag{x02}[c][c][1][0]{$0.05$}
		\psfrag{x03}[c][c][1][0]{$0.10$}
		\psfrag{x04}[c][c][1][0]{$0.15$}
		\psfrag{x05}[c][c][1][0]{$0.20$}
		\psfrag{x06}[c][c][1][0]{$0.25$}
		
		\psfrag{v01}[r][c][1][0]{$0$ }
		\psfrag{v02}[r][c][1][0]{$0.05$}
		\psfrag{v03}[r][c][1][0]{$0.10$}
		\psfrag{v04}[r][c][1][0]{$0.15$}
		\psfrag{v05}[r][c][1][0]{$0.20$}
		\psfrag{v06}[r][c][1][0]{$0.25$}
		\psfrag{v07}[r][c][1][0]{$0.30$}
		\psfrag{v08}[r][c][1][0]{$0.35$}
		\psfrag{s10}{ }
		\psfrag{s11}{ }
		\psfrag{s13}[l][l][1][0]{$\Delta_a= 0$ }
		\psfrag{s14}[l][l][1][0]{$\Delta_a= 0.8011 \times10^{-3}$ }
		\psfrag{s15}[l][l][1][0]{$\Delta_a= 1.6639 \times10^{-3}$ }
		\psfrag{s05}[c][b][1.2][0]{$\frac{\Delta_\gamma}{T}$}

        \epsfxsize=3.8in
        \epsfbox{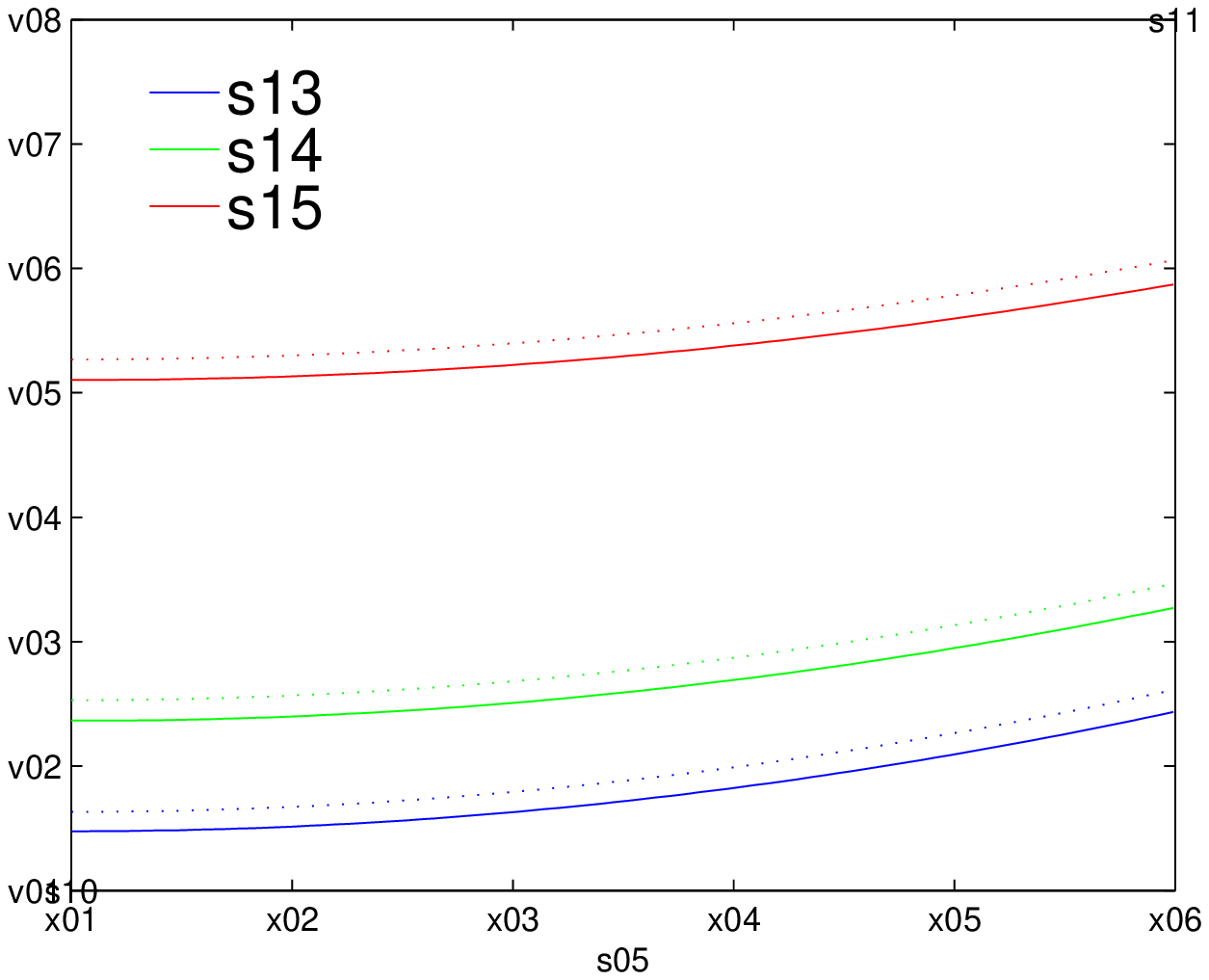}
    \end{center}
	\caption{Simulated error and derived upper bound vs. $\frac{\Delta_\gamma}{T}$ for fixed values of $\Delta_a$ using time-shift/Doppler-shift model.  Simulated error is plotted with solid lines while the derived bound is plotted with dotted lines.}
    \label{fig:dopp_vs_gamma}
\end{figure}

\begin{figure}[htbp]
	\begin{center}
		\psfrag{x01}[c][c][1][0]{$0$}
		\psfrag{x02}[c][c][1][0]{$0.2$}
		\psfrag{x03}[c][c][1][0]{$0.4$}
		\psfrag{x04}[c][c][1][0]{$0.6$}
		\psfrag{x05}[c][c][1][0]{$0.8$}
		\psfrag{x06}[c][c][1][0]{$1.0$}
		\psfrag{x07}[c][c][1][0]{$1.2$}
		\psfrag{x08}[c][c][1][0]{$1.4$}
		\psfrag{x09}[c][c][1][0]{$1.6$}
		
		\psfrag{v01}[r][c][1][0]{$0$ }
		\psfrag{v02}[r][c][1][0]{$0.05$}
		\psfrag{v03}[r][c][1][0]{$0.10$}
		\psfrag{v04}[r][c][1][0]{$0.15$}
		\psfrag{v05}[r][c][1][0]{$0.20$}
		\psfrag{v06}[r][c][1][0]{$0.25$}

		\psfrag{s10}{ }
		\psfrag{s11}{ }
		\psfrag{s13}[l][l][1][0]{$\Delta_\gamma/T= 0$ }
		\psfrag{s14}[l][l][1][0]{$\Delta_\gamma/T=0.1674$ }
		\psfrag{s15}[l][l][1][0]{$\Delta_\gamma/T=0.2496$ }
		\psfrag{s05}[c][b][1][0]{$\Delta_a \times 10^3$}

        \epsfxsize=3.8in
        \epsfbox{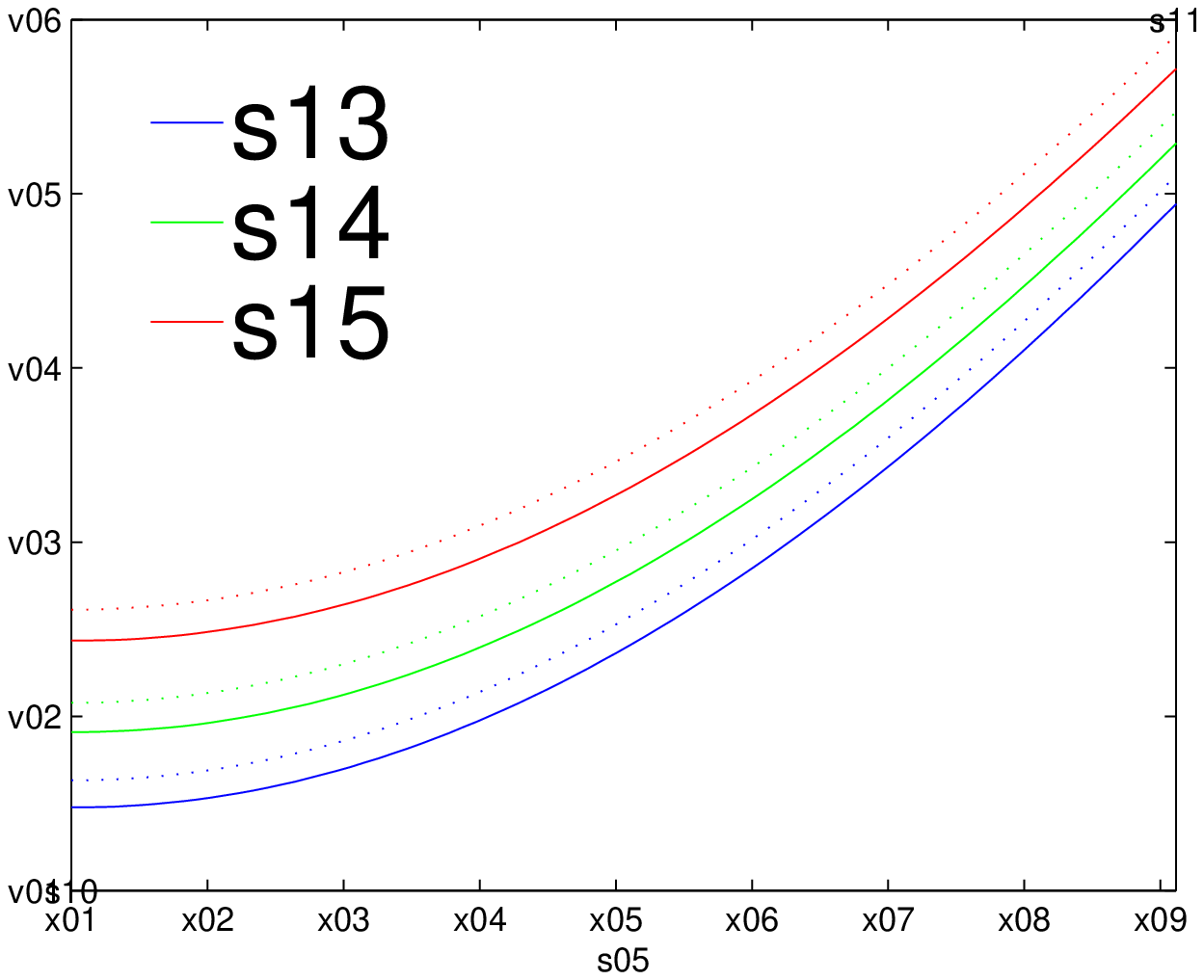}
    \end{center}
	\caption{Simulated error and derived upper bound vs. $\Delta_a$ for fixed values of $\frac{\Delta_\gamma}{T}$ using time-shift/Doppler-shift model.  Simulated error is plotted with solid lines while the derived bound is plotted with dotted lines. }
    \label{fig:dopp_vs_a}
\end{figure}

\Figref{dopp_bound} plots this upper bound while \Figref{dopp_sim} plots the average simulated error.  \Figref{dopp_contour} shows the two together, and it can be seen that the simulated error is less than the upper bound we derived.  These results are also shown in \Figref{dopp_vs_gamma} and \Figref{dopp_vs_a}.  \Figref{dopp_vs_gamma} plots the simulated error and the derived bound for fixed values of $\Delta_a$, while \Figref{dopp_vs_a} plots the simulated error and the derived bound  for fixed values of $\Delta_\gamma$.  It all cases it can be seen that the bound derived in \eqref{doppbound} does indeed hold and appears to be relatively tight.

\section{Doppler-shift Model vs. Scale-shift Model} \label{sec:dopp_vs_ts}

A primary motivation of this thesis is to analyze the cost of using the Doppler approximation.  In \chref{error} we derived an expression \eqref{ts_final} which upper bounds the average estimation error when a model composed of the effects scale-shifts is used.  Then in \chref{doppler} we derived a similar expression \eqref{doppbound} which upper bounds the average estimation error when a model composed of the effects of Doppler-shifts is used.  Through simulation, we have shown that both these upper bounds are relatively tight. From these simulations, we can also see that there is an additional cost occurred when using the Doppler approximation.   For instance, by comparing \Figref{scale_vs_gamma} with \Figref{dopp_vs_gamma}, we see that when $\Delta_\gamma = \Delta_a = 0$, the simulated error is 0 when the scale-shift model is used, but when a Doppler-shift model is used, the simulated error is roughly 0.024.

We now analyze the effects of the fractional bandwidth $\frac{1}{f_c T}$ and rate of change of the channel paths $\{ \al \}^L_{l=1}$ on the error resulting from the Doppler approximation.  To do this, we consider the case where $L=1$ and $M \gg 1$ to approximate the upper bound of \eqref{ts_final} as 
\begin{eqnarray}
  \mc{E}_{\text{wideband}}
  &=&  |\theta\of{l}|^2 + \frac{1.2011}{M}|\theta\of{l}|^2 - 
	\frac{1}{4}|\theta\of{l}|^2 \cos^2\bigg(\frac{B\Delta_\gamma}{2T}\bigg)
  \nonumber\\&&\mbox{}\times 
	\bigg[\sinc\bigg( \bigg( \pi f_c T \Big(1+\frac{1}{1-\al}\Big) 
		+ \frac{B}{1-\al}\bigg)\frac{M \Delta_a}{4\pi} \bigg)
  \nonumber\\&&\mbox{}+
		\sinc\bigg( \bigg( \pi f_c T \Big(1+\frac{1}{1-\al}\Big) 
		- \frac{B}{1-\al}\bigg)\frac{M \Delta_a}{4\pi} \bigg)
		\bigg]^2,			\label{eq:wide1}
\end{eqnarray}
and the upper bound of \eqref{doppbound} as
\begin{eqnarray}
  \mc{E}_{\text{Doppler}} 
  &=&  |\theta\of{l}|^2 + \frac{1.2011}{M}|\theta\of{l}|^2 - 
	\frac{1}{4}|\theta\of{l}|^2 \cos^2\bigg(\frac{B\Delta_\gamma}{2T}\bigg)
  \nonumber\\&&\mbox{}\times 
	\bigg[\sinc\bigg( \pi f_c T \Big(1+\frac{1}{1-\al}\Big)
		\frac{M \Delta_a}{4\pi} + \frac{B}{1-\al}\frac{M \al}{2\pi} \bigg) 
  \nonumber\\&&\mbox{}+
	\bigg[\sinc\bigg( \pi f_c T \Big(1+\frac{1}{1-\al}\Big)
		\frac{M \Delta_a}{4\pi} - \frac{B}{1-\al}\frac{M \al}{2\pi} \bigg) 
		\bigg]^2.			\label{eq:dopp1}
\end{eqnarray}

By considering the difference between $\mc{E}_{\text{wideband}}$ and $\mc{E}_{\text{Doppler}}$ for practical (relatively small) values of $\mc{E}_{\text{wideband}}$, we can analyze the cost of the Doppler approximation.   Notice that the scale-error contribution to $\mc{E}_{\text{wideband}}$ is proportional to the magnitude of
\begin{eqnarray}
  \phi_{\text{wideband}}
  &\defn& \bigg( \pi f_c T \Big(1+\frac{1}{1-\al}\Big) 
	+ \frac{B}{1-\al}\bigg)\frac{M \Delta_a}{4\pi} .
						\label{eq:phi_wide}
\end{eqnarray}
In other words, when $|\phi_{\text{wideband}}|$ is very small, the
scale-error contribution to $\mc{E}_{\text{wideband}}$ is also very small.  Similarly, the scale-error contribution to $\mc{E}_{\text{Dopper}}$ is proportional to the magnitude of
\begin{eqnarray}
  \phi_{\text{Doppler}}
  &\defn& \pi f_c T \Big(1+\frac{1}{1-\al}\Big)
	\frac{M \Delta_a}{4\pi} + \frac{B}{1-\al}\frac{M \al}{2\pi}.
						\label{eq:phi_dopp}
\end{eqnarray}
Now we consider a fixed value of $\phi_{\text{wideband}}$ and solve for $\phi_{\text{Doppler}}$ in terms of $\phi_{\text{wideband}}$.  Solving for $\frac{M \Delta_a}{4\pi}$ in \eqref{phi_wide} and plugging the result into \eqref{phi_dopp}, we can write
\begin{eqnarray}
  \phi_{\text{Doppler}} - \phi_{\text{wideband}}
  &=&  \frac{B}{1-a}\bigg(
	\frac{M \al}{2\pi} - \frac{\phi_{\text{wideband}}}{
  	\pi f_c T \Big(1+\frac{1}{1-\al}\Big) + \frac{B}{1-\al}} \bigg) .
						\label{eq:phi_dopp2}
\end{eqnarray}
From \eqref{phi_dopp2}, we notice that as $f_c T\rightarrow \infty$, we get
\begin{eqnarray}
  \phi_{\text{Doppler}} -\phi_{\text{wideband}}
  &\rightarrow&  \frac{B}{1-a}
	\frac{M a}{2\pi} .
						\label{eq:phi_dopp3}
\end{eqnarray}
This shows that there two conditions in which the Doppler approximation is accurate.  One is that the signal should have a small fractional bandwidth (i.e. large $f_c T$). The other is that $a \ll \phi_{\text{wideband}}\frac{2\pi}{BM} = \phi_{\text{wideband}}\frac{\pi}{B} \frac{2}{(MT)\times (1/T)}$ where $MT$ can be recognized as the signal duration and $1/T$ can be recognized as the signal bandwidth.

\begin{figure}[htbp]
	\begin{center}
		\psfrag{x01}[c][c][1][0]{$5$}
		\psfrag{x02}[c][c][1][0]{$10$}
		\psfrag{x03}[c][c][1][0]{$15$}
		\psfrag{x04}[c][c][1][0]{$20$}
		\psfrag{x05}[c][c][1][0]{$25$}
		\psfrag{x06}[c][c][1][0]{$30$}
		\psfrag{x07}[c][c][1][0]{$35$}
		\psfrag{x08}[c][c][1][0]{$40$}
		\psfrag{x09}[c][c][1][0]{$45$}
		\psfrag{x10}[c][c][1][0]{$50$}
		\psfrag{v01}[r][rc][1][0]{$0.0123$ }
		\psfrag{v02}[r][c][1][0]{$0.0124$}
		\psfrag{v03}[r][c][1][0]{$0.0125$}
		\psfrag{v04}[r][c][1][0]{$0.0126$}
		\psfrag{v05}[r][c][1][0]{$0.0127$}
		\psfrag{v06}[r][c][1][0]{$0.0128$}
		\psfrag{v07}[r][c][1][0]{$0.0129$}
		\psfrag{v08}[r][c][1][0]{$0.0130$}
		
		\psfrag{s10}{ }
		\psfrag{s11}{ }
	
		\psfrag{s13}[l][l][1][0]{$\mc{E}_{\text{wideband}}$ }
		\psfrag{s14}[l][l][1][0]{$\mc{E}_{\text{Doppler}}$ }
		\psfrag{s05}[c][b][1][0]{$f_c T$}

        \epsfxsize=3.8in
        \epsfbox{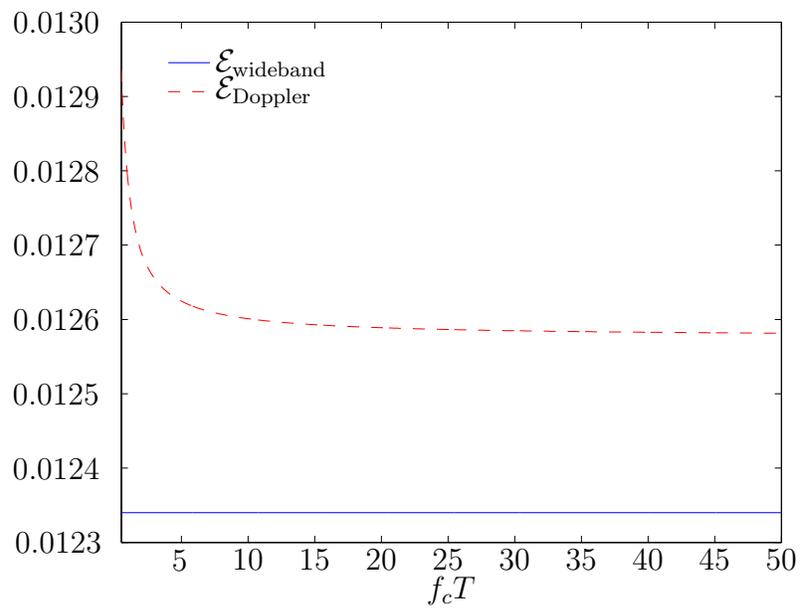}
    \end{center}
	\caption{Upper bound of \eqref{ts_final} and upper bound of \eqref{doppbound} vs. $f_c T$ }
    \label{fig:dopp_vs_scale}
\end{figure}

To further illustrate the effects of $f_c T$ on the error resulting from the Doppler approximation, we plot the upper bounds given in \eqref{wide1} and \eqref{dopp1} over many values of $f_c T$ in \Figref{dopp_vs_scale}.  For this plot, we set $L=1$, $\al=.0008$, $B=1.8128$, $M=100$, $\Delta_\gamma=0$, and chose $\Delta_a$ so that $\mc{E}_{\text{wideband}}$ was constant for all $f_c T$.  As can be seen, the approximation becomes more accurate as $f_c T$ increases.

\chapter{Experimental Results and Conclusion} 
\label{ch:matlab}

\section{Simulated Channel}

\begin{figure}[htbp]
	\begin{center}
        \epsfxsize=4.0in
        \epsfbox{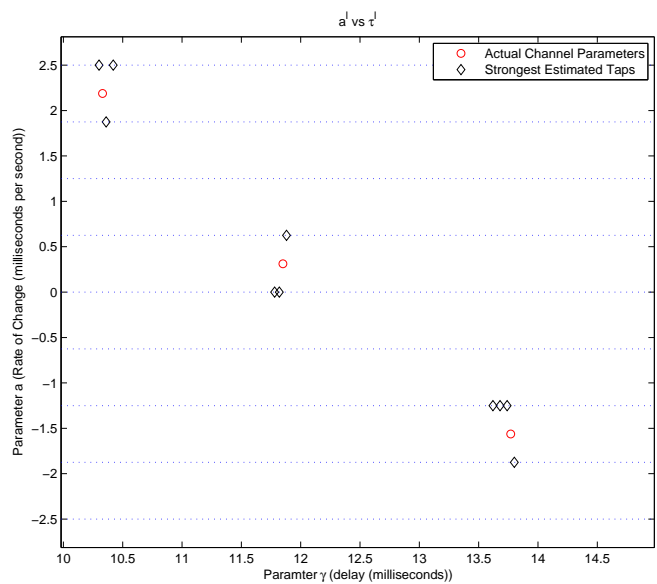}
    \end{center}
	\caption{Plot of true channel parameters (red circles) and the grid restricted parameter estimates (black diamonds).  The bottom axis ranges from $\gamma$=10ms to $\gamma$=15 ms.  The left axis ranges from $a$=-2.5 ms/s to $a$=2.5 ms/s.  \Figref{notexamp2} shows another view of this channel estimate.}
    \label{fig:notexamp1}
\end{figure}

\begin{figure}[htbp]
	\begin{center}
        \epsfxsize=4.0in
        \epsfbox{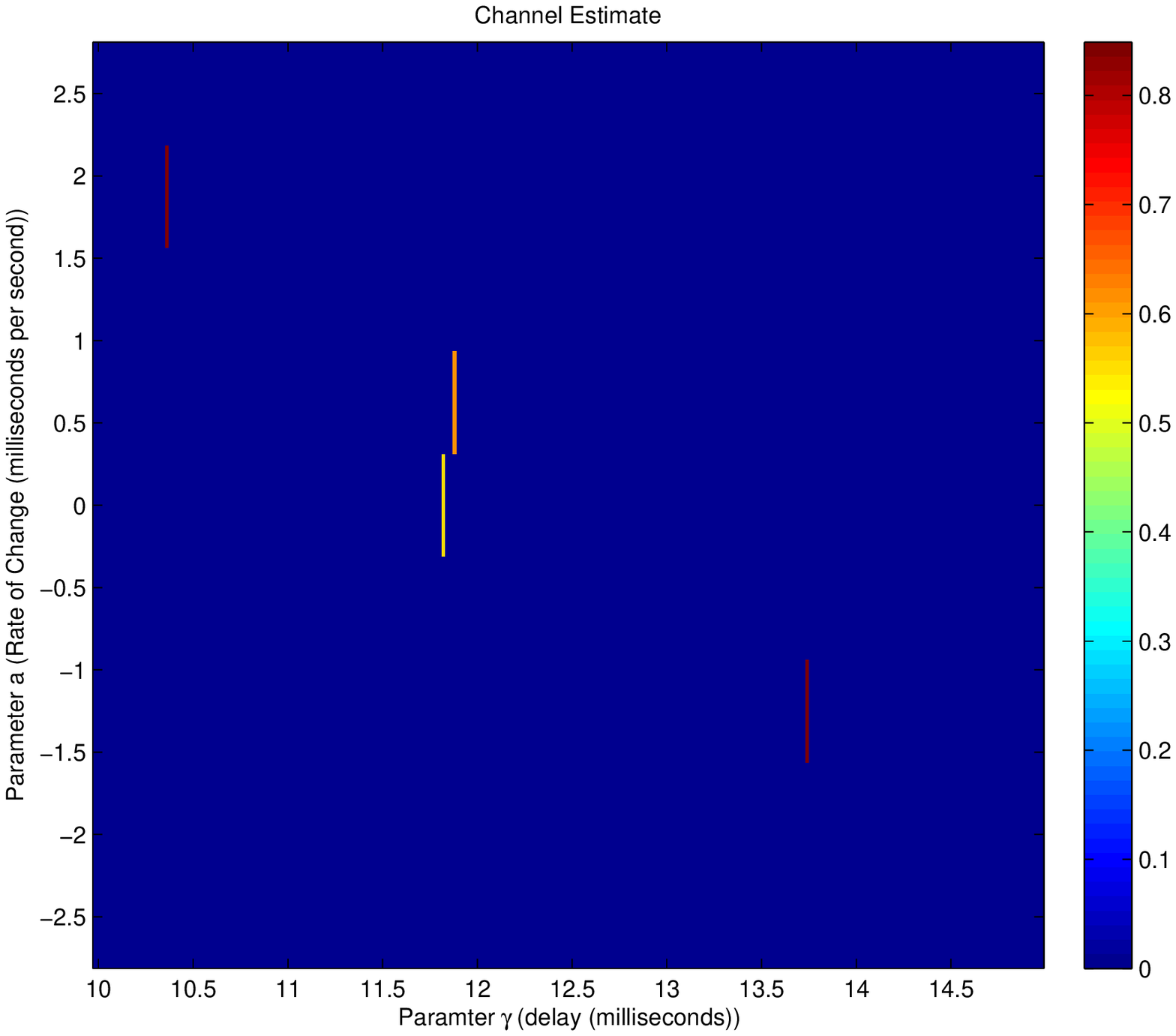}
    \end{center}
	\caption{Another view of the channel estimate shown in \figref{notexamp1}}
    \label{fig:notexamp2}
\end{figure}

For further analysis, we simulate the estimation strategy described in \chref{estimation_strategy} over a three path channel.  In this simulation, the training signal we use has a carrier frequency of 10 kHz, a baud rate of 10 kHz, and a sampling frequency of 30 kHz.  The transmitted training signal is modulated from 200 random QPSK symbols.  In order to estimate the channel parameters, an under-determined Matrix composed of the effects of time-shifts and scale-shifts on the training signal is constructed.  We simulate the worst case scenario where the true channel parameters are half a grid spacing away from the parameter estimates (i.e. $\ahat= \al +\frac{\Delta_a}{2}$ and $\gammahat= \gammal +\frac{\Delta_\gamma}{2}$ for $l=1,..,L$).  \Figref{notexamp1} and \Figref{notexamp2} show the true channel parameters and parameter estimates found using Orthogonal Matching Pursuit (OMP) \cite{P1993}.

To analyze the estimation strategy further, we define two error metrics.  The first we call the training error which we define as $\frac{\vectornorm{\vec{z}-\vec{A}\vec{\hat{\theta}} }^2}{\vectornorm{\vec{z}}^2}$, where $\vec{z}$ is a vector of noisy measurements of the received training signal, $\vec{A}$ is a matrix composed of the effects of time-shifts and scale-shifts of the training signal, and $\vec{\hat{\theta}}$ is the sparse estimate of the channel gains.  The second we call the the data error which we define as $\frac{\vectornorm{\vec{y}-\vec{B}\vec{\hat{\theta}} }^2}{\vectornorm{\vec{y}}^2}$, where $\vec{y}$ is the received data signal (noiseless), and the columns of $\vec{B}$ are composed of the effects of time-shifts and scale-shifts on transmitted data signal.  In this simulation, the channel seen as the data signal is sent is the same as the channel seen as the training signal is sent.   However, we know that algorithms, such as least squares, can minimize the training error with inaccurate (noise corrupted) estimates \cite{C2007}.   For this reason, we are most interested in the data error for our analysis. 

\begin{figure}[htbp]
	\begin{center}
		\psfrag{x01}[c][c][1][0]{$0$}
		\psfrag{x02}[c][c][1][0]{$10$}
		\psfrag{x03}[c][c][1][0]{$20$}
		\psfrag{x04}[c][c][1][0]{$30$}
		\psfrag{x05}[c][c][1][0]{$40$}
		\psfrag{x06}[c][c][1][0]{$50$}
		\psfrag{x07}[c][c][1][0]{$60$}
		\psfrag{x08}[c][c][1][0]{$70$}
		\psfrag{x09}[c][c][1][0]{$80$}
		\psfrag{x10}[c][c][1][0]{$90$}
		\psfrag{x11}[c][c][1][0]{$100$}
		
		\psfrag{v01}[r][rc][1][0]{$-45$}
		\psfrag{v02}[r][rc][1][0]{$-40$}
		\psfrag{v03}[r][rc][1][0]{$-35$}
		\psfrag{v04}[r][rc][1][0]{$-30$}
		\psfrag{v05}[r][rc][1][0]{$-25$}
		\psfrag{v06}[r][rc][1][0]{$-20$}
		\psfrag{v07}[r][rc][1][0]{$-15$}
		\psfrag{v08}[r][rc][1][0]{$-10$}
		\psfrag{v09}[r][rc][1][0]{$-5$}
		\psfrag{v10}[r][rc][1][0]{$0$}
		
		\psfrag{s06}[r][c][1][270]{dB}
		\psfrag{s05}[c][b][1][0]{OMP iterations}
		\psfrag{s11}{}
		\psfrag{s13}[l][c][1][0]{Training Error}
		\psfrag{s14}[l][c][1][0]{Data Error}
		\psfrag{s07}{}
		\psfrag{s10}{}
        \epsfxsize=4.0in
        \epsfbox{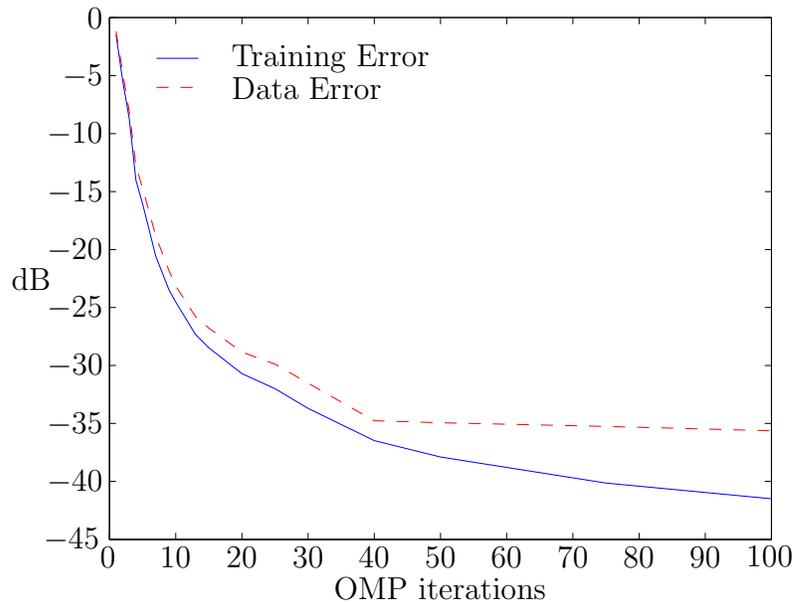}
    \end{center}
	\caption{Estimation error vs OMP iterations (noiseless case).}
    \label{fig:notomp1}
\end{figure}
\begin{figure}[htbp]
	\begin{center}
		\psfrag{x01}[c][c][1][0]{$0$}
		\psfrag{x02}[c][c][1][0]{$10$}
		\psfrag{x03}[c][c][1][0]{$20$}
		\psfrag{x04}[c][c][1][0]{$30$}
		\psfrag{x05}[c][c][1][0]{$40$}
		\psfrag{x06}[c][c][1][0]{$50$}
		\psfrag{x07}[c][c][1][0]{$60$}
		\psfrag{x08}[c][c][1][0]{$70$}
		\psfrag{x09}[c][c][1][0]{$80$}
		\psfrag{x10}[c][c][1][0]{$90$}
		\psfrag{x11}[c][c][1][0]{$100$}
		
		\psfrag{v01}[r][rc][1][0]{$-15$}
		\psfrag{v02}[r][rc][1][0]{$-10$}
		\psfrag{v03}[r][rc][1][0]{$-5$}
		\psfrag{v04}[r][rc][1][0]{$0$}
		
		\psfrag{s06}[r][c][1][270]{dB}
		\psfrag{s05}[c][b][1][0]{OMP iterations}
		\psfrag{s11}{}
		\psfrag{s13}[l][c][1][0]{Training Error}
		\psfrag{s14}[l][c][1][0]{Data Error}
		\psfrag{s07}{}
		\psfrag{s10}{}
        \epsfxsize=4.0in
        \epsfbox{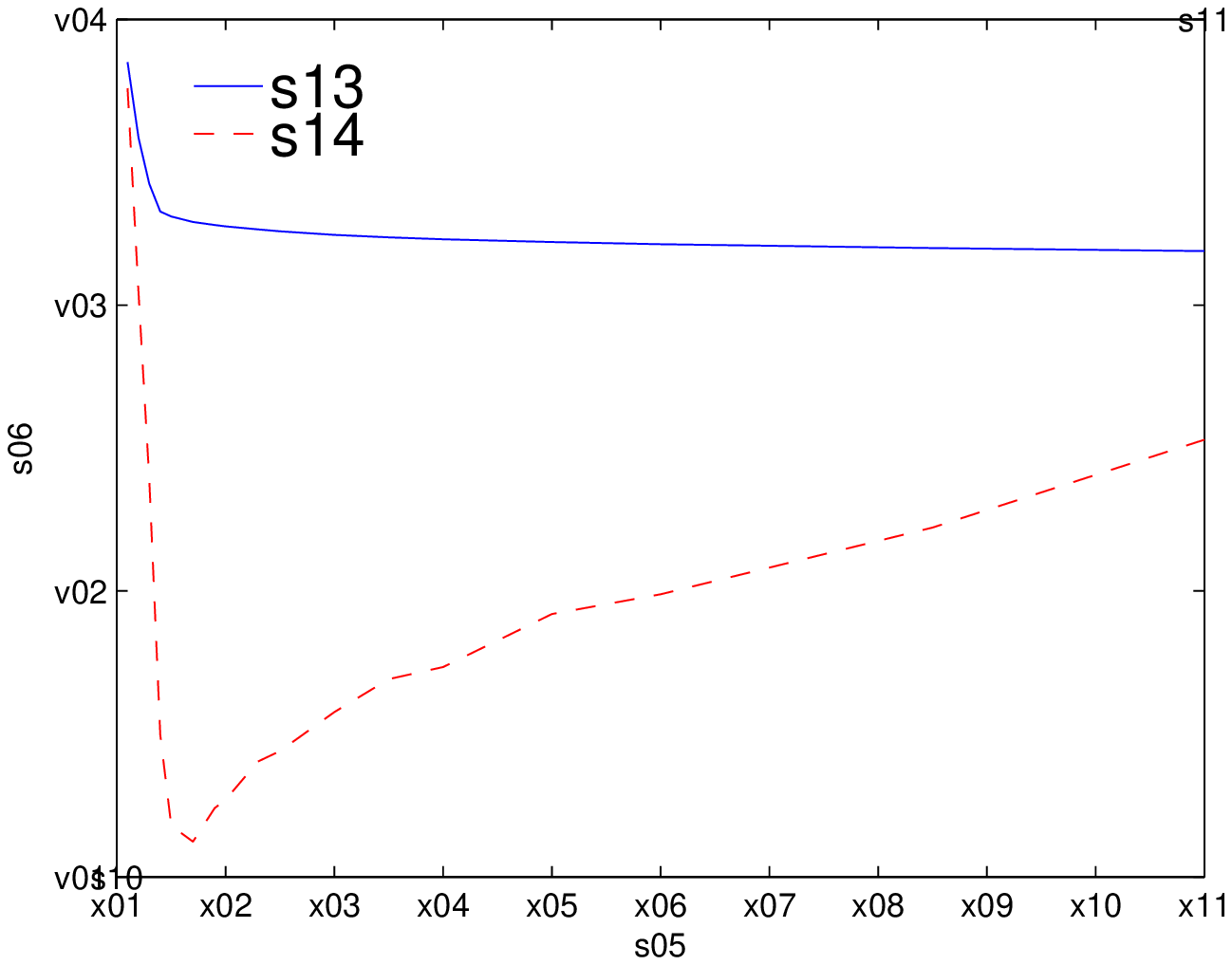}
    \end{center}
	\caption{Estimation error vs OMP iterations (SNR = 1 dB).}
    \label{fig:notomp1db}
\end{figure}
\begin{figure}[htbp]
	\begin{center}
		\psfrag{x01}[c][c][1][0]{$0$}
		\psfrag{x02}[c][c][1][0]{$10$}
		\psfrag{x03}[c][c][1][0]{$20$}
		\psfrag{x04}[c][c][1][0]{$30$}
		\psfrag{x05}[c][c][1][0]{$40$}
		\psfrag{x06}[c][c][1][0]{$50$}
		\psfrag{x07}[c][c][1][0]{$60$}
		\psfrag{x08}[c][c][1][0]{$70$}
		\psfrag{x09}[c][c][1][0]{$80$}
		\psfrag{x10}[c][c][1][0]{$90$}
		\psfrag{x11}[c][c][1][0]{$100$}
		
		\psfrag{v01}[r][rc][1][0]{$-16$}
		\psfrag{v02}[r][rc][1][0]{$-14$}
		\psfrag{v03}[r][rc][1][0]{$-12$}
		\psfrag{v04}[r][rc][1][0]{$-10$}
		\psfrag{v05}[r][rc][1][0]{$-8$}
		\psfrag{v06}[r][rc][1][0]{$-6$}
		\psfrag{v07}[r][rc][1][0]{$-4$}
		\psfrag{v08}[r][rc][1][0]{$-2$}
		\psfrag{v09}[r][rc][1][0]{$0$}
		
		\psfrag{s06}[r][c][1][270]{dB}
		\psfrag{s05}[c][b][1][0]{OMP iterations}
		\psfrag{s11}{}
		\psfrag{s13}[l][c][1][0]{Training Error}
		\psfrag{s14}[l][c][1][0]{Data Error}
		\psfrag{s07}{}
		\psfrag{s10}{}
        \epsfxsize=4.0in
        \epsfbox{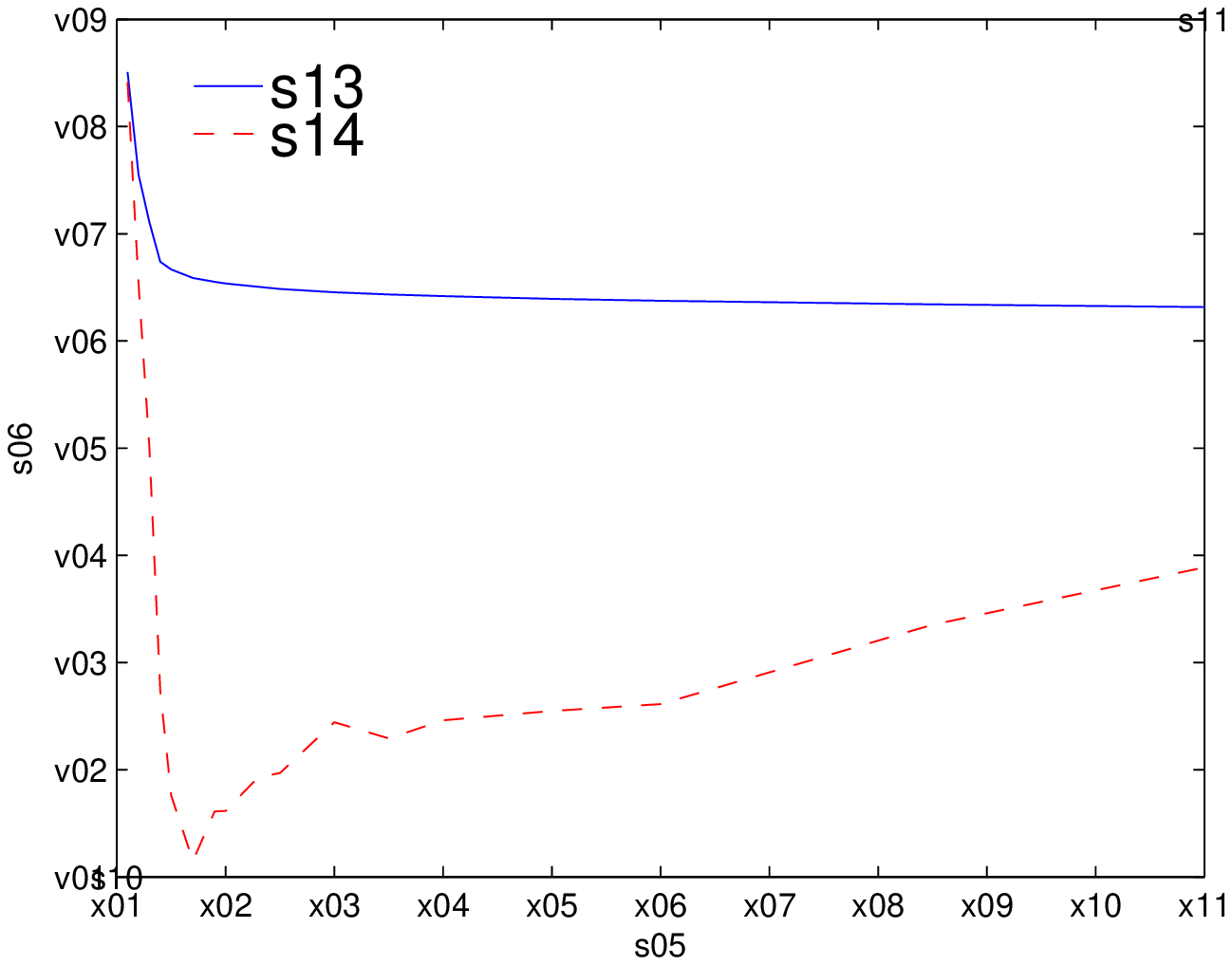}
    \end{center}
	\caption{Estimation error vs OMP iterations (SNR = 3 dB).}
    \label{fig:notomp3db}
\end{figure}
\begin{figure}[htbp]
	\begin{center}
		\psfrag{x01}[c][c][1][0]{$0$}
		\psfrag{x02}[c][c][1][0]{$10$}
		\psfrag{x03}[c][c][1][0]{$20$}
		\psfrag{x04}[c][c][1][0]{$30$}
		\psfrag{x05}[c][c][1][0]{$40$}
		\psfrag{x06}[c][c][1][0]{$50$}
		\psfrag{x07}[c][c][1][0]{$60$}
		\psfrag{x08}[c][c][1][0]{$70$}
		\psfrag{x09}[c][c][1][0]{$80$}
		\psfrag{x10}[c][c][1][0]{$90$}
		\psfrag{x11}[c][c][1][0]{$100$}
		
		\psfrag{v01}[r][rc][1][0]{$-25$}
		\psfrag{v02}[r][rc][1][0]{$-20$}
		\psfrag{v03}[r][rc][1][0]{$-15$}
		\psfrag{v04}[r][rc][1][0]{$-10$}
		\psfrag{v05}[r][rc][1][0]{$-5$}
		\psfrag{v06}[r][rc][1][0]{$0$}
		
		\psfrag{s06}[r][c][1][270]{dB}
		\psfrag{s05}[c][b][1][0]{OMP iterations}
		\psfrag{s11}{}
		\psfrag{s13}[l][c][1][0]{Training Error}
		\psfrag{s14}[l][c][1][0]{Data Error}
		\psfrag{s07}{}
		\psfrag{s10}{}
        \epsfxsize=4.0in
        \epsfbox{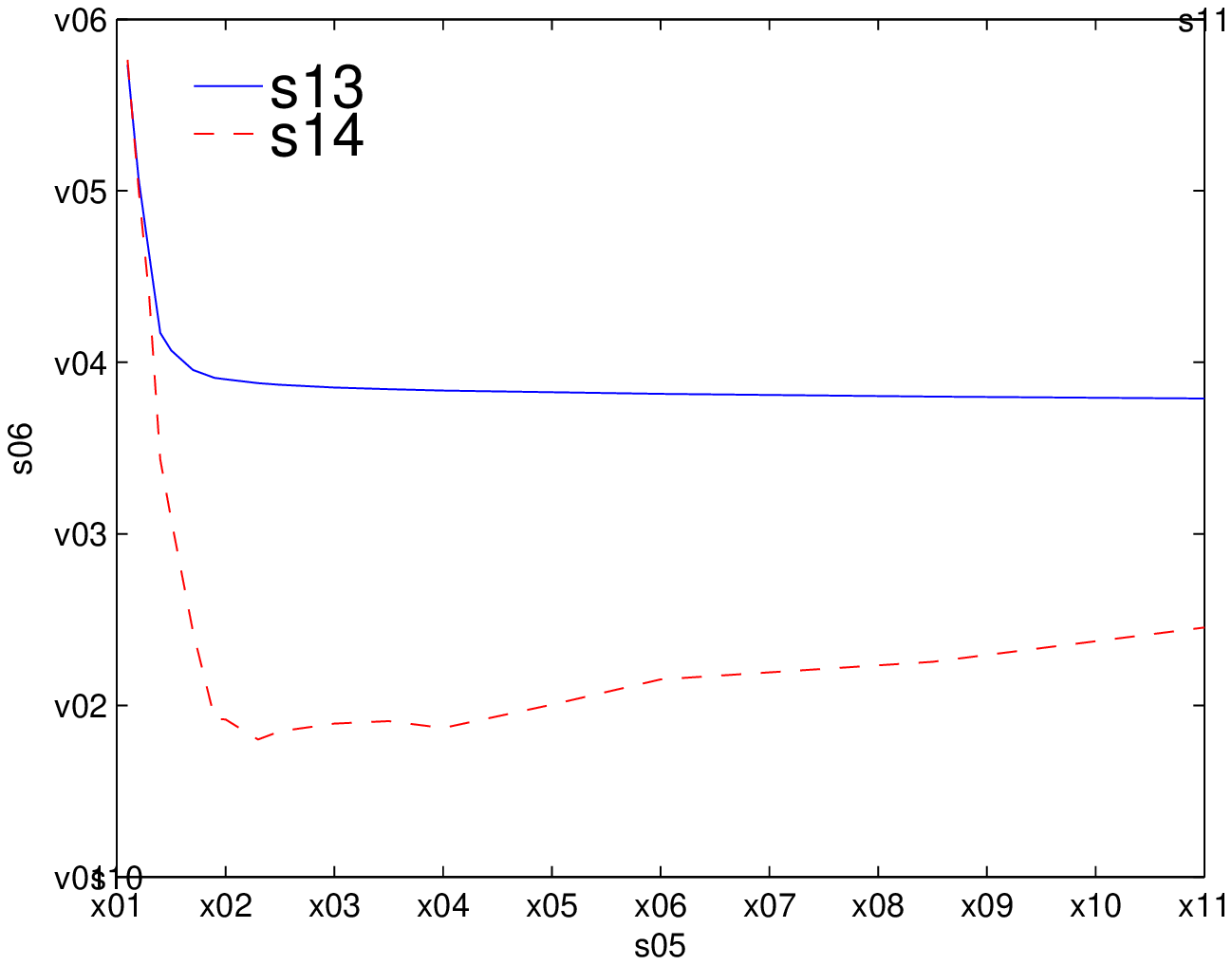}
    \end{center}
	\caption{Estimation error vs OMP iterations (SNR = 10 dB).}
    \label{fig:notomp10db}
\end{figure}

\Figref{notomp1} shows both the training error and the data error vs the number of iterations using OMP for the noiseless case.  We see that both the training error and the data error decrease as the number of OMP iterations increases.  However, when the received training signal $\vec{z}$ is noisy, we find this is not the case.  \Figref{notomp1db}, \Figref{notomp3db}, and \Figref{notomp10db} plot the results for different signal-to-noise ratios (SNRs).  As can be seen, the data error begins to increase as the number of OMP iterations increases.  The lower the SNR, the smaller the number of OMP iterations which minimizes the data error.  This illustrates how a sparse solution (few OMP iterations) can give a better solution which is less corrupted by noise.


\begin{figure}[htbp]
	\begin{center}
		\psfrag{x01}[c][c][1][0]{$0$}
		\psfrag{x02}[c][c][1][0]{$0.5$}
		\psfrag{x03}[c][c][1][0]{$1$}
		\psfrag{x04}[c][c][1][0]{$1.5$}
		\psfrag{x05}[c][c][1][0]{$2$}
		\psfrag{x06}[c][c][1][0]{$2.5$}
		\psfrag{x07}[c][c][1][0]{$3$}
		\psfrag{x08}[c][c][1][0]{$3.5$}
		\psfrag{x09}[c][c][1][0]{$4$}
				
		\psfrag{v01}[r][rc][1][0]{$-40$}
		\psfrag{v02}[r][rc][1][0]{$-35$}
		\psfrag{v03}[r][rc][1][0]{$-30$}
		\psfrag{v04}[r][rc][1][0]{$-25$}
		\psfrag{v05}[r][rc][1][0]{$-20$}
		\psfrag{v06}[r][rc][1][0]{$-15$}
		\psfrag{v07}[r][rc][1][0]{$-10$}
		\psfrag{v08}[r][rc][1][0]{$-5$}
		\psfrag{v09}[r][rc][1][0]{$0$}
		\psfrag{s06}[r][lc][1][270]{dB}
		\psfrag{s05}[c][b][1][0]{$\lambda$}
		\psfrag{s11}{}
		\psfrag{s13}[l][c][1][0]{Training Error}
		\psfrag{s14}[l][c][1][0]{Data Error}
		\psfrag{s07}{}
		\psfrag{s10}{}

        \epsfxsize=4.0in
        \epsfbox{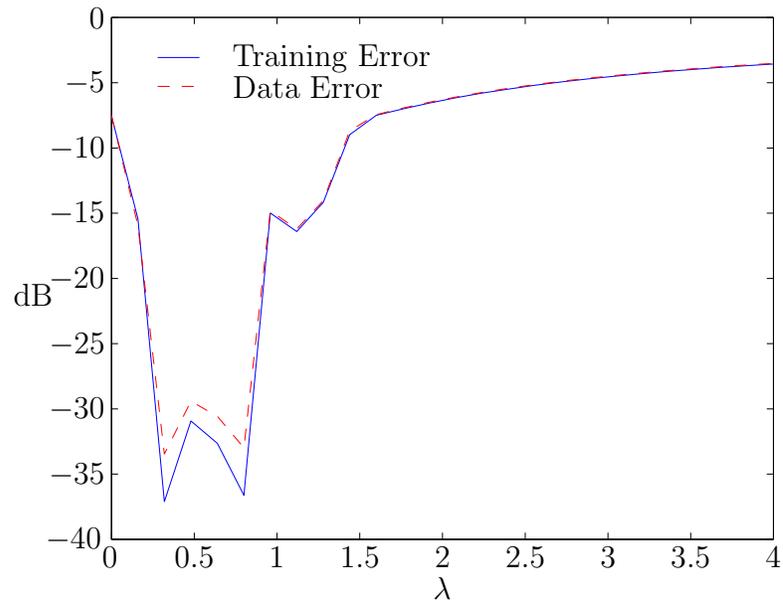}
    \end{center}
	\caption{Estimation error occurred using Basis Pursuit (noiseless case).}
    \label{fig:newbpnonoise}
\end{figure}

\begin{figure}[htbp]
	\begin{center}
		\psfrag{x01}[c][c][1][0]{$0$}
		\psfrag{x02}[c][c][1][0]{$0.5$}
		\psfrag{x03}[c][c][1][0]{$1$}
		\psfrag{x04}[c][c][1][0]{$1.5$}
		\psfrag{x05}[c][c][1][0]{$2$}
		\psfrag{x06}[c][c][1][0]{$2.5$}
		\psfrag{x07}[c][c][1][0]{$3$}
		\psfrag{x08}[c][c][1][0]{$3.5$}
		\psfrag{x09}[c][c][1][0]{$4$}
				
		\psfrag{v01}[r][rc][1][0]{$-7$}
		\psfrag{v02}[r][rc][1][0]{$-6$}
		\psfrag{v03}[r][rc][1][0]{$-5$}
		\psfrag{v04}[r][rc][1][0]{$-4$}
		\psfrag{v05}[r][rc][1][0]{$-3$}
		\psfrag{v06}[r][rc][1][0]{$-2$}
		\psfrag{v07}[r][rc][1][0]{$-1$}
		\psfrag{v08}[r][rc][1][0]{$0$}
		\psfrag{v09}[r][rc][1][0]{$1$}
		
		\psfrag{s06}[r][c][1][270]{dB}
		\psfrag{s05}[c][b][1][0]{$\lambda$}
		\psfrag{s11}{}
		\psfrag{s13}[l][c][1][0]{Training Error}
		\psfrag{s14}[l][c][1][0]{Data Error}
		\psfrag{s07}{}
		\psfrag{s10}{}
        \epsfxsize=4.0in
        \epsfbox{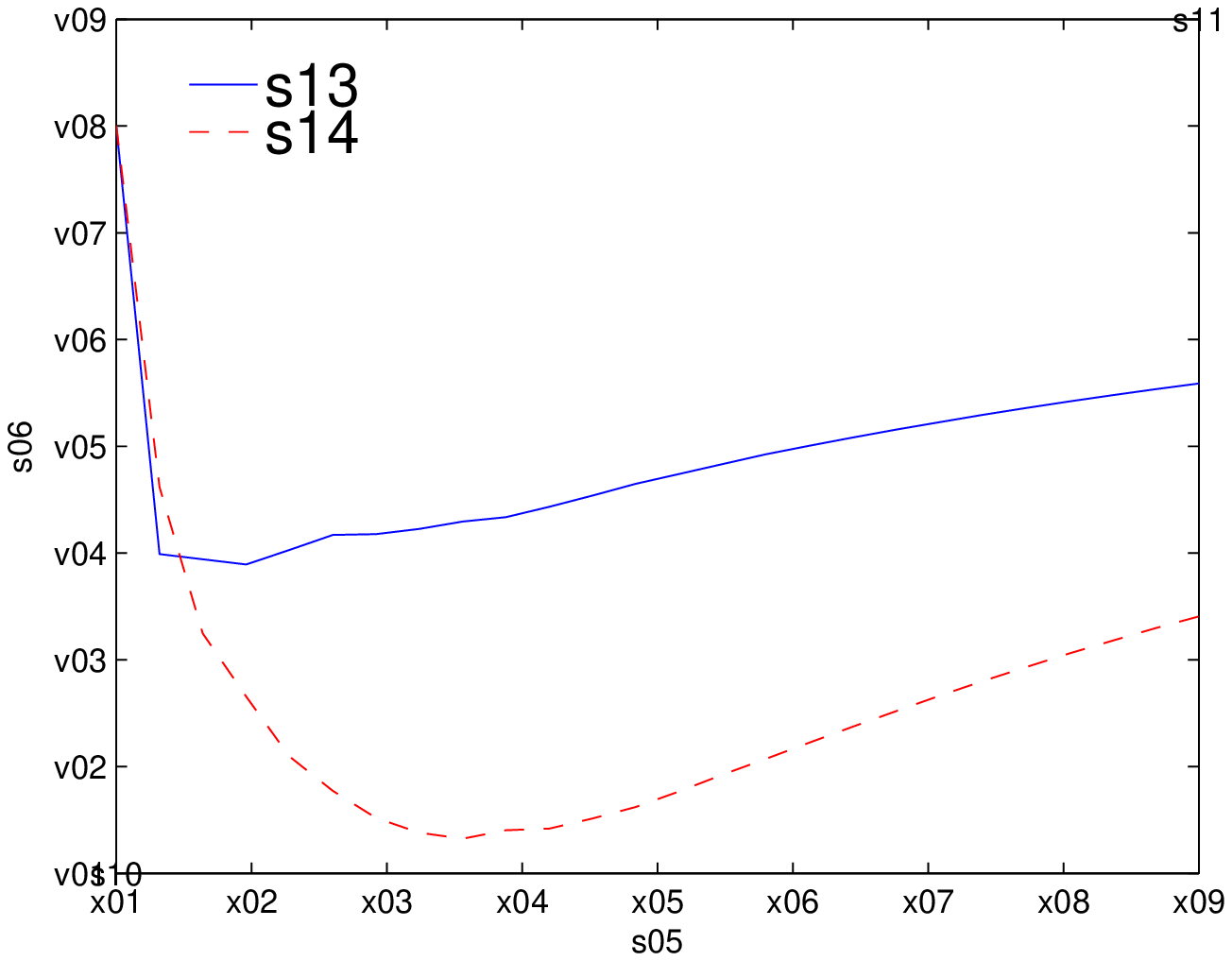}
    \end{center}
	\caption{Estimation error occurred using Basis Pursuit (SNR = 1 dB).}
    \label{fig:newbp1db}
\end{figure}

\begin{figure}[htbp]
	\begin{center}
		\psfrag{x01}[c][c][1][0]{$0$}
		\psfrag{x02}[c][c][1][0]{$0.5$}
		\psfrag{x03}[c][c][1][0]{$1$}
		\psfrag{x04}[c][c][1][0]{$1.5$}
		\psfrag{x05}[c][c][1][0]{$2$}
		\psfrag{x06}[c][c][1][0]{$2.5$}
		\psfrag{x07}[c][c][1][0]{$3$}
		\psfrag{x08}[c][c][1][0]{$3.5$}
		\psfrag{x09}[c][c][1][0]{$4$}
				
		\psfrag{v01}[r][rc][1][0]{$-9$}
		\psfrag{v02}[r][rc][1][0]{$-8$}
		\psfrag{v03}[r][rc][1][0]{$-7$}
		\psfrag{v04}[r][rc][1][0]{$-6$}
		\psfrag{v05}[r][rc][1][0]{$-5$}
		\psfrag{v06}[r][rc][1][0]{$-4$}
		\psfrag{v07}[r][rc][1][0]{$-3$}
		\psfrag{v08}[r][rc][1][0]{$-2$}
		\psfrag{v09}[r][rc][1][0]{$-1$}
		\psfrag{v10}[r][rc][1][0]{$0$}

		\psfrag{s06}[r][c][1][270]{dB}
		\psfrag{s05}[c][b][1][0]{$\lambda$}
		\psfrag{s11}{}
		\psfrag{s13}[l][c][1][0]{Training Error}
		\psfrag{s14}[l][c][1][0]{Data Error}
		\psfrag{s07}{}
		\psfrag{s10}{}

        \epsfxsize=4.0in
        \epsfbox{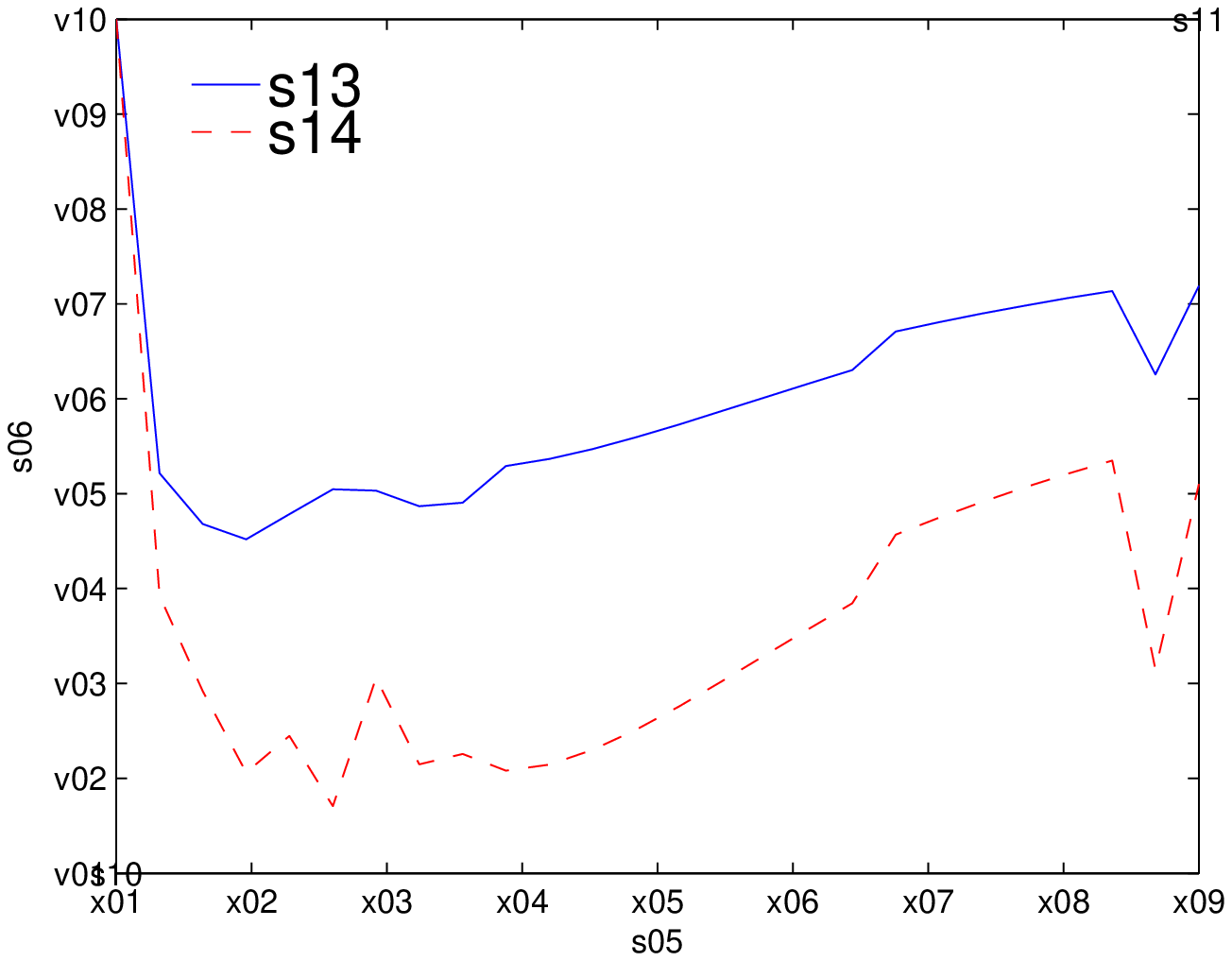}
    \end{center}
	\caption{Estimation error occurred using Basis Pursuit (SNR = 3 dB).}
    \label{fig:newbp3db}
\end{figure}

\begin{figure}[htbp]
	\begin{center}
		\psfrag{x01}[c][c][1][0]{$0$}
		\psfrag{x02}[c][c][1][0]{$0.5$}
		\psfrag{x03}[c][c][1][0]{$1$}
		\psfrag{x04}[c][c][1][0]{$1.5$}
		\psfrag{x05}[c][c][1][0]{$2$}
		\psfrag{x06}[c][c][1][0]{$2.5$}
		\psfrag{x07}[c][c][1][0]{$3$}
		\psfrag{x08}[c][c][1][0]{$3.5$}
		\psfrag{x09}[c][c][1][0]{$4$}
				
		\psfrag{v01}[r][rc][1][0]{$-16$}
		\psfrag{v02}[r][rc][1][0]{$-14$}
		\psfrag{v03}[r][rc][1][0]{$-12$}
		\psfrag{v04}[r][rc][1][0]{$-10$}
		\psfrag{v05}[r][rc][1][0]{$-8$}
		\psfrag{v06}[r][rc][1][0]{$-6$}
		\psfrag{v07}[r][rc][1][0]{$-4$}
		\psfrag{v08}[r][rc][1][0]{$-2$}
		\psfrag{v09}[r][rc][1][0]{$0$}

		\psfrag{s06}[r][c][1][270]{dB}
		\psfrag{s05}[c][b][1][0]{$\lambda$}
		\psfrag{s11}{}
		\psfrag{s13}[l][c][1][0]{Training Error}
		\psfrag{s14}[l][c][1][0]{Data Error}
		\psfrag{s07}{}
		\psfrag{s10}{}
        \epsfxsize=4.0in
        \epsfbox{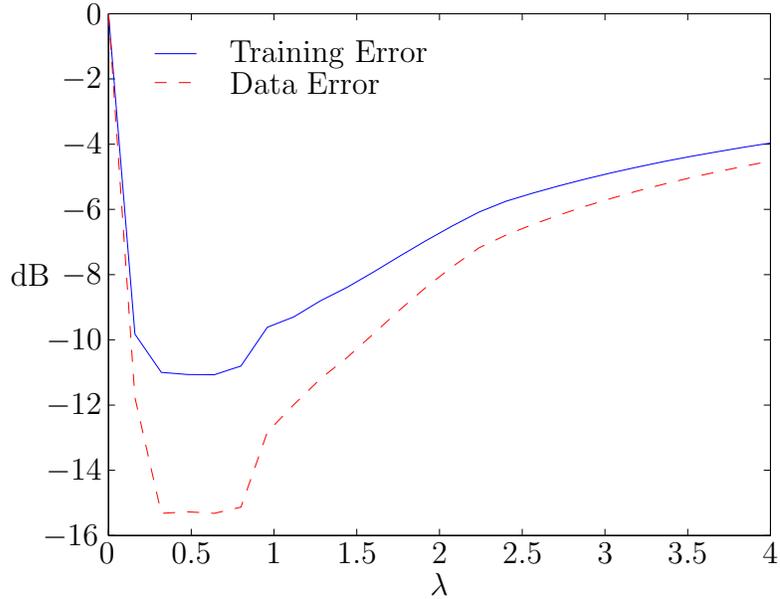}
    \end{center}
	\caption{Estimation error occurred using Basis Pursuit (SNR = 10 dB).}
    \label{fig:newbp10db}
\end{figure}

In addition to OMP, we also test Basis Pursuit (BP) \cite{C1998} which finds the solution to   
$$
\min_{\vec{\hat{\theta}}} \lambda |\vec{\hat{\theta}}|_1 +1/2\vectornorm{\vec{z}-\vec{A}\vec{\hat{\theta}}}^2.
$$ 
For this algorithm, we measure the data error and the training error over several values $\lambda$.  \Figref{newbpnonoise}, \Figref{newbp1db}, \Figref{newbp3db}, \Figref{newbp10db} plot the results for four different SNRs.  From these figures we see that the lower the SNR, the higher the value of $\lambda$ which minimizes the data error.

\begin{figure}[htbp]
	\begin{center}
		\psfrag{x01}[c][c][1][0]{$0$}
		\psfrag{x02}[c][c][1][0]{$0.02$}
		\psfrag{x03}[c][c][1][0]{$0.04$}
		\psfrag{x04}[c][c][1][0]{$0.06$}
		\psfrag{x05}[c][c][1][0]{$0.08$}
		\psfrag{x06}[c][c][1][0]{$0.1$}
		\psfrag{x07}[c][c][1][0]{$0.12$}
		\psfrag{x08}[c][c][1][0]{$0.14$}
		\psfrag{x09}[c][c][1][0]{$0.16$}
		\psfrag{x10}[c][c][1][0]{$0.18$}
		\psfrag{x11}[c][c][1][0]{$0.2$}
				
		\psfrag{v01}[r][rc][1][0]{$-66.5$}
		\psfrag{v02}[r][rc][1][0]{$-66$}
		\psfrag{v03}[r][rc][1][0]{$-65.5$}
		\psfrag{v04}[r][rc][1][0]{$-65$}
		\psfrag{v05}[r][rc][1][0]{$-64.5$}
		\psfrag{v06}[r][rc][1][0]{$-64$}
		\psfrag{v07}[r][rc][1][0]{$-63.5$}
		\psfrag{v08}[r][rc][1][0]{$-63$}
		
		\psfrag{s06}[r][c][1][270]{dB}
		\psfrag{s05}[c][b][1][0]{$p$}
		\psfrag{s11}{}
		\psfrag{s13}[l][c][1][0]{Training Error}
		\psfrag{s14}[l][c][1][0]{Data Error}
		\psfrag{s07}{}
		\psfrag{s10}{}
        \epsfxsize=4.0in
        \epsfbox{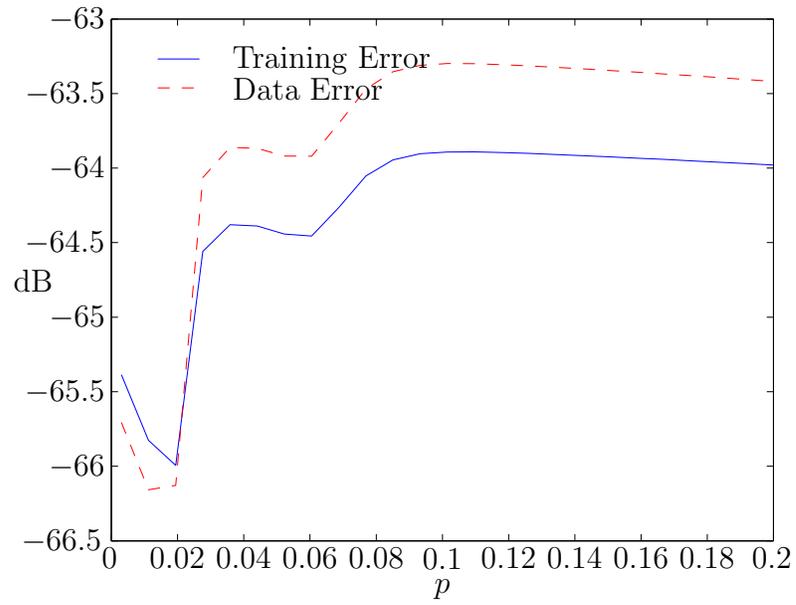}
    \end{center}
	\caption{Estimation error occurred using FBMP (noiseless case).}
    \label{fig:good_fbmp_nonoise}
\end{figure}

\begin{figure}[htbp]
	\begin{center}
		\psfrag{x01}[c][c][1][0]{$0$}
		\psfrag{x02}[c][c][1][0]{$0.02$}
		\psfrag{x03}[c][c][1][0]{$0.04$}
		\psfrag{x04}[c][c][1][0]{$0.06$}
		\psfrag{x05}[c][c][1][0]{$0.08$}
		\psfrag{x06}[c][c][1][0]{$0.1$}
		\psfrag{x07}[c][c][1][0]{$0.12$}
		\psfrag{x08}[c][c][1][0]{$0.14$}
		\psfrag{x09}[c][c][1][0]{$0.16$}
		\psfrag{x10}[c][c][1][0]{$0.18$}
		\psfrag{x11}[c][c][1][0]{$0.2$}
				
		\psfrag{v01}[r][rc][1][0]{$-20$}
		\psfrag{v02}[r][rc][1][0]{$-18$}
		\psfrag{v03}[r][rc][1][0]{$-16$}
		\psfrag{v04}[r][rc][1][0]{$-14$}
		\psfrag{v05}[r][rc][1][0]{$-12$}
		\psfrag{v06}[r][rc][1][0]{$-10$}
		\psfrag{v07}[r][rc][1][0]{$-8$}
		\psfrag{v08}[r][rc][1][0]{$-6$}
		\psfrag{v09}[r][rc][1][0]{$-4$}
		\psfrag{v10}[r][rc][1][0]{$-2$}
		\psfrag{v11}[r][rc][1][0]{$0$}
		\psfrag{v12}[r][rc][1][0]{$2$}
		
		\psfrag{s06}[r][c][1][270]{dB}
		\psfrag{s05}[c][b][1][0]{$p$}
		\psfrag{s11}{}
		\psfrag{s13}[l][c][1][0]{Training Error}
		\psfrag{s14}[l][c][1][0]{Data Error}
		\psfrag{s07}{}
		\psfrag{s10}{}

        \epsfxsize=4.0in
        \epsfbox{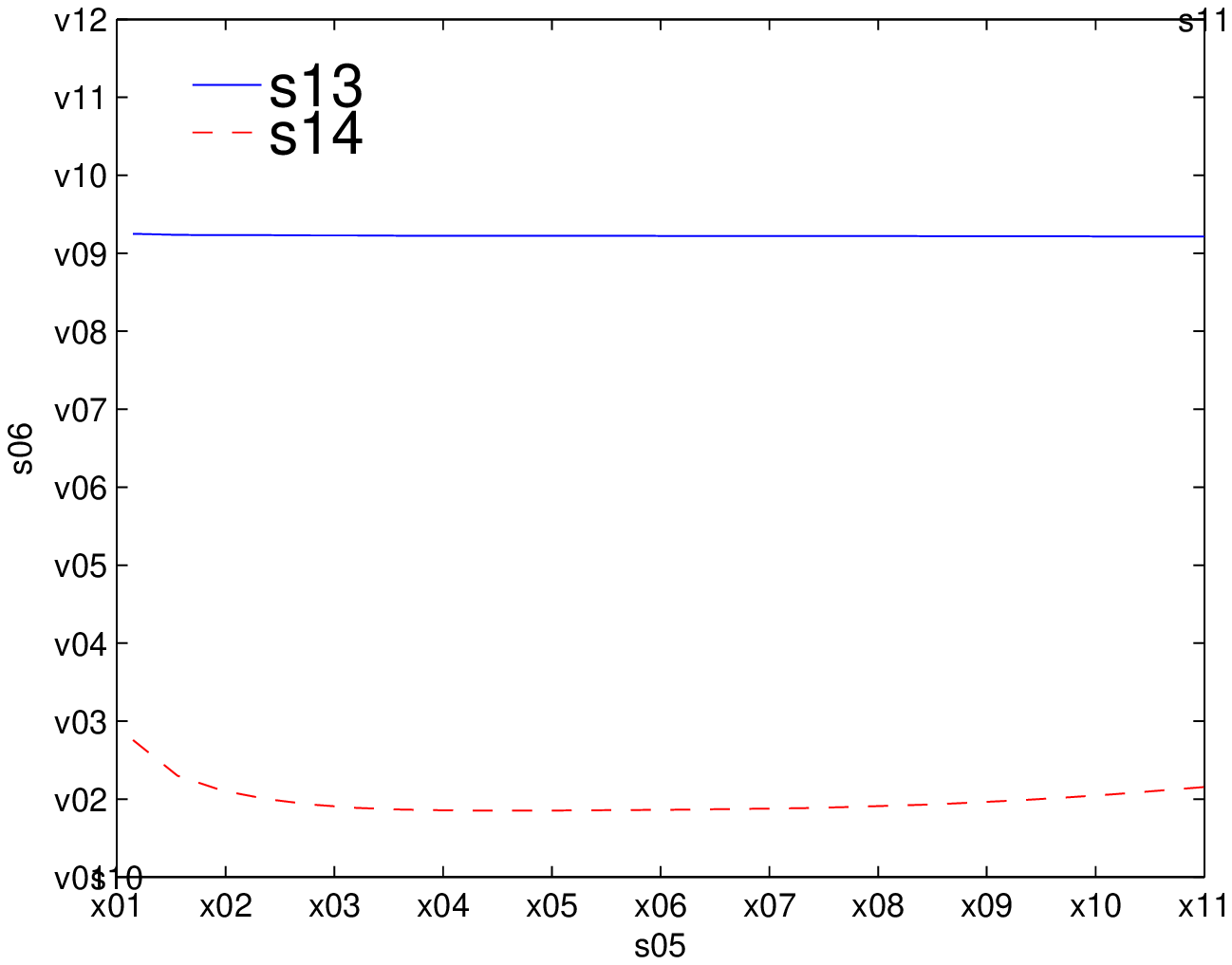}
    \end{center}
	\caption{Estimation error occurred using FBMP (SNR = 1dB).}
    \label{fig:good_fbmp1db}
\end{figure}

\begin{figure}[htbp]
	\begin{center}
		\psfrag{x01}[c][c][1][0]{$0$}
		\psfrag{x02}[c][c][1][0]{$0.02$}
		\psfrag{x03}[c][c][1][0]{$0.04$}
		\psfrag{x04}[c][c][1][0]{$0.06$}
		\psfrag{x05}[c][c][1][0]{$0.08$}
		\psfrag{x06}[c][c][1][0]{$0.1$}
		\psfrag{x07}[c][c][1][0]{$0.12$}
		\psfrag{x08}[c][c][1][0]{$0.14$}
		\psfrag{x09}[c][c][1][0]{$0.16$}
		\psfrag{x10}[c][c][1][0]{$0.18$}
		\psfrag{x11}[c][c][1][0]{$0.2$}
				
		\psfrag{v01}[r][rc][1][0]{$-22$}
		\psfrag{v02}[r][rc][1][0]{$-20$}
		\psfrag{v03}[r][rc][1][0]{$-18$}
		\psfrag{v04}[r][rc][1][0]{$-16$}
		\psfrag{v05}[r][rc][1][0]{$-14$}
		\psfrag{v06}[r][rc][1][0]{$-12$}
		\psfrag{v07}[r][rc][1][0]{$-10$}
		\psfrag{v08}[r][rc][1][0]{$-8$}
		\psfrag{v09}[r][rc][1][0]{$-6$}
		\psfrag{v10}[r][rc][1][0]{$-4$}
		\psfrag{v11}[r][rc][1][0]{$-2$}
		\psfrag{v12}[r][rc][1][0]{$0$}
		
		\psfrag{s06}[r][c][1][270]{dB}
		\psfrag{s05}[c][b][1][0]{$p$}
		\psfrag{s11}{}
		\psfrag{s13}[l][c][1][0]{Training Error}
		\psfrag{s14}[l][c][1][0]{Data Error}
		\psfrag{s07}{}
		\psfrag{s10}{}
        \epsfxsize=4.0in
        \epsfbox{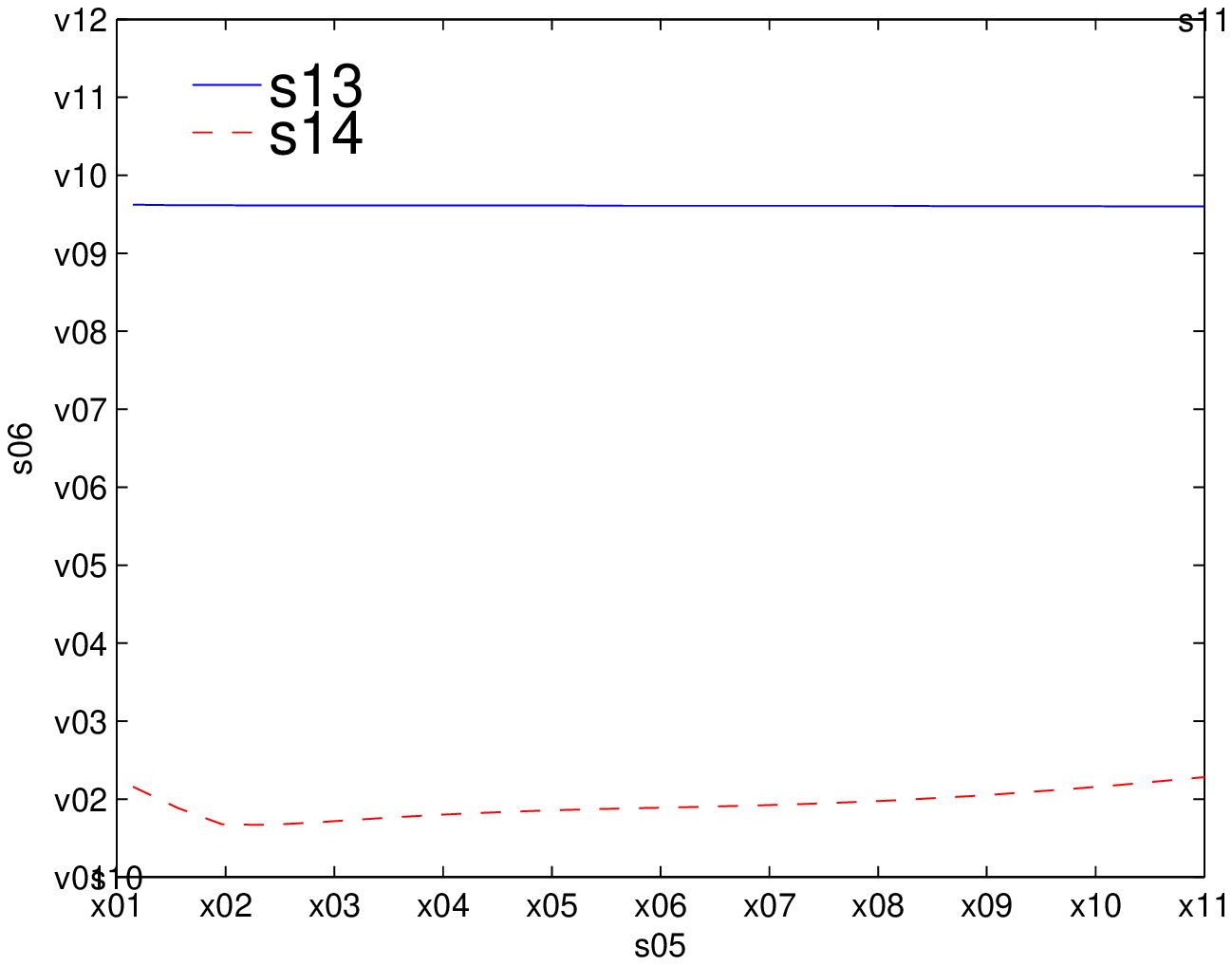}
    \end{center}
	\caption{Estimation error occurred using FBMP (SNR = 3 dB).}
    \label{fig:good_fbmp3db}
\end{figure}

\begin{figure}[htbp]
	\begin{center}
		\psfrag{x01}[c][c][1][0]{$0$}
		\psfrag{x02}[c][c][1][0]{$0.02$}
		\psfrag{x03}[c][c][1][0]{$0.04$}
		\psfrag{x04}[c][c][1][0]{$0.06$}
		\psfrag{x05}[c][c][1][0]{$0.08$}
		\psfrag{x06}[c][c][1][0]{$0.1$}
		\psfrag{x07}[c][c][1][0]{$0.12$}
		\psfrag{x08}[c][c][1][0]{$0.14$}
		\psfrag{x09}[c][c][1][0]{$0.16$}
		\psfrag{x10}[c][c][1][0]{$0.18$}
		\psfrag{x11}[c][c][1][0]{$0.2$}
				
		\psfrag{v01}[r][rc][1][0]{$-30$}
		\psfrag{v02}[r][rc][1][0]{$-25$}
		\psfrag{v03}[r][rc][1][0]{$-20$}
		\psfrag{v04}[r][rc][1][0]{$-15$}
		\psfrag{v05}[r][rc][1][0]{$-10$}
		\psfrag{v06}[r][rc][1][0]{$-5$}

		\psfrag{s06}[r][c][1][270]{dB}
		\psfrag{s05}[c][b][1][0]{$p$}
		\psfrag{s11}{}
		\psfrag{s13}[l][c][1][0]{Training Error}
		\psfrag{s14}[l][c][1][0]{Data Error}
		\psfrag{s07}{}
		\psfrag{s10}{}
        \epsfxsize=4.0in
        \epsfbox{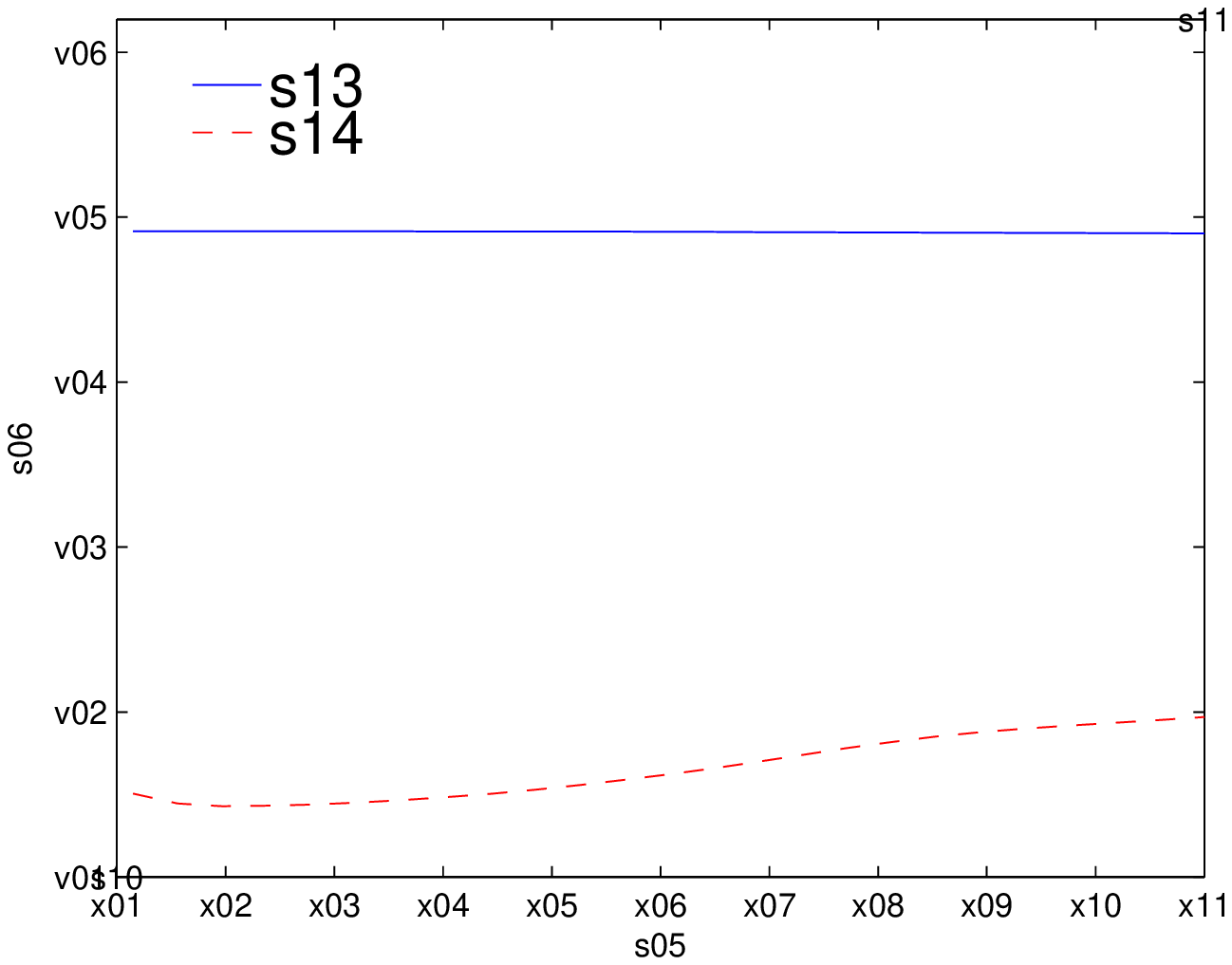}
    \end{center}
	\caption{Estimation error occurred using FBMP (SNR = 10 dB).}
    \label{fig:good_fbmp10db}
\end{figure}

Finally we test Fast Bayesian Matching Pursuit (FBMP) \cite{S2009}.  This algorithm has a user defined input $p$ which corresponds to the probability of a non-zero tap.  \Figref{good_fbmp_nonoise}, \Figref{good_fbmp1db}, \Figref{good_fbmp3db}, and \Figref{good_fbmp10db} plot training error and data error vs $p$ for different SNRs.  We find that FBMP tends to outperform both OMP and BP for this application.

\section{Unknown Channel}

\begin{figure}[htbp]
	\begin{center}
        \epsfxsize=6.0in
        \epsfbox{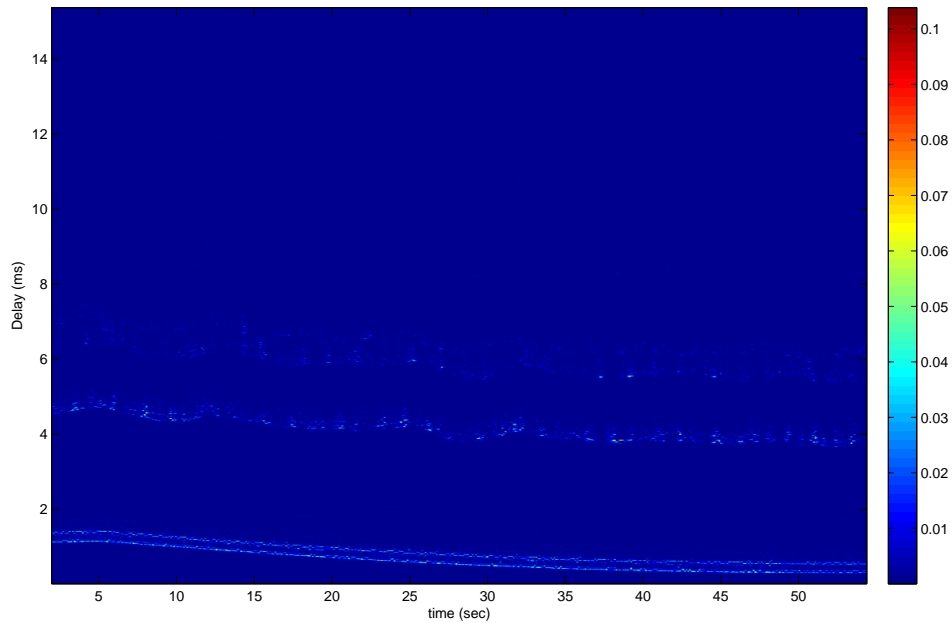}
    \end{center}
	\caption{Channel Estimate using OMP and a LTI model.  Bottom axis represents time in seconds and the left axis represents the channel delay in milliseconds}
    \label{fig:newlti_sim1}
\end{figure}

\begin{figure}[htbp]
	\begin{center}
        \epsfxsize=6.0in
        \epsfbox{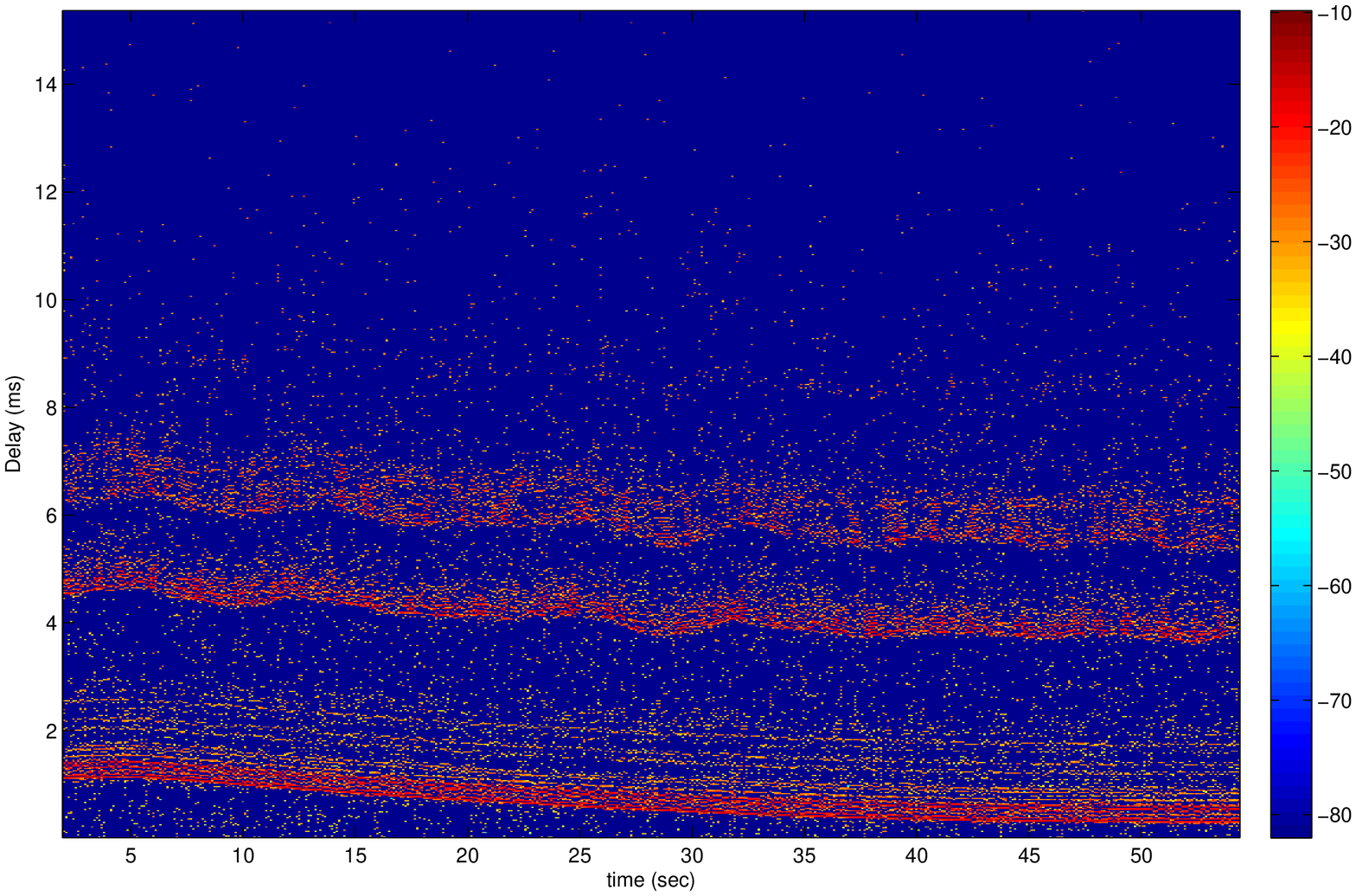}
    \end{center}
	\caption{Estimate shown in \figref{newlti_sim1} on log scale}
    \label{fig:newlti_logsim1}
\end{figure}

To study the estimation strategy further, we implemented it on experimental data collected from the Woods Hole Oceanographic Institution.  Specifically, we used a transmitted passband signal and a received passband signal to estimate an unknown channel.  The transmitted signal had a carrier frequency of 13 kHz, a bandwidth of 10 kHz, and a time duration of about 60 seconds.  Both the transmitted signal and the received signal were sampled a frequency of 39062.5 Hz (roughly 4 times the signal bandwidth).  For the first channel estimate, only the effects of time-shifts were considered as done in \secref{error_lti}.  Estimates were repeatedly made using OMP over several time periods.  \figref{newlti_sim1} and \figref{newlti_logsim1} plot the resulting estimate .  As can be seen, the channel is multi path and varies with time.  

Since the true channel parameters are of course unknown, we can not use the same error metrics we used in previous sections.  For this reason, we us the following metric, similar to one used in \cite{L2007}.  We denote this metric as the prediction error and define it as follows

\begin{equation}
\epsilon_z^2 [i+1] \defn | z[i+1] -\hat{z}[i + 1|i ] |^2,
\end{equation}
where $\hat{z}[i + 1~| i]$ is the prediction of the received signal at time $i+1$ based on the signal received up to time $i$.  

\begin{figure}[htbp]
	\begin{center}
		\psfrag{x01}[c][c][1][0]{$0$}
		\psfrag{x02}[c][c][1][0]{$0.5$}
		\psfrag{x03}[c][c][1][0]{$1$}
		\psfrag{x04}[c][c][1][0]{$1.5$}
		\psfrag{x05}[c][c][1][0]{$2$}
		\psfrag{x06}[c][c][1][0]{$2.5$}

		\psfrag{v01}[r][rc][1][0]{$0$}
		\psfrag{v02}[r][rc][1][0]{$0.1$}
		\psfrag{v03}[r][rc][1][0]{$0.2$}
		\psfrag{v04}[r][rc][1][0]{$0.3$}
		\psfrag{v05}[r][rc][1][0]{$0.4$}
		\psfrag{v06}[r][rc][1][0]{$0.5$}
		\psfrag{v07}[r][rc][1][0]{$0.6$}
		\psfrag{v08}[r][rc][1][0]{$0.7$}
		\psfrag{v09}[r][rc][1][0]{$0.8$}
		\psfrag{v10}[r][rc][1][0]{$0.9$}

		\psfrag{s04}[c][b][1][0]{$\frac{\Delta_\gamma}{T}$}

        \epsfxsize=4.0in
        \epsfbox{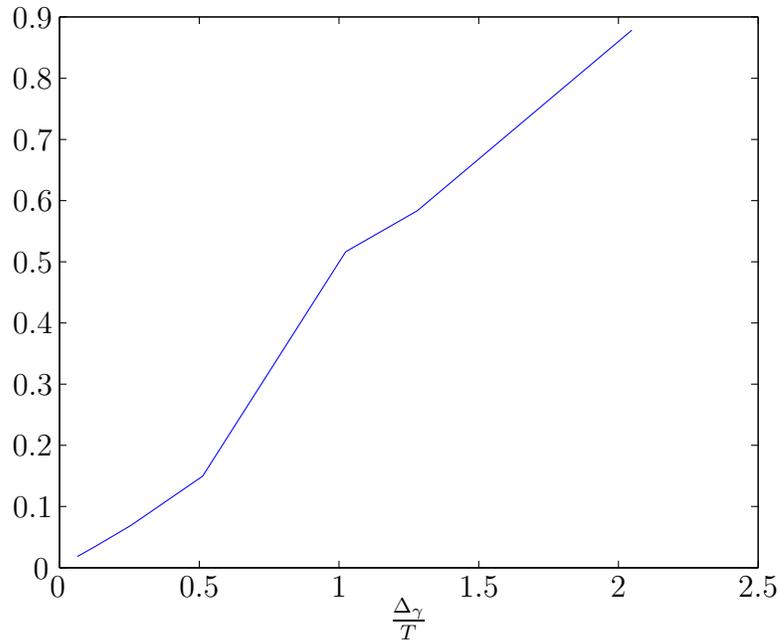}
    \end{center}
	\caption{Normalized prediction error vs. $\frac{\Delta_\gamma}{T}$}
    \label{fig:grid_sim1}
\end{figure}

\Figref{grid_sim1} plots this prediction error for various grid spacings.  For this simulation, a LTI invariant model was constructed to estimate $201^{th}$ sample $z[201]$ based on the previous 200 samples $z[1:200]$.  This process was repeated until estimates were made for $z[201:300]$.  We normalized the prediction error so that \Figref{grid_sim1} plots $\vectornorm{z[201:300]-\hat{z}[201:300]}^2 / \vectornorm{z[201:300]}^2$ vs $\frac{\Delta_\gamma}{T}$.  As can be seen, the prediction error increases with increasing grid spacing.

\section{Conclusion}

In this thesis, we have analyzed the use of spare reconstruction algorithms for the application of channel estimation.  In particular, we examined the effects of the choice of grid spacing on the average estimation error.  We derived expressions which upper bound the average estimation error in terms of the choice of grid spacing for a a linear time invariant model, a time varying narrowband model, and a time varying wideband model.  Additionally, we explored the cost of using a narrow band model to approximate a wide band model.  We then quantified the cost of this approximation in terms of the fractional bandwidth of the signal and the rate of change of channel.  Finally, we compared through simulation the use of three different sparse reconstruction algorithms (OMP, BP, and FBMP) for this application.

\appendix
\chapter{Expected Value Evaluation}
\label{app:3to1}
In this appendix, we simplify $\frac{1}{M}\Expec{\vectornorm{\thetahat  \dhat(t) - \thetal \dl(t) }^2}$ as follows
\begin{eqnarray}
\lefteqn{\frac{1}{M}\Expec{\vectornorm{\thetahat  \dhat(t) - \thetal \dl(t) }^2}} \\
&=& \frac{1}{M} \Expecl{ \int \abss{\thetahat} \abss{\dhat(t)}  -\thetahat \dhat(t)
\thetalc \dlc(t) } \nonumber \\
& & \Expecr{ -\mbox{}\thetahatc \dhatc(t) \thetal \dl(t) + \abss{\thetal} \abss{\dl(t)} \,dt } \\
&=& \frac{\abss{\thetal}}{M} \Expecl{ \int \abss{\tht} \abss{\dhat(t)} - \tht \dhat(t) \dlc(t) } \nonumber \\
& & \Expecr{ -\mbox{} \frac{\thetahatc}{\thetalc} \dhatc(t) \dl(t) + \abss{\dl(t)} \,dt } \label{eq:buttold}
\end{eqnarray}
Equation \eqref{buttold} can be reduced further by noticing that the sum of the second term\\ $ \tht \dhat(t) \dlc(t)$ and third term $ \frac{\thetahatc}{\thetalc} \dhatc(t) \dl(t)$ is equal to $2 \real{ \tht \dhat(t) \dlc(t) }$.  This is because for any $G,F \in \field{C}$, $G^* F + G F^* = 2 \real{G F^*}$.  We thus rewrite \eqref{buttold} as
\begin{eqnarray}
\lefteqn{ \frac{1}{M} \Expec{ \vectornorm{\thetahat  \dhat(t) - \thetal \dl(t)}^2 } } \nonumber \\
&=& \frac{\abss{\thetal}}{M} \Expec{ \int \abss{\tht} \abss{\dhat(t)} -2 \real{ \tht \dhat(t) \dlc(t) }  + \abss{\dl(t)}  \,dt } \label{eq:butt}
\end{eqnarray}
Equation \eqref{butt} can be split into 3 different terms as follows
\begin{eqnarray}
\frac{1}{M}\Expec{\vectornorm{\thetahat  \dhat(t) - \thetal \dl(t)}^2} 
&=& \frac{\abss{\thetal}}{M} \Expec{\int \abss{\tht} \abss{\dhat(t)} \,dt} \nonumber \\
& & -\mbox{} \frac{2\abss{\thetal}}{M} \Expec{\int \real{ \tht \dhat(t) \dlc(t)  } \,dt} \nonumber \\
& & +\mbox{} \frac{\abss{\thetal}}{M} \Expec{\int \abss{\dl(t)} \,dt}. \label{eq:1error_old}
\end{eqnarray}

\chapter{Upper bound on $\alphal$}
\label{app:alphabound}

In this appendix, we derive an expression $\alphal$, defined in \eqref{alpha}, in terms of $N$.  To do this, we first define $\deltal \defn \frac{1}{T} \{ \gammal - \gammahat \}$.  We assume that the $|\gammal -\gammahat | \leq \frac{T}{4}$, so that $|\deltal| \leq 0.25$.  Now we use the change of variables $m\defn q-r$ to write
\begin{eqnarray}
 \alphal 
 &=& \frac{1}{N^2} \sum_{m=-N+1,m\neq 0}^{N-1}
 	(N-|m|)~R^2( mT +\deltal T). \label{eq:alpha_1}
\end{eqnarray}
Defining $\alpha_R$ has the roll off factor of the raise-cosine pulse $R$, we can write 
\begin{eqnarray}
 R(xT)
 &\defn& \frac{\cos(\pi\alpha_R x)}{1-(2\alpha_R x)^2}
 	\frac{\sin(\pi x)}{\pi x} .
\end{eqnarray}
By noticing that 
\begin{eqnarray}
 \left| \frac{\cos(\pi\alpha_R x)}{1-(2\alpha_R x)^2} \right|
 &\leq& 1~~~\text{for all $\alpha_R, x$}
\end{eqnarray}
and that $|\sin(\pi x)|\leq 1$, we can write
\begin{eqnarray}
 R^2(xT)
 &\leq& \frac{1}{\pi^2 x^2}. \label{eq:r_2_bound}
\end{eqnarray}
Putting \eqref{alpha_1} together with \eqref{r_2_bound}, we get
\begin{eqnarray}
 \alphal 
 &\leq& \frac{1}{\pi^2 N^2} \sum_{m=-N+1,m\neq 0}^{N-1}
 	\frac{N-|m|}{(m+\deltal)^2}\\
 &=& \frac{1}{\pi^2 N^2} \sum_{m=1}^{N-1} \frac{N-m}{(m+\deltal)^2}
  	+ \frac{1}{\pi^2 N^2} \sum_{m=1}^{N-1} \frac{N-m}{(m-\deltal)^2}\\
 &\leq& \frac{2}{\pi^2 N^2} \sum_{m=1}^{N-1} \frac{N-m}{(m-0.25)^2}
 						\label{eq:del} \\
 &=& \frac{2}{\pi^2 N^2}\left( \sum_{m=1}^{N-1} \frac{N-0.25}{(m-0.25)^2} 
 	- \sum_{m=1}^{N-1} \frac{m-0.25}{(m-0.25)^2} \right)\\
 &=& \frac{2N-0.5}{\pi^2 N^2} \sum_{m=1}^{N-1} \frac{1}{(m-0.25)^2} 
 	- \frac{2}{\pi^2 N^2} \sum_{m=1}^{N-1} \frac{1}{m-0.25} ,
						\label{eq:am}
\end{eqnarray}
where \eqref{del} is due to $|\deltal|<0.25$.
We can upper bound the first term in \eqref{am} as follows:
\begin{eqnarray}
  \sum_{m=1}^{N-1} \frac{1}{(m-0.25)^2}
  &=& \frac{16}{9} + \sum_{m=2}^{N-1} \frac{1}{(m-0.25)^2} \\
  &\leq& \frac{16}{9} + \sum_{m=2}^{N-1} \frac{1}{(m-1)^2} \\
  &=& \frac{16}{9} + \sum_{m=1}^{N-2} \frac{1}{m^2} \\
  &\leq& \frac{16}{9} + \frac{\pi^2}{6},			\label{eq:zeta}
\end{eqnarray}
using the fact that $\sum_{m=1}^{\infty} \frac{1}{m^2}=\frac{\pi^2}{6}$ 
and noting that all the terms in the series $\sum_{m=1}^\infty\frac{1}{m^2}$ 
are positive.
To upper bound the second term in \eqref{am}, we lower bound its negative:
\begin{eqnarray}
  \sum_{m=1}^{N-1} \frac{1}{m-0.25}
  &\geq & \sum_{m=1}^{N-1} \frac{1}{m} \\
  &\geq & \ln(N).				\label{eq:harm}
\end{eqnarray}
Plugging \eqref{zeta} and \eqref{harm} into \eqref{am}, we get
\begin{eqnarray}
 \alphal
 &\leq& \frac{2N-0.5}{\pi^2 N^2} (\frac{16}{9}+\frac{\pi^2}{6}) - \frac{2\ln(N)}{\pi^2 N^2} 
						\label{eq:am2}\\
 &\leq& \frac{0.694}{N}.
						\label{eq:am3}
\end{eqnarray}
Note that, when $N$ is large, $2N-0.5\approx 2N$ and $(\frac{16}{9}+\pi^2/6) N-\ln(N) 
\approx (\frac{16}{9}+\pi^2/6)N$.
Thus, although the bound in \eqref{am2} is a tiny bit more accurate than 
the one in \eqref{am3}, the latter is much more intuitive.

\chapter{Inner Product Evaluation} 
\label{app:sshat}

To simplify the inner product $\langle \ssl(t), \shat(t) \rangle $, we first express it using \eqref{innerprod}, \eqref{sdef}, and \eqref{shatdef}.
\begin{eqnarray}
\lefteqn{ \langle \ssl(t), \shat(t) \rangle } \nonumber \\
 &=& \int \ssl(t) \shatc(t) \,dt  \\
&=&\int e^{j 2 \pi f_c ((1-\ahat )\gammahat - (1-\al )\gammal +(\ahat-\al )t)} \nonumber \\
& & \times \mbox{ } \sqrt{1-\al} \displaystyle\sum_{r=-M/2}^{M/2-1} b_r p((1-\al)t -r T -(1-\al)\gammal ) \nonumber \\
& &  \times \mbox{ } \sqrt{1-\ahat} \displaystyle\sum_{q=-M/2}^{M/2-1} b_q^* p((1-\ahat)t -q T -(1-\ahat)\gammahat ) \,dt  \\ \label{eq:wideband1}
&=&  \sqrt{1-\al} \sqrt{1-\ahat} e^{ j 2 \pi f_c ((1-\ahat )\gammahat - (1-\al )\gammal ) } \displaystyle\sum_{r=-M/2}^{M/2-1} \displaystyle\sum_{q=-M/2}^{M/2-1} b_r b_q^* \nonumber \\
& & \times \mbox{ } \int e^{j 2 \pi f_c (\ahat - \al ) t  } p((1-\al)t -r T -(1-\al)\gammal ) \nonumber \\
& & \times \mbox{ }  p((1-\ahat)t -q T -(1-\ahat)\gammahat )    \,dt \label{eq:hardint}
\end{eqnarray}
To evaluate \eqref{hardint}, it is helpful to approximate the term $ e^{ j 2 \pi f_c (\ahat - \al ) t }$ with a time invariant constant in order to evaluate the integral.  To do this, we note that the term $p((1-\ahat)t -q T -(1-\ahat)\gammahat )$ has its maximum value when $t= \frac{qT}{1-\ahat} + \gammahat$, but diminishes to 0 as $\abs{(1-\ahat)t -q T -(1-\ahat)\gammahat}$ increases.  Likewise, the term $p((1-\al)t -r T -(1-\al)\gammal )$ has its maximum value when $t= \frac{rT}{1-\al}+\gammal$ and diminishes to 0 as $\abs{(1-\al)t -r T -(1-\al)\gammal}$ increases.  We define $\hat{F}_{q,r}\of{l}$ as the value of $ e^{ j 2 \pi f_c (\ahat - \al ) t }$ when $t$ is half way between the peak values $t_1 \defn \frac{qT}{1-\ahat} + \gammahat$ and $t_2 \defn \frac{rT}{1-\al}+\gammal$.
\begin{equation}
\hat{F}_{q,r}\of{l} \defn e^{j 2 \pi f_c (\ahat - \al ) ( \frac{qT}{2(1-\ahat )} +  \frac{rT}{2(1-\al )} +\frac{\gammahat + \gammal}{2}  )} \label{eq:Fhatdef}
\end{equation}
The term $ e^{ j 2 \pi f_c (\ahat - \al ) t }$ in \eqref{hardint} can be well approximated by $\hat{F}_{q,r}\of{l}$ over the periods of $t$ where both $p((1-\ahat)t -q T -(1-\ahat)\gammahat )$ and $p((1-\al)t -r T -(1-\al)\gammal ) $ are not negligible.  This in part is because we make the mild assumptions that the value of $\al$ to be relatively small (i.e. $\al < 0.01$) and that $\ahat \approx \al$.  It is also because we have assumed that the signal has a high fractional bandwidth (i.e.  $\frac{1}{T f_c} \approx 1$).  By using $\hat{F}_{q,r}\of{l}$ to approximate $e^{ j 2 \pi f_c (\ahat - \al ) t }$, we can express \eqref{hardint} as follows
\begin{eqnarray}
\lefteqn{ \langle \ssl(t), \shat(t) \rangle } \nonumber \\
&=&  \sqrt{1-\al} \sqrt{1-\ahat} e^{ j 2 \pi f_c ((1-\ahat )\gammahat - (1-\al )\gammal ) } \displaystyle\sum_{r=-M/2}^{M/2-1} \displaystyle\sum_{q=-M/2}^{M/2-1} b_r b_q^* \nonumber \\
& & \times \mbox{ } \hat{F}_{q,r}\of{l} \int  p((1-\al)t -r T -(1-\al)\gammal ) \nonumber \\
& & \times \mbox{ }  p((1-\ahat)t -q T -(1-\ahat)\gammahat )    \,dt \label{eq:wideband1x}
\end{eqnarray}
Now if we make the substitution $v=(1-\ahat)t -q T -(1-\ahat)\gammahat$, \eqref{wideband1x} becomes
\begin{eqnarray}
\lefteqn{ \frac{\sqrt{1-\al}}{\sqrt{1-\ahat}} e^{j 2 \pi f_c ((1-\ahat )\gammahat -(1-\al )\gammal )} \displaystyle\sum_{r=-M/2}^{M/2-1} \displaystyle\sum_{q=-M/2}^{M/2-1} b_r b_q^* \hat{F}_{q,r}\of{l}  } \nonumber \\
\lefteqn{  \times \mbox{ } \int  p(v) p(\frac{1-\al}{1-\ahat}v +\frac{1-\al}{1-\ahat}qT-rT+(1-\al)(\gammahat -\gammal)) \,dv } \nonumber \\
&=& \sqrt{\betal} e^{j 2 \pi f_c ((1-\ahat )\gammahat -(1-\al )\gammal )} \displaystyle\sum_{r=-M/2}^{M/2-1} \displaystyle\sum_{q=-M/2}^{M/2-1}  b_r b_q^* \hat{F}_{q,r}\of{l}  \nonumber \\
& & \times \mbox{ } \int \ p(v) p(\betal v +\betal qT-rT+(1-\al)(\gammahat -\gammal)) \,dv , \label{eq:wideband2}
\end{eqnarray}
where $\betal \defn \frac{1-\al}{1-\ahat}$.  To simplify \eqref{wideband2} further, we use the Wide Band Ambiguity function defined in \cite{S1981}
\begin{equation}
W(\lambda,\beta) \defn \sqrt{\beta} \int p(t) p(\beta (t+\lambda)) \,dt. \label{eq:wbdef}
\end{equation}
Using this function, we can express \eqref{wideband2} as
\begin{eqnarray}
\langle \ssl(t), \shat(t) \rangle &=& \displaystyle\sum_{r=-M/2}^{M/2-1} \displaystyle\sum_{q=-M/2}^{M/2-1} b_r b_q^* \hat{F}_{q,r}\of{l} e^{j 2 \pi f_c ((1-\ahat )\gammahat -(1-\al )\gammal )} \nonumber \\
 & & \times W(qT - \frac{rT}{\betal}+(1-\ahat )(\gammahat-\gammal ),\betal ). \label{eq:thingxxx_old}
\end{eqnarray}

\chapter{Upper bound on $\omegal$}
\label{app:omegabound}

In this appendix, we derive an expression $\omegal$ (defined in \eqref{omegadef}) in terms of $M$.  To do this, we first note that the wideband ambiguity function (defined in \eqref{wbdef}) can be simplified as follows
\begin{eqnarray}
  W(\lambda,\beta)
  &\defn& \sqrt{\beta} \int p(t) p(\beta(t+\lambda)) dt\\
  &=& \sqrt{\beta} \int \int P(f_1)e^{j2\pi f_1t} df_1 
  	\int P(f_2) e^{j2\pi f_2 \beta(t+\lambda)}df_2 dt\\
  &=& \frac{1}{\sqrt{\beta}} \int \int P(f_1)e^{j2\pi f_1t} df_1 
  	\int P(f/\beta) e^{j2\pi f(t+\lambda)}df dt\\
  &=& \frac{1}{\sqrt{\beta}} \int \int P(f_1) 
  	P(f/\beta) e^{j2\pi f\lambda} \underbrace{
	\int e^{j2\pi (f_1+f) t}dt }_{\delta(f_1+f)} df df_1\\
  &=& \frac{1}{\sqrt{\beta}} \int P(-f) 
  	P(f/\beta) e^{j2\pi f\lambda} df.
\end{eqnarray}
We assume that $p(t)$ is the SRRC pulse with rolloff $\alpha_R=0$ so that
\begin{eqnarray}
  P(f)
  &=& \begin{cases}
  	\sqrt{T} & |f|\leq \frac{1}{2T} \\
	0 & \text{else}
	\end{cases} ,
\end{eqnarray}
in which case
\begin{eqnarray}
  W(\lambda,\beta)
  &=& \frac{T}{\sqrt{\beta}} \int_{-\frac{\min(1,\beta)}{2T}}^{\frac{\min(1,\beta)}{2T}}
  	e^{j2\pi f\lambda} df \\
  &=& \frac{T}{\sqrt{\beta}} \frac{\min(1,\beta)}{T}\sinc(\frac{\min(1,\beta)}{T}\lambda) \\
  &=& \frac{T}{\sqrt{\beta}} 
  	\frac{\min(1,\beta)}{T}
  	\frac{\sin(\pi\frac{\min(1,\beta)}{T}\lambda)}
  	{\pi\frac{\min(1,\beta)}{T}\lambda} \\
  &=& \frac{T}{\pi\lambda\sqrt{\beta}} 
  	\sin(\pi\frac{\min(1,\beta)}{T}\lambda) \\
  W(\lambda,\beta)^2
  &\leq& \frac{T^2}{\pi^2\lambda^2\beta} \label{eq:wub}
\end{eqnarray}
Using \eqref{wub}, the term $\omegal$ can be upper bounded as follows
\begin{eqnarray}
\omegal &\defn & \frac{1}{M^2} \sum_{r =-M/2}^{M/2-1} \sum_{e \neq r =-M/2}^{M/2-1} W^2(\lambda_{e,r}\of{l},\betal ) \\
& \leq & \frac{1}{M^2} \sum_{r =-M/2}^{M/2-1} \sum_{e \neq r =-M/2}^{M/2-1} \frac{T^2}{\pi^2\hat{\lambda}_{e,r}^{(l)2} \betal} \label{eq:omegaub1}
\end{eqnarray}
We assume that $\al \leq \frac{1}{2M}$ and $\ahat \leq \frac{1}{2M}$ so that for large $M$, we can make the approximation $\frac{1}{\betal}= \frac{1-\ahat}{1-\al} \approx 1+\al-\ahat$.  Using this approximation, we can express $\lambdal_{e,r}$ (defined in \eqref{lambda}) as follows
\begin{equation}
\lambdal_{e,r} = rT - eT +(\ahat -\al )eT -(1-\ahat )T \deltal \label{eq:lambdaapprox}
\end{equation}
where $\deltal \defn \frac{1}{T} \{ \gammal - \gammahat \} \leq \frac{1}{4}$.  Using the change of variables $m \defn e-r$ and combining \eqref{omegaub1} and \eqref{lambdaapprox}, it follows that
\begin{eqnarray}
\omegal &\leq& \frac{1}{M^2 \pi^2 \betal} \sum_{r =-M/2}^{M/2-1} \sum_{e \neq r =-M/2}^{M/2-1} \frac{1}{ (r - e +(\ahat -\al )e -(1-\ahat ) \deltal  )^2} \\
&=& \frac{1}{M^2 \pi^2 \betal} \sum_{m=1}^{M-1}
  \sum_{e=-M/2+m}^{M/2-1} \frac{1}{\big(
  	m + (\ahat-\al)e + (1-\hat{a}\of{l})\delta\of{l}\big)^2}
	\nonumber\\&&\mbox{}
  +\frac{1}{M^2 \pi^2 \betal} \sum_{m=1-M}^{-1}
  \sum_{e=-M/2}^{M/2+m} \frac{1}{\big(
  	m + (\ahat-\al)e + (1-\hat{a}\of{l})\delta\of{l}\big)^2}\\
  &\leq& \frac{2}{M^2 \pi^2 \betal} \sum_{m=1}^{M-1}
  \sum_{e=-M/2}^{M/2-1} \frac{1}{\big(
  	m + (\ahat-\al)e + (1-\hat{a}\of{l})\delta\of{l}\big)^2}
\end{eqnarray}
Notice that, since $\abs{e} \leq \frac{M}{2}$, $\deltal \leq \frac{1}{4}$, and $\ahat \leq \frac{1}{2M}$ then 
\begin{eqnarray}
  \big| (\ahat-\al)e + (1-\ahat)\delta\of{l} \big|
  &\leq& \big| (\ahat-\al)e \big|
  	+ \big|(1-\hat{a}\of{l})\delta\of{l} \big| \\
  &=& |\ahat-\al|~|e| 
  	+ |1-\hat{a}\of{l}|~|\delta\of{l}| \\
  &\leq& |\ahat-\al|~\frac{M}{2}
  	+ (1+\frac{1}{2M})~\frac{1}{4} \\
  &=& |\ahat-\al|~\frac{M}{2}
  	+ (1+\frac{1}{2M})~\frac{1}{4}.
\end{eqnarray}
We assume that
$|\ahat-\al| \leq \frac{1}{2M}(1-\frac{1}{2M})$.
With this assumption in mind, it immediately follows that
$\big| (\ahat-\al)e + (1-\hat{a}\of{l})\delta\of{l} \big| 
< \frac{1}{2}$.
In this case
\begin{eqnarray}
  \omegal
  &\leq& \frac{2}{M \pi^2 \betal} \sum_{m=1}^{M-1} 
  	\frac{1}{( m - 0.5 )^2} \\
  &=& \frac{2}{M \pi^2 \betal} \left( 4+\sum_{m=2}^{M-1} \frac{1}{(m-0.5)^2} \right) \\
  &\leq& \frac{2}{M \pi^2 \betal} \left( 4+\sum_{m=1}^{M-2} \frac{1}{m^2} \right) \\
  &\leq& \frac{2 (4+\pi^2/6)}{M \pi^2 \betal} 
\end{eqnarray}
using the fact that $\displaystyle\sum_{m=1}^{M} \frac{1}{m^2} \leq \frac{\pi^2}{6}$.
Finally, since 
\begin{eqnarray}
  \frac{1}{\betal} 
  &=& \frac{1-\hat{a}\of{l}}{1-a\of{l}} \\
  &\approx&  1-\hat{a}\of{l}+a\of{l} \\
  &\leq& 1 + |\hat{a}\of{l}-a\of{l}| \\
  &\leq& 1 + \frac{1}{2M} (1-\frac{1}{2M})\\
  &\leq& 1 + \frac{1}{2M} 
  ~=~ \frac{2M+1}{2M} ,
\end{eqnarray}
we have
\begin{eqnarray}
  \omegal
  &\leq& \frac{2 (4+\pi^2/6)}{M \pi^2}~\frac{2M+1}{2M} 
\end{eqnarray}
Now if we say that $M\geq 10$, we know that $\frac{2M+1}{2M}\leq 
\frac{21}{20}$, so that
\begin{eqnarray}
  \omegal
  &\leq& \frac{2 (4+\pi^2/6)}{M \pi^2}~\frac{21}{20} \leq \frac{1.2011}{M}.
\end{eqnarray}

\chapter{Lower bound on $\abss{ \displaystyle\sum_{n=-M/2}^{M/2-1} \hat{F}_{n,n}\of{l} W(\lambdal_{n,n},\betal ) }$}
\label{app:fwbound}
The primary goal of \secref{error_tv} is to find an insightful expression which relates an upper bound of the error metric $\metric$ with the the grid spacing ($\Delta_a$ and $\Delta_\gamma$).  With this goal in mind, it is necessary to find a lower bound on the expression $\abss{ \displaystyle\sum_{n=-M/2}^{M/2-1} \hat{F}_{n,n}\of{l} W(\lambdal_{n,n},\betal ) } $ in terms of the grid spacing.  For our analysis, we assume that $|a\of{l}|\leq\frac{1}{2M}$ and $|\hat{a}\of{l}|\leq\frac{1}{2M}$.  We also assume that the bounds on the quantities $|\hat{a}\of{l}-a\of{l}| $ and $|\hat{\gamma}\of{l}-\gamma\of{l}|$ are slightly tighter than the bounds assumed in \appref{omegabound}, so that 
\begin{eqnarray}
  |\hat{a}\of{l}-a\of{l}|
  &\leq& \textstyle \frac{1}{2M}(1-\frac{1}{2M})\frac{1}{2f_cT} \label{eq:aa}\\
  |\hat{\gamma}\of{l}-\gamma\of{l}|
  &\leq& \textstyle \frac{T}{4}(1+\frac{1}{2M})^{-1} . 
\end{eqnarray}
Note the presence of the $f_c T$ term in \eqref{aa},
and recall that $f_c \geq \frac{1}{2T}$ (assuming that the carrier
frequency is centered in the signal passband) which implies that
$2f_c T \geq 1$.

We start by recalling the definitions of $\hat{F}\of{l}_{n,n} $ and $W(\lambdal_{n,n},\betal)$ given in \eqref{Fhatdef} and \eqref{wbdef}, respectively.  We now notice that 
\begin{eqnarray}
  \left| \sum_{n=-M/2}^{M/2-1} \hat{F}\of{l}_{n,n} 
  	W(\lambdal_{n,n},\betal) \right|^2
  &=& \left| \sum_{n=-M/2}^{M/2-1} \hat{K}\of{l}_{n} 
  	W(\lambdal_{n,n},\betal) \right|^2
\end{eqnarray}
where 
\begin{eqnarray}
  \hat{K}_{n}\of{l}
  &\defn& e^{j \hat{\chi}_n\of{l}} \\
  \hat{\chi}_n\of{l}
  &\defn& \textstyle \pi f_c(\hat{a}\of{l}-a\of{l})
  	(\frac{nT}{1-\hat{a}\of{l}}+\frac{nT}{1-a\of{l}}) .
\end{eqnarray}
Using the Euler expansion 
$\hat{K}_{n}\of{l} = \cos(\hat{\chi}_n\of{l})+j\sin(\hat{\chi}_n\of{l})$, we get
\begin{eqnarray}
  \left| \sum_{n=-M/2}^{M/2-1} \hat{K}\of{l}_{n} 
  	W(\lambda\of{l}_{n,n},\betal) \right|^2
  &=&\left| \sum_{n=-M/2}^{M/2-1} W(\lambdal_{n,n},\betal) 
	[\cos(\hat{\chi}_n\of{l})+j\sin(\hat{\chi}_n\of{l})]\right|^2 \\
  &\geq& \left( \sum_{n=-M/2}^{M/2-1} W(\lambdal_{n,n},\betal) 
	\cos(\hat{\chi}_n\of{l})\right)^2 ,		\label{eq:bnd1}
\end{eqnarray}
where the inequality follows because 
$W(\lambdal_{n,n},\betal)\in\Real$. 
The latter bound is actually quite tight because the quantity
$W(\lambdal_{n,n},\betal)$ is symmetric around $n=0$,
the quantity $\sin(\hat{\chi}_n\of{l})$ is antisymmetric around $n=0$,
and the summation is almost symmetric around $n=0$.

Next we note that, under our assumptions, we have
\begin{eqnarray}
  |\lambdal_{n,n}|
  &=& \textstyle |\frac{1}{1-a\of{l}}nT(1-a\of{l}-1+\hat{a}\of{l})
          +(1-\hat{a}\of{l})(\hat{\gamma}\of{l}-\gamma\of{l})| \\
  &\leq& \textstyle \frac{1}{1-\frac{1}{2M}}|n|T \cdot
  	\frac{1}{2M}(1-\frac{1}{2M})\frac{1}{2f_cT}
          +(1+\frac{1}{2M})\cdot\frac{T}{4}(1+\frac{1}{2M})^{-1} \\
  & =  & \textstyle  \frac{|n|T}{2M\cdot 2f_cT} +\frac{T}{4} ~\leq \frac{T}{2}			\label{eq:lam}\\
  |\hat{\chi}_n\of{l}|
  &\leq& \textstyle \pi f_c \cdot \frac{1}{2M}(1-\frac{1}{2M})\frac{1}{2f_cT}
  	\cdot \frac{2|n|T}{1-\frac{1}{2M}} 
  ~=~ \textstyle \frac{\pi|n|}{2M} \\
  &\leq& \textstyle \frac{\pi}{4} .
\end{eqnarray}
Due to the range of $|\hat{\chi}_n\of{l}|$, the term $\cos(\hat{\chi}_n\of{l})$ is
clearly non-negative over the range of $n$ in \eqref{bnd1}. 
We now show that $W(\lambdal_{n,n},\betal)= \frac{\min(1,\betal)}{\sqrt{\betal}} \sinc(\frac{\min(1,\betal)}{T}\lambdal_{n,n})$
is also non-negative over those $n$.
First, notice that 
\begin{eqnarray}
\textstyle
\frac{\min(1,\betal)}{\sqrt{\betal}}
~=~\min(\frac{1}{\sqrt{\betal}},\sqrt{\betal}) 
&\geq& \textstyle
\sqrt{\frac{1-\frac{1}{M}}{1+\frac{1}{M}}} ~=~ \sqrt{\frac{M-1}{M+1}}
\end{eqnarray}
because $\frac{1-\frac{1}{M}}{1+\frac{1}{M}} \leq \betal
\leq \frac{1+\frac{1}{M}}{1-\frac{1}{M}}$.
Then, notice that the argument of the $\sinc$ obeys
\begin{eqnarray}
  \textstyle \frac{\min(1,\betal)}{T}\lambdal_{n,n}
  &\leq& \textstyle \frac{\lambdal_{n,n}}{T} 
  ~\leq~ \textstyle \frac{1}{2} .
\end{eqnarray}
Because both $\cos(\hat{\chi}_n\of{l})$ and $W(\lambdal_{n,n},\betal)$ 
are non-negative over the range of $n$ in \eqref{bnd1}, 
we can lower-bound \eqref{bnd1} by plugging 
in lower bounds for  $W(\lambdal_{n,n},\betal)$ in the summation.
Next, we derive this bound.

\begin{figure}[htbp]
	\begin{center}
		\psfrag{x01}[c][c][1][0]{$-0.5$}
		\psfrag{x02}[c][c][1][0]{$-0.4$}
		\psfrag{x03}[c][c][1][0]{$-0.3$}
		\psfrag{x04}[c][c][1][0]{$-0.2$}
		\psfrag{x05}[c][c][1][0]{$-0.1$}
		\psfrag{x06}[c][c][1][0]{$0$}
		\psfrag{x07}[c][c][1][0]{$0.1$}
		\psfrag{x08}[c][c][1][0]{$0.2$}
		\psfrag{x09}[c][c][1][0]{$0.3$}
		\psfrag{x10}[c][c][1][0]{$0.4$}
		\psfrag{x11}[c][c][1][0]{$0.5$}
		\psfrag{v01}[c][c][1][0]{$0.65$}
		\psfrag{v02}[c][c][1][0]{$0.7$}
		\psfrag{v03}[c][c][1][0]{$0.75$}
		\psfrag{v04}[c][c][1][0]{$0.8$}
		\psfrag{v05}[c][c][1][0]{$0.85$}
		\psfrag{v06}[c][c][1][0]{$0.9$}
		\psfrag{v07}[c][c][1][0]{$0.95$}
		\psfrag{v08}[c][c][1][0]{$1$}
		\psfrag{s05}[c][c][1][0]{$x$}
		\psfrag{s10}{}
		\psfrag{s11}{}
		\psfrag{s13}[c][c][1][0]{$\sinc(x)$}
		\psfrag{s14}[c][c][1][0]{$\cos(Bx) $}

        \epsfxsize=4.0in
        \epsfbox{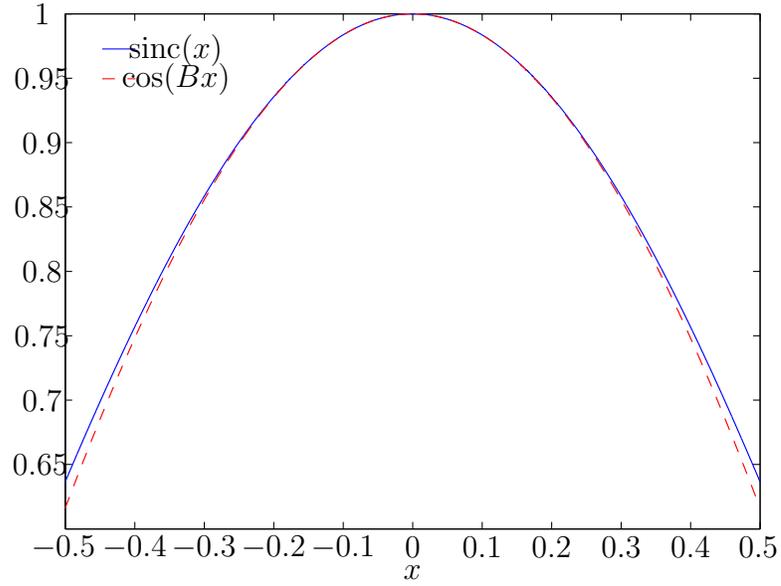}
    \end{center}
	\caption{Plot of $\sinc(x)$ and $\cos(Bx)$ when $B=1.8138$ }
    \label{fig:sinc_vs_cosB}
\end{figure}

Because  $\sinc(x)$ decreases as $x$ increases for $|x|\leq \frac{1}{2}$
\begin{equation}
  W(\lambdal_{n,n},\betal)  \geq \textstyle \sqrt{\frac{M-1}{M+1}} \sinc\Big(\frac{\lambdal_{n,n}}{T} \Big).
\end{equation}
Applying this inequality,
\begin{eqnarray}
  \lefteqn{ 
  	\left( \sum_{n=-M/2}^{M/2-1} W(\lambdal_{n,n},\betal) 
	\cos(\hat{\chi}_n\of{l})\right)^2 	}\nonumber\\
  &\geq& \frac{M-1}{M+1} \Bigg( \sum_{n=-M/2}^{M/2-1} \textstyle
  	\sinc\Big( \frac{\lambdal_{n,n}}{T} \Big) \cos\Big( \frac{2}{(1-\frac{1}{M})} \pi f_c T (\hat{a}\of{l}-a\of{l}) n  \Big) \Bigg)^2 . \label{eq:Wbound1new}
\end{eqnarray}
We now observe that for particular positive values of $B \leq \pi$, $\sinc(x) \gtrsim \cos(Bx)$ for $|x| \leq \frac{1}{2}$ (see \figref{sinc_vs_cosB}).  Since $| \frac{\lambdal_{n,n}}{T} | \leq \frac{1}{2}$, we can rewrite \eqref{Wbound1new} as 
\begin{eqnarray}
  \lefteqn{ 
  	\left( \sum_{n=-M/2}^{M/2-1} W(\lambdal_{n,n},\betal) 
	\cos(\hat{\chi}_n\of{l})\right)^2 	}\nonumber\\
  &\geq& \frac{M-1}{M+1} \Bigg( \sum_{n=-M/2}^{M/2-1} \textstyle
  	\cos\Big( \frac{B \lambdal_{n,n}}{T} \Big) \cos\Big(  \pi f_c T (\hat{a}\of{l}-a\of{l}) n \left( \frac{1}{1-\ahat}+\frac{1}{1-\al}\right)  \Big) \Bigg)^2 \nonumber \\
   &=& \frac{M-1}{M+1} \Bigg( \sum_{n=-M/2}^{M/2-1} \textstyle
  	\cos\Big( \frac{B}{1-a\of{l}}n(\hat{a}\of{l}-a\of{l})+\frac{B}{T}(1-\hat{a}\of{l})(\hat{\gamma}\of{l}-\gamma\of{l}) \Big) \nonumber \\
   & & \times \mbox{ } \cos\Big( \pi f_c T (\hat{a}\of{l}-a\of{l}) n \left( \frac{1}{1-\ahat}+\frac{1}{1-\al}\right) \Big) \Bigg)^2 \\
   &=& \frac{M-1}{M+1} \left( \sum_{n=-M/2}^{M/2-1} \textstyle
  	\cos\Big(C\of{l} n + D\of{l}  \Big) 
  	\cos\Big( E\of{l} n  \Big) 
	\right)^2, \label{eq:hardpart1}
\end{eqnarray}
where
\begin{eqnarray}
C\of{l} &\defn& \frac{B}{1-a\of{l}}(\hat{a}\of{l}-a\of{l}) \label{eq:Cdef}  \\
D\of{l} &\defn& \frac{B}{T}(1-\hat{a}\of{l})(\hat{\gamma}\of{l}-\gamma\of{l}) \label{eq:Ddef} \\
E\of{l} &\defn& \pi f_c T (\hat{a}\of{l}-a\of{l}) \left( \frac{1}{1-\ahat}+\frac{1}{1-\al}\right)    . \label{eq:Edef}  
\end{eqnarray}
Equation \eqref{hardpart1} can be recognized as a Riemann-sum approximation of the following continuous integral
\begin{eqnarray}
\lefteqn{ \frac{M-1}{M+1} \left( \int_{-M/2}^{M/2} \textstyle
  	\cos\Big(C\of{l} t + D\of{l}  \Big) 
  	\cos\Big( E\of{l} t  \Big) \,dt   \right)^2    } \nonumber \\
&=& \frac{M-1}{M+1} \left( \int_{-M/2}^{M/2} \textstyle
  	\left( \cos\Big(C\of{l} t\Big) \cos\Big(D\of{l} \Big)  \right. \right. \nonumber \\
& &  \left. \left. - \mbox{ } \sin\Big(C\of{l} t\Big) \sin\Big(D\of{l} \Big)\right) \cos\Big( E\of{l} t  \Big) \,dt   \right)^2 \label{eq:trigtrick1} \\
&=& \frac{M-1}{M+1} \left( \int_{-M/2}^{M/2} \textstyle
  	 \cos\Big(C\of{l} t\Big) \cos\Big(D\of{l} \Big)  \cos\Big( E\of{l} t  \Big) \,dt   \right)^2 	\label{eq:hardpart2}
\end{eqnarray}
Equation \eqref{trigtrick1} follows from the trig identity $\cos(x+y) = \cos(x)\cos(y)-\sin(x)\sin(y) $ and \eqref{hardpart2} follows from the fact that $\int_{-M/2}^{M/2} \textstyle \sin\Big(C\of{l} t\Big) \cos\Big( E\of{l} t  \Big) \,dt   = 0$.  We can now use the trig identity $\cos(x)\cos(y)=\frac{1}{2}\cos(x+y)+\frac{1}{2}\cos(x-y) $ to rewrite \eqref{hardpart2} as follows
\begin{eqnarray}
\lefteqn{ \frac{M-1}{M+1} \left( \int_{-M/2}^{M/2} \textstyle
  	 \cos\Big(C\of{l} t\Big) \cos\Big(D\of{l} \Big)  \cos\Big( E\of{l} t  \Big) \,dt   \right)^2  } \nonumber \\
&=& \frac{M-1}{M+1} \textstyle \cos^2(D\of{l}) \displaystyle \left( \int_{-M/2}^{M/2} \textstyle \frac{1}{2}\cos\Big( (E\of{l}+C\of{l} )t \Big) +\frac{1}{2}\cos\Big( (E\of{l}-C\of{l} )t \Big) \,dt     \right)^2 \nonumber \\
&=&  \frac{M-1}{M+1} \textstyle \cos^2(D\of{l}) \left( \frac{ \sin\Big( (E\of{l}+C\of{l} )\frac{M}{2} \Big)}{E\of{l}+C\of{l}} + \frac{ \sin\Big( (E\of{l}-C\of{l} )\frac{M}{2} \Big)}{E\of{l}-C\of{l}} \right)^2  \label{eq:hardpart3}  \\
&=&  \frac{M-1}{M+1} \textstyle \cos^2(D\of{l}) \left(\frac{M}{2} \sinc\Big( (E\of{l}+C\of{l} )\frac{M}{2\pi} \Big) + \frac{M}{2} \sinc\Big( (E\of{l}-C\of{l} )\frac{M}{2\pi} \Big)  \right)^2  \\
&=&  \frac{M-1}{M+1}  \cdot \frac{M^2}{4} \textstyle \cos^2(D\of{l}) \left( \sinc\Big( (E\of{l}+C\of{l} )\frac{M}{2\pi} \Big) +  \sinc\Big( (E\of{l}-C\of{l} )\frac{M}{2\pi} \Big)  \right)^2        \label{eq:hardpart4}
\end{eqnarray}
Finally, by combining \eqref{bnd1}, \eqref{hardpart1}, and \eqref{hardpart4} and plugging back in defined expressions for $C\of{l}$,$D\of{l}$, and $E\of{l}$ (equations \eqref{Cdef}, \eqref{Ddef}, and \eqref{Edef}), we find
\begin{eqnarray}
\lefteqn{\abss{ \displaystyle\sum_{n=-M/2}^{M/2-1} \hat{F}_{n,n}\of{l} W(\lambdal_{n,n},\betal ) }} \nonumber \\
&\geq& \frac{M-1}{M+1} \cdot \frac{M^2}{4} \textstyle \cos^2(\frac{B}{T}(1-\hat{a}\of{l})(\hat{\gamma}\of{l}-\gamma\of{l}) ) \nonumber \\
&     & \times \mbox{ } \left(  \sinc\left( \left( \frac{\pi f_c T}{1-\ahat}+\frac{\pi f_c T + B}{1-\al} \right) \frac{M ( \ahat-\al )}{2\pi} \right)  \right. \nonumber \\
&     & \left.  + \mbox{ }  \sinc\left( \left( \frac{\pi f_c T}{1-\ahat}+\frac{\pi f_c T - B}{1-\al} \right) \frac{M ( \ahat-\al )}{2\pi} \right)  \right)^2.
\end{eqnarray}
Since $\cos(x)$ and $\sinc(x)$ are symmetric around $x=0$ and we have assumed  $|\ahat| \leq \frac{1}{2M}$, $|\ahat -\al |\leq \frac{\Delta_a}{2}$, and $|\gammahat-\gammal | \leq \frac{\Delta_\gamma}{2}$, we can now write
\begin{eqnarray}
\lefteqn{\abss{ \displaystyle\sum_{n=-M/2}^{M/2-1} \hat{F}_{n,n}\of{l} W(\lambdal_{n,n},\betal ) }} \nonumber \\
&\geq& \frac{M-1}{M+1} \cdot \frac{M^2}{4} \textstyle \cos^2(\frac{B\Delta_\gamma}{2T}(1+\frac{1}{M}) ) \nonumber \\
&     & \times \mbox{ } \left( \sinc\left( \left( \frac{\pi f_c T}{1-\frac{1}{2M}}+\frac{\pi f_c T + B}{1-\al} \right) \frac{M  \Delta_a }{4\pi} \right)  \right. \nonumber \\
&     & \left.  + \mbox{ }  \sinc\left( \left( \frac{\pi f_c T}{1-\frac{1}{2M}}+\frac{\pi f_c T - B}{1-\al} \right) \frac{M  \Delta_a }{4\pi} \right)  \right)^2.
\end{eqnarray}

\chapter{Inner Product Evaluation (Doppler Case)} 
\label{app:ssdopp}

To simplify the inner product $\langle \ssl(t), \sdopp(t) \rangle $, we first express it using \eqref{innerprod}, \eqref{sdef}, and \eqref{sdopp}.
\begin{eqnarray}
\lefteqn{ \langle \ssl(t), \sdopp(t) \rangle } \nonumber \\
 &=& \int \ssl(t) \sdoppc(t) \,dt  \nonumber \\
&=&\int e^{j 2 \pi f_c ((1-\adopp )\gammadopp - (1-\al )\gammal +(\adopp-\al )t)} \sqrt{1-\al} \nonumber \\
& & \times \mbox{ }  \displaystyle\sum_{r=-M/2}^{M/2-1} b_r p((1-\al)t -r T -(1-\al)\gammal )  \displaystyle\sum_{q=-M/2}^{M/2-1} b_q^* p(t -q T -\gammadopp ) \,dt \nonumber \\ \label{eq:wideband1_dopp}
&=&  \sqrt{1-\al} e^{ j 2 \pi f_c ((1-\adopp )\gammadopp - (1-\al )\gammal ) } \displaystyle\sum_{r=-M/2}^{M/2-1} \displaystyle\sum_{q=-M/2}^{M/2-1} b_r b_q^* \nonumber \\
& & \times \mbox{ } \int e^{j 2 \pi f_c (\adopp - \al ) t  } p((1-\al)t -r T -(1-\al)\gammal )  p(t -q T -\gammadopp )    \,dt \label{eq:hardint_dopp}
\end{eqnarray}
To further simplify \eqref{hardint_dopp}, it is helpful to approximate the term $ e^{ j 2 \pi f_c (\adopp - \al ) t }$ with a time invariant constant in order to evaluate the integral.  To do this, well note that the term $p(t -q T -\gammadopp )$ has its maximum value when $t= qT + \gammadopp$, but diminishes to 0 as $\abs{t -q T -\gammadopp}$ increases.  Likewise, the term $p((1-\al)t -r T -(1-\al)\gammal )$ has its maximum value when $t= \frac{rT}{1-\al}+\gammal$ and diminishes to 0 as $\abs{(1-\al)t -r T -(1-\al)\gammal}$ increases.  We define $\tilde{F}_{q,r}\of{l}$ as the value of $ e^{ j 2 \pi f_c (\adopp - \al ) t }$ when $t$ is half way between the peak values $t_1 \defn qT+ \gammadopp$ and $t_2 \defn \frac{rT}{1-\al}+\gammal$.
\begin{equation}
\tilde{F}_{q,r}\of{l} \defn e^{j 2 \pi f_c (\adopp - \al ) ( \frac{qT}{2} +  \frac{rT}{2(1-\al )} +\frac{\gammadopp + \gammal}{2}  )} \label{eq:Ftildef}
\end{equation}
The term $ e^{ j 2 \pi f_c (\adopp - \al ) t }$ in \eqref{hardint_dopp} can be well approximated by $\tilde{F}_{q,r}\of{l}$ over the periods of $t$ where both $p(t -q T -\gammadopp )$ and $p((1-\al)t -r T -(1-\al)\gammal ) $ are not negligible.  This in part is because we make the mild assumptions that the value of $\al$ to be relatively small (i.e. $\al < 0.01$) and that $\ahat \approx \al$.  It is also because we have assumed that the signal has a high fractional bandwidth (i.e.  $\frac{1}{T f_c} \approx 1$).  By using $\tilde{F}_{q,r}\of{l}$ to approximate $e^{ j 2 \pi f_c (\adopp - \al ) t }$, we can express \eqref{hardint_dopp} as follows
\begin{eqnarray}
\lefteqn{ \langle \ssl(t), \sdopp(t) \rangle } \nonumber \\
&=&  \sqrt{1-\al}  e^{ j 2 \pi f_c ((1-\adopp )\gammadopp - (1-\al )\gammal ) } \displaystyle\sum_{r=-M/2}^{M/2-1} \displaystyle\sum_{q=-M/2}^{M/2-1} b_r b_q^* \nonumber \\
& & \times \mbox{ } \tilde{F}_{q,r} \int  p((1-\al)t -r T -(1-\al)\gammal ) p(t -q T -\gammadopp )    \,dt.\label{eq:wideband1x_dopp}
\end{eqnarray}
Now if we make the substitution $v=t -q T -\gammadopp$, \eqref{wideband1x_dopp} becomes
\begin{eqnarray}
\lefteqn{ \sqrt{1-\al} e^{j 2 \pi f_c ((1-\adopp )\gammadopp -(1-\al )\gammal )} \displaystyle\sum_{r=-M/2}^{M/2-1} \displaystyle\sum_{q=-M/2}^{M/2-1} b_r b_q^* \tilde{F}_{q,r}  } \nonumber \\
\lefteqn{  \times \mbox{ } \int  p(v) p((1-\al )v +(1-\al ) qT-rT +(1-\al ) (\gammadopp -\gammal ) ) \,dv } \nonumber \\
&=& \sqrt{\betaldopp} e^{j 2 \pi f_c ((1-\adopp )\gammadopp -(1-\al )\gammal )} \displaystyle\sum_{r=-M/2}^{M/2-1} \displaystyle\sum_{q=-M/2}^{M/2-1}  b_r b_q^* \tilde{F}_{q,r}  \nonumber \\
& & \times \mbox{ } \int \ p(v) p(\betaldopp v +\betaldopp qT-rT+\betaldopp(\gammadopp -\gammal)) \,dv, \label{eq:wideband2_dopp}
\end{eqnarray}
where $\betaldopp \defn 1-\al$.  To simplify \eqref{wideband2_dopp} further, we use the wide band ambiguity function defined in \eqref{wbdef}, so that we can express \eqref{wideband2_dopp} as 
\begin{eqnarray}
\langle \ssl(t), \sdopp(t) \rangle &=& \displaystyle\sum_{r=-M/2}^{M/2-1} \displaystyle\sum_{q=-M/2}^{M/2-1} b_r b_q^* \tilde{F}_{q,r} e^{j 2 \pi f_c ((1-\adopp )\gammadopp -(1-\al )\gammal )} \nonumber \\
 & & \times W(qT - \frac{rT}{\betaldopp}+\gammadopp-\gammal,\betaldopp ). \label{eq:thingxxx_old_dopp}
\end{eqnarray}

\chapter{Lower bound on $\abss{ \displaystyle\sum_{n=-M/2}^{M/2-1} \tilde{F}_{n,n}\of{l} W(\lambdaldopp_{n,n},\betaldopp ) }$}
\label{app:fwdoppbound}
The primary goal of \secref{error_doppler} is to find an insightful expression which relates an upper bound of the error metric $\metricdopp$ with the the grid spacing ($\Delta_a$ and $\Delta_\gamma$).  With this goal in mind, it is necessary to find a lower bound on the expression $\abss{ \displaystyle\sum_{n=-M/2}^{M/2-1} \tilde{F}_{n,n}\of{l} W(\lambdaldopp_{n,n},\betaldopp ) } $ in terms of the grid spacing.  For our analysis, we assume that $|a\of{l}|\leq\frac{1}{2M}$ and $|\hat{a}\of{l}|\leq\frac{1}{2M}$.  We also assume that the bounds on the quantities $|\hat{a}\of{l}-a\of{l}| $ and $|\hat{\gamma}\of{l}-\gamma\of{l}|$ are slightly tighter than the bounds assumed in \appref{omegabound}, so that 
\begin{eqnarray}
  |\tilde{a}\of{l}-a\of{l}|
  &\leq& \textstyle \frac{1}{2M}(1-\frac{1}{2M})\frac{1}{2f_cT} \label{eq:aadopp}\\
  |\tilde{\gamma}\of{l}-\gamma\of{l}|
  &\leq& \textstyle \frac{T}{4}~\frac{2M-2}{2M-1} . 
\end{eqnarray}
Note the presence of the $f_c T$ term in \eqref{aadopp},
and recall that $f_c \geq \frac{1}{2T}$ (assuming that the carrier
frequency is centered in the signal passband) which implies that
$2f_c T \geq 1$.

We start by recalling the definitions of $\tilde{F}\of{l}_{n,n} $ and $W(\lambdaldopp_{n,n},\betaldopp)$ given in \eqref{Ftildef} and \eqref{wbdef}, respectively.  We now notice that 
\begin{eqnarray}
  \left| \sum_{n=-M/2}^{M/2-1} \tilde{F}\of{l}_{n,n} 
  	W(\lambdaldopp_{n,n},\betaldopp) \right|^2
  &=& \left| \sum_{n=-M/2}^{M/2-1} \tilde{K}\of{l}_{n} 
  	W(\lambdaldopp_{n,n},\betaldopp) \right|^2
\end{eqnarray}
where 
\begin{eqnarray}
  \tilde{K}_{n}\of{l}
  &\defn& e^{j \tilde{\chi}_n\of{l}} \\
  \tilde{\chi}_n\of{l}
  &\defn& \textstyle \pi f_c(\tilde{a}\of{l}-a\of{l})
  	(nT+\frac{nT}{1-a\of{l}}) .
\end{eqnarray}
Using the Euler expansion 
$\tilde{K}_{n}\of{l} = \cos(\tilde{\chi}_n\of{l})+j\sin(\tilde{\chi}_n\of{l})$, we get
\begin{eqnarray}
  \left| \sum_{n=-M/2}^{M/2-1} \tilde{K}\of{l}_{n} 
  	W(\lambdaldopp_{n,n},\betaldopp) \right|^2
  &=&\left| \sum_{n=-M/2}^{M/2-1} W(\lambdaldopp_{n,n},\betaldopp) 
	[\cos(\tilde{\chi}_n\of{l})+j\sin(\tilde{\chi}_n\of{l})]\right|^2 \\
  &\geq& \left( \sum_{n=-M/2}^{M/2-1} W(\lambdaldopp_{n,n},\betaldopp) 
	\cos(\tilde{\chi}_n\of{l})\right)^2 ,		\label{eq:bnd1dopp}
\end{eqnarray}
where the inequality follows because 
$W(\lambdaldopp_{n,n},\betaldopp)\in\Real$. 
The latter bound is actually quite tight because the quantity
$W(\lambdaldopp_{n,n},\betaldopp)$ is symmetric around $n=0$,
the quantity $\sin(\tilde{\chi}_n\of{l})$ is antisymmetric around $n=0$,
and the summation is almost symmetric around $n=0$.

Next we note that, under our assumptions, we have
\begin{eqnarray}
  |\lambdaldopp_{n,n}|
  &=& \textstyle |\frac{1}{1-a\of{l}}nT(1-a\of{l}-1)
          +(\hat{\gamma}\of{l}-\gamma\of{l})| \\
  &\leq& \textstyle \frac{1}{1-\frac{1}{2M}}|n|T \cdot
  	\frac{1}{2M}
          +\frac{T}{4}~\frac{2M-2}{2M-1} \\
  &\leq& \textstyle \frac{T}{2}			\label{eq:lamdopp}\\
  |\tilde{\chi}_n\of{l}|
  &\leq& \textstyle \pi f_c \cdot \frac{1}{2M}(1-\frac{1}{2M})\frac{1}{2f_cT}
  	\cdot |n|T(1+\frac{1}{1-\frac{1}{2M}})
  ~\leq~ \textstyle \frac{\pi|n|}{2M} \\
  &\leq& \textstyle \frac{\pi}{4} , \label{eq:obydopp}
\end{eqnarray}
where \eqref{lamdopp} and \eqref{obydopp} follow from the fact that $|n| \leq \frac{M}{2}$.  Due to the range of $|\tilde{\chi}_n\of{l}|$, the term $\cos(\tilde{\chi}_n\of{l})$ is
clearly non-negative over the range of $n$ in \eqref{bnd1dopp}. 
We now show that $W(\lambdaldopp_{n,n},\betaldopp)= \frac{\min(1,\betaldopp)}{\sqrt{\betaldopp}} \sinc(\frac{\min(1,\betaldopp)}{T}\lambdaldopp_{n,n})$
is also non-negative over those $n$.
First, notice that 
\begin{eqnarray}
\textstyle
\frac{\min(1,\betaldopp)}{\sqrt{\betaldopp}}
~=~\min(\frac{1}{\sqrt{\betaldopp}},\sqrt{\betaldopp}) 
&\geq& \textstyle
\sqrt{1-\frac{1}{M}}
\end{eqnarray}
because $1-\frac{1}{M} \leq \betaldopp
\leq 1+\frac{1}{M}$.
Then, notice that the argument of the $\sinc$ obeys
\begin{eqnarray}
  \textstyle \frac{\min(1,\betaldopp)}{T}\lambda\of{l}_{n,n}
  &\leq& \textstyle \frac{\lambda\of{l}_{n,n}}{T} 
  ~\leq~ \textstyle \frac{1}{2} .
\end{eqnarray}
Because both $\cos(\tilde{\chi}_n\of{l})$ and $W(\lambdaldopp_{n,n},\betaldopp)$ 
are non-negative over the range of $n$ in \eqref{bnd1dopp}, 
we can lower-bound \eqref{bnd1dopp} by plugging 
in a lower bound for $W(\lambdaldopp_{n,n},\betaldopp)$ in the summation.
Next, we derive this bound.

Because $\sinc(x)$ decreases as $x$ increases for $|x|\leq \frac{1}{2}$
\begin{equation}
  W(\lambdaldopp,\betaldopp)  \geq \textstyle \sqrt{1-\frac{1}{M}} \sinc\Big(\frac{\lambdaldopp_{n,n}}{T} \Big).
\end{equation}

Applying this inequality,
\begin{eqnarray}
  \lefteqn{ 
  	\left( \sum_{n=-M/2}^{M/2-1} W(\lambdaldopp_{n,n},\betaldopp) 
	\cos(\tilde{\chi}_n\of{l})\right)^2 	}\nonumber\\
  &\geq& \left( 1-\frac{1}{M} \right) \Bigg( \sum_{n=-M/2}^{M/2-1} \textstyle
  	\sinc\Big( \frac{\lambdaldopp_{n,n}}{T} \Big) \cos\Big( \pi f_c(\tilde{a}\of{l}-a\of{l})nT
  	(1+\frac{1}{1-\al}) \Big) \Bigg)^2 . \label{eq:Wbound1newdopp}
\end{eqnarray}
As we did in \appref{fwbound}, we approximate $\sinc(x) \approx \cos(Bx)$, so that we can write \eqref{Wbound1newdopp} as 
\begin{eqnarray}
  \lefteqn{ 
  	\left( \sum_{n=-M/2}^{M/2-1} W(\lambdaldopp_{n,n},\betaldopp) 
	\cos(\tilde{\chi}_n\of{l})\right)^2 	}\nonumber\\
  &\geq& \left( 1-\frac{1}{M} \right) \Bigg( \sum_{n=-M/2}^{M/2-1} \textstyle
  	\cos\Big( \frac{\Bdopp \lambdaldopp_{n,n}}{T} \Big) \cos\Big( \pi f_c(\tilde{a}\of{l}-a\of{l})nT
  	(1+\frac{1}{1-\al}) \Big) \Bigg)^2 \\
   &=& \left( 1-\frac{1}{M} \right) \Bigg( \sum_{n=-M/2}^{M/2-1} \textstyle
  	\cos\Big( \frac{-\Bdopp n \al}{1-a\of{l}}+\frac{\Bdopp}{T}(\gammadopp-\gammal) \Big) \nonumber \\
   & & \times \mbox{ } \cos\Big(  \pi f_c(\tilde{a}\of{l}-a\of{l})nT
  	(1+\frac{1}{1-\al})  \Big) \Bigg)^2 \\
   &=& \left( 1-\frac{1}{M} \right) \left( \sum_{n=-M/2}^{M/2-1} \textstyle
  	\cos\Big(\Cdopp n + \Ddopp  \Big) 
  	\cos\Big( \Edopp n  \Big) 
	\right)^2, \label{eq:hardpart1dopp}
\end{eqnarray}

where
\begin{eqnarray}
\Cdopp &\defn& \frac{-\Bdopp \al}{1-a\of{l}} \label{eq:Cdefdopp}  \\
\Ddopp &\defn& \frac{\Bdopp}{T}(\gammadopp-\gammal) \label{eq:Ddefdopp} \\
\Edopp &\defn& \pi f_c(\tilde{a}\of{l}-a\of{l})T
  	(1+\frac{1}{1-\al})     . \label{eq:Edefdopp}  
\end{eqnarray}
Equation \eqref{hardpart1dopp} can be recognized as a Riemann-sum approximation of the following continuous integral
\begin{eqnarray}
\lefteqn{ \left( 1-\frac{1}{M} \right) \left( \int_{-M/2}^{M/2} \textstyle
  	\cos\Big(\Cdopp t + \Ddopp  \Big) 
  	\cos\Big( \Edopp t  \Big) \,dt   \right)^2    } \nonumber \\
&=& \left( 1-\frac{1}{M} \right) \left( \int_{-M/2}^{M/2} \textstyle
  	\left( \cos\Big(\Cdopp t\Big) \cos\Big(\Ddopp \Big) \right. \right. \nonumber \\
& & \left. \left. -\mbox{ } \sin\Big(\Cdopp t\Big) \sin\Big(\Ddopp \Big)\right) \cos\Big( \Edopp t  \Big) \,dt   \right)^2 \label{eq:trigtrick1dopp} \\
&=& \left( 1-\frac{1}{M} \right) \left( \int_{-M/2}^{M/2} \textstyle
  	 \cos\Big(\Cdopp t\Big) \cos\Big(\Ddopp \Big)  \cos\Big( \Edopp t  \Big) \,dt   \right)^2 	\label{eq:hardpart2dopp}
\end{eqnarray}
Equation \eqref{trigtrick1dopp} follow from the trig identity $\cos(x+y) = \cos(x)\cos(y)-\sin(x)\sin(y) $ and \eqref{hardpart2} follows from the fact that $\int_{-M/2}^{M/2} \textstyle \sin\Big(\Cdopp t\Big) \cos\Big( \Edopp t  \Big) \,dt   = 0$.  We can now use the trig identity $\cos(x)\cos(y)=\frac{1}{2}\cos(x+y)+\frac{1}{2}\cos(x-y) $ to rewrite \eqref{hardpart2dopp} as follows
\begin{eqnarray}
\lefteqn{ \left( 1-\frac{1}{M} \right) \left( \int_{-M/2}^{M/2} \textstyle
  	 \cos\Big(\Cdopp t\Big) \cos\Big(\Ddopp \Big)  \cos\Big( \Edopp t  \Big) \,dt   \right)^2  } \nonumber \\
&=& \left( 1-\frac{1}{M} \right) \textstyle \cos^2(\Ddopp) \displaystyle \left( \int_{-M/2}^{M/2} \textstyle \frac{1}{2}\cos\Big( (\Edopp+\Cdopp )t \Big) +\frac{1}{2}\cos\Big( (\Edopp-\Cdopp )t \Big) \,dt     \right)^2 \nonumber \\
&=&  \left( 1-\frac{1}{M} \right) \textstyle \cos^2(\Ddopp) \left( \frac{ \sin\Big( (\Edopp+\Cdopp )\frac{M}{2} \Big)}{\Edopp+\Cdopp} + \frac{ \sin\Big( (\Edopp-\Cdopp )\frac{M}{2} \Big)}{\Edopp-\Cdopp} \right)^2  \nonumber \label{eq:hardpart3dopp}  \\
&=&  \left( 1-\frac{1}{M} \right) \textstyle \cos^2(\Ddopp) \left(\frac{M}{2} \sinc\Big( (\Edopp+\Cdopp )\frac{M}{2\pi} \Big) + \frac{M}{2} \sinc\Big( (\Edopp-\Cdopp )\frac{M}{2\pi} \Big)  \right)^2 \nonumber  \\
&=&  \left( 1-\frac{1}{M} \right)~\frac{M^2}{4} \textstyle \cos^2(\Ddopp) \nonumber \\
& & \times \mbox{ }\left(\sinc\Big( (\Edopp+\Cdopp )\frac{M}{2\pi} \Big) + \sinc\Big( (\Edopp-\Cdopp )\frac{M}{2\pi} \Big)  \right)^2  \label{eq:hardpart4dopp}
\end{eqnarray}
Finally, by combining \eqref{bnd1dopp}, \eqref{hardpart1dopp}, and \eqref{hardpart4dopp} and plugging back in defined expressions for $\Cdopp$,$\Ddopp$, and $\Edopp$ (equations \eqref{Cdefdopp}, \eqref{Ddefdopp}, and \eqref{Edefdopp}), we find
\begin{eqnarray}
\lefteqn{\abss{ \displaystyle\sum_{n=-M/2}^{M/2-1} \tilde{F}_{n,n}\of{l} W(\lambdaldopp,\betaldopp ) }} \nonumber \\
&\geq& \left( 1-\frac{1}{M} \right) ~\frac{M^2}{4} \textstyle \cos^2(\frac{\Bdopp}{T}(\gammadopp-\gammal) ) \nonumber \\
&     & \times \mbox{ } \left(  \sinc\left( \left( \pi f_c (\adopp-\al)T \left(1+\frac{1}{1-\al} \right) -\frac{\Bdopp \al}{1-\al}  \right) \frac{M}{2\pi} \right)  \right. \nonumber \\
&     & \left.  + \mbox{ }  \sinc\left( \left( \pi f_c (\adopp-\al)T \left(1+\frac{1}{1-\al} \right) +\frac{\Bdopp \al}{1-\al}  \right) \frac{M}{2\pi} \right)  \right)^2.
\end{eqnarray}
Since $\cos(x)$ and $\sinc(x)$ are symmetric around $x=0$ and we have assumed\\ $|\ahat -\al |\leq \frac{\Delta_a}{2}$ and $|\gammahat-\gammal | \leq \frac{\Delta_\gamma}{2}$, we can now write
\begin{eqnarray}
\lefteqn{\abss{ \displaystyle\sum_{n=-M/2}^{M/2-1} \tilde{F}_{n,n}\of{l} W(\lambdaldopp,\betaldopp ) }} \nonumber \\
&\geq& \left( 1-\frac{1}{M} \right) ~\frac{M^2}{4} \textstyle \cos^2(\frac{\Bdopp\Delta_\gamma}{2T} ) \nonumber \\
&     & \times \mbox{ } \left(  \sinc\left( \left( \frac{\pi f_c \Delta_a T}{2} \left(1+\frac{1}{1-\al} \right) -\frac{\Bdopp \al}{1-\al}  \right) \frac{M}{2\pi} \right)  \right. \nonumber \\
&     & \left.  + \mbox{ }  \sinc\left( \left( \frac{\pi f_c \Delta_a T}{2} \left(1+\frac{1}{1-\al} \right) +\frac{\Bdopp \al}{1-\al}  \right) \frac{M}{2\pi} \right)  \right)^2.
\end{eqnarray}

\chapter{Upper bound on $\omegaldopp$}
\label{app:omegabounddopp}

In this appendix, we derive an expression $\omegaldopp$ (defined in \eqref{omegadefdopp}) in terms of $M$.  To do this we first note that, using \eqref{wub}, we can upperbound $\omegaldopp$ as follows

\begin{eqnarray}
\omegal &\defn & \frac{1}{M^2} \sum_{r =-M/2}^{M/2-1} \sum_{e \neq r =-M/2}^{M/2-1} W^2(\lambdaldopp_{e,r},\betaldopp ) \\
& \leq & \frac{1}{M^2} \sum_{r =-M/2}^{M/2-1} \sum_{e \neq r =-M/2}^{M/2-1} \frac{T^2}{\pi^2\tilde{\lambda}_{e,r}^{(l)2} \betal} \label{eq:omegaub1dopp}
\end{eqnarray}

Using our assumption $\al \leq \frac{1}{2M}$ and $\ahat \leq \frac{1}{2M}$, we can approximate $\frac{1}{\betaldopp} = \frac{1}{1-\al} \approx 1 + \al$ for large $M$.  Using this approximation, we can express $\lambdaldopp_{e,r}$ (defined in \eqref{lambda_dopp}) as follows
\begin{equation}
\lambdal_{e,r} = rT - eT -\al eT - T \deltaldopp, \label{eq:lambdaapproxdopp}
\end{equation}
where $\deltaldopp \defn \frac{1}{T} \{ \gammal - \gammadopp \} \leq \frac{1}{4}$.  Using the change of variables $m \defn e-r$ and combining \eqref{omegaub1dopp} and \eqref{lambdaapproxdopp}, it follows that
\begin{eqnarray}
\omegaldopp &\leq& \frac{1}{M^2 \pi^2 \betaldopp} \sum_{r =-M/2}^{M/2-1} \sum_{e \neq r =-M/2}^{M/2-1} \frac{1}{ (r - e -\al e - \deltaldopp  )^2} \\
&=& \frac{1}{M^2 \pi^2 \betaldopp} \sum_{m=1}^{M-1}
  \sum_{e=-M/2+m}^{M/2-1} \frac{1}{\big(
  	m - \al e + \deltaldopp \big)^2}
	\nonumber\\&&\mbox{}
  +\frac{1}{M^2 \pi^2 \betaldopp} \sum_{m=1-M}^{-1}
  \sum_{e=-M/2}^{M/2+m} \frac{1}{\big(
  	m - \al e + \deltaldopp \big)^2}\\
  &\leq& \frac{2}{M^2 \pi^2 \betaldopp} \sum_{m=1}^{M-1}
  \sum_{e=-M/2}^{M/2-1} \frac{1}{\big(
  	m - \al e + \deltaldopp \big)^2}.
\end{eqnarray}
Notice that, since $\abs{e} \leq \frac{M}{2}$, $\deltaldopp \leq \frac{1}{4}$, and $\ahat \leq \frac{1}{2M}$ then 
\begin{eqnarray}
  \big| -\al e + \deltaldopp \big|
  &\leq& \big| -\al e \big|
  	+ \big|\deltaldopp \big|
  = |\al|~|e| 
  	+ |\delta\of{l}| \\
  &\leq& \frac{1}{2M}~\frac{M}{2}
  	+ \frac{1}{4}
  = \frac{1}{2}.
\end{eqnarray}

In this case
\begin{eqnarray}
  \omegaldopp
  &\leq& \frac{2}{M \pi^2 \betaldopp} \sum_{m=1}^{M-1} 
  	\frac{1}{( m - 0.5 )^2} \\
  &=& \frac{2}{M \pi^2 \betaldopp} \left( 4+\sum_{m=2}^{M-1} \frac{1}{(m-0.5)^2} \right) \\
  &\leq& \frac{2}{M \pi^2 \betaldopp} \left( 4+\sum_{m=1}^{M-2} \frac{1}{m^2} \right) \\
  &\leq& \frac{2 (4+\pi^2/6)}{M \pi^2 \betaldopp} 
\end{eqnarray}
using the fact that $\displaystyle\sum_{m=1}^{M} \frac{1}{m^2} \leq \frac{\pi^2}{6}$.
Finally, since 
\begin{eqnarray}
  \frac{1}{\betaldopp} 
  &=& \frac{1}{1-\al} \\
  &\approx&  1+\al \\
  &\leq& 1 + \frac{1}{2M} ~=~\frac{2M+1}{2M},
\end{eqnarray}
we have
\begin{eqnarray}
  \omegal
  &\leq& \frac{2 (4+\pi^2/6)}{M \pi^2} ~ \frac{2M+1}{2M}
\end{eqnarray}
Now if we say that $M\geq 10$, we know that $\frac{2M+1}{2M}\leq 
\frac{21}{20}$, so that
\begin{eqnarray}
  \omegal
  &\leq& \frac{2 (4+\pi^2/6)}{M \pi^2}~\frac{21}{20} \leq \frac{1.2011}{M}.
\end{eqnarray}

\bibliography{Brian_bib}
\bibliographystyle{unsrt}

\end{document}